\documentclass[aps,10pt,eqsecnum,preprint,nofootinbib,superscriptaddress,a4paper]{revtex4}
\usepackage{amssymb,amsmath,amsthm,graphicx,amscd,ulem}
\usepackage[mathscr]{eucal}
\usepackage{mathtools, enumerate,color,comment, cancel}
\usepackage{bm,slashed}
\usepackage[hidelinks]{hyperref}

\usepackage{orcidlink}

\usepackage{xcolor}

\DeclareMathSymbol{\shortminus}{\mathbin}{AMSa}{"39}
\newcommand{\shortpm}{\scalebox{0.6}[0.7]{$\pm$}}

\begin{document}

\title{Graviton physics: Quantum field theory of gravitons, graviton noise and gravitational decoherence --  a concise tutorial}    
\author{Jen-Tsung Hsiang\orcidlink{0000-0002-9801-208X}}
\email{cosmology@gmail.com}
\affiliation{College of Electrical Engineering and Computer Science, National Taiwan University of Science and Technology, Taipei City, Taiwan 106, R.O.C.}
\author{Hing-Tong Cho\orcidlink{0000-0002-8497-1490}}
\email{htcho@mail.tku.edu.tw}
\affiliation{Department of Physics, Tamkang University, New Taipei City, Taiwan 251, R.O.C.}
\author{Bei-Lok Hu\orcidlink{0000-0003-2489-9914}}
\email{blhu@umd.edu}
\affiliation{Maryland Center for Fundamental Physics and Joint Quantum Institute,  University of Maryland, College Park, Maryland 20742, USA}

\begin{abstract}
The detection of gravitational waves in 2015 ushered in a  new era of gravitational wave (GW) astronomy capable of probing into the strong field dynamics of black holes and neutron stars. It has opened up an exciting new window for laboratory and space tests of Einstein's theory of {\it classical general relativity} (GR). In recent years there are two interesting proposals aimed at revealing the quantum natures of perturbative gravity: 1) theoretical predictions in how graviton noise from the early universe after the vacuum of the gravitational field  was strongly squeezed by inflationary expansion; 2) experimental proposals using the {\it quantum entanglement} between two masses each in a superposition (gravitational cat, or gravcat)  state. The first proposal invokes the stochastic properties of {\it quantum fields} (QF), the second invokes a key concept of {\it quantum information} (QI). An equally basic and interesting idea is to ask whether and how gravity might be responsible for a quantum system becoming classical in appearance, known as {\it gravitational decoherence}. Decoherence due to gravity is of special interest because gravity is universal, meaning, gravitational interaction is present for all massive objects.   This is an important issue in macroscopic quantum phenomena (MQP) which also underlies many proposals  in alternative quantum theories (AQT).  To fully appreciate or conduct research in these exciting developments requires a working knowledge in classical GR, QF theory and QI plus some familiarity with stochastic processes (SP), namely, noise in quantum fields and in decohering environments. Traditionally a new researcher may be conversant in one or two of  these four subjects: GR, QFT, QI, SP, depending on his/her background.  This tutorial attempts to provide the necessary connective tissues between them, helping an engaging reader from any one of these four subjects to leapfrog to the frontier of these interdisciplinary research topics. In the present version we shall treat the three topics listed in the title, save gravitational entanglement, because, despite the high attention some recent experimental proposals have drawn,  its nature and implications proclaimed in relation to quantum gravity still contain many controversial elements.

\end{abstract}

\maketitle

\baselineskip=18pt
\allowdisplaybreaks

\tableofcontents

\section*{Matching the main themes with different readers' backgrounds and needs}

We assume three different communities of readers with a common goal, that of trying to understand the newly arisen important topics of graviton noise and gravitational decoherence in relation to gravitational radiation. These readers may come with different backgrounds, such as in  A) classical general relativity,  specializing in gravitational wave physics or astronomy, who are curious about what LIGO / LISA types of experiments can be of use for the detection of these effects;  B) quantum field or particle theorists who know field theory well but wonder how quantum noise arises and how these stochastic quantities and  related effects can impact on possibly observable gravitational quantum phenomena; C) atomic-optical and condensed matter physicists with  working knowledge in open quantum systems and quantum information issues who are curious about how this knowhow can be applied to the investigation of  gravitating systems, possibly even bearing on quantum gravity issues.   

As remarked in the Abstract, graviton physics invokes GR and QFT, gravitational decoherence invokes GR (we are not considering alternative quantum or gravity theories here) and concepts and techniques of open quantum systems (OQS). Thus we can see some crisscrossing of the four ingredients: classical gravity (GR), quantum field (QFT),  quantum information (QI) and quantum noise (OQS).  In the environment-induced decoherence (QI) conceptual scheme, the noise in the  graviton field (GR + QFT) acting as an environment to a quantum system is what induces decoherence of that quantum system.  Thus quantum - classical transition or correspondence is one of our underlying themes. A full range of coherent topics would be starting from graviton noise, i.e., fluctuations in the quantized (perturbative) gravitational field and ending with classical gravitational radiation, with radiation reaction considerations. This issue has been addressed for scalar and electromagnetic fields so we shall only mention the conceptual pathways here without going into the details.  

After this dissection of the two main themes, readers with different backgrounds can  see from the Tables of Contents which sections or topics they can skip over and which ones are more useful for their specific purposes. 

Contentswise,  we start with gravitational wave physics based on perturbation theory (not fluctuations)  off of the Minkowski spacetime. (Note the often ignored yet important difference between perturbations which obey deterministic equations of motion, and fluctuations/noises which are stochastic variables). Then we quantize the field where spin-2 gravitons appear. We show that each of the two polarizations obeys an equation of the same form as a massless minimally-coupled scalar field. We then analyze the Green's functions and the quantum states of this field. We point out the two most important Green's functions for our problems are the retarded and the Hadamard functions, the latter being responsible for noise in the quantum field. We discuss four types of {Gaussian} quantum states: the vacuum, coherent, squeezed vacuum and squeezed coherent states.  We then argue how to go from quantum field theory to classical radiation theory, via the coherent state,  but hasten to point out not to ignore the important quantum features of coherent states in the consideration of quantum to classical transition. Hereby we show how noise and fluctuations can be gleaned off from a full quantum field theoretical treatment, discouraging anything put in by hand at will as doing so will violate the innate structure of the theory such as manifested in fluctuation-dissipation relations. 

As a tutorial we do away with the explanation of motivations for specific topics and the vast background literature but dive right into developing the necessary ingredients. We shall refer to the textbooks and research papers where our materials come from, where the reader can trace back for a  wider coverage or for further details.  However, a  broader perspective is necessary to avoid misinterpretations of the bigger issues. Therefore,  in the final section we discuss how the topics presented here bear on graviton physics, the quantum nature of gravity and quantum gravity. These are old issues of fundamental significance but they have taken on a new life with the additional consideration of quantum information issues and stochastic properties of quantum fields. The focus is on the difference between {\bf the physics of gravitons} which are the quantized weak perturbations of spacetime and {\bf quantum gravity} which refers to theories concerning the basic constituents and fundamental structures of spacetime. Gravitons have been present since the time when the manifold structure of spacetime began to take shape, as they are the quanta of its weak perturbations and are in principle detectable at today's low energy, whereas quantum gravity is nonperturbative,  background independent and involves the physics at the Planck energy and above. Their consequences can only be deduced  indirectly  from observations of the early universe or from black holes' quantum processes below the Planck energy. We want the readers to understand there is a big conceptual divide and technical challenge between graviton physics and quantum gravity. In particular, recent proposals of laboratory experiments using setups aided by quantum information considerations such as entanglement can help to understand the quantum nature of perturbative gravity or graviton physics but bears little with quantum gravity proper. 

Stylistically, we try to present just enough materials from each base, in some cases starting from familiar textbook materials,  to enable an uninitiated reader to comprehend these interdisciplinary subjects with some economy in effort. For this reason our presentation may appear crisp here, compact there, even pedestrian for some, depending on what topics the reader is more familiar with, or less so,  but there is no sacrifice of rigor or accuracy. We hope the pedestrian bridges we build here between these scenic attractions are smoothly paved and can enable the curious explorers or serious researchers to take an easy stroll crisscrossing over them while enjoying the pleasant views.     

\newpage

\section{Gravitational Waves: Metric perturbations off Minkowski spacetime}\label{S:bgfkdf}
Here we include a short review of the classical gravitational wave, adapted from the classic tome  ``Gravitation" by Misner, Thorne and Wheeler (MTW)~\cite{MTW73}.

\subsection{linear perturbations}
We consider a simpler case that the metric perturbations $h_{\mu\nu}$ are excited in background Minkowski spacetime, whose metric, denoted by $\eta_{\mu\nu}$, has the signature $(-,+,+,+)$. Thus we can write the full metric $g_{\mu\nu}$ as
\begin{equation}
	g_{\mu\nu}=\eta_{\mu\nu}+h_{\mu\nu}\,,
\end{equation}
in which we have supposed that the metric perturbation to be small $\lvert h_{\mu\nu}\rvert\ll1$. The condition $g_{\mu\alpha}g^{\alpha\nu}=\delta_{\mu}{}^{\nu}$ gives
\begin{align}
	g^{\mu\nu}&=\eta^{\mu\nu}-h^{\mu\nu}\,,&&\text{with}&h^{\mu\nu}&=\eta^{\mu\alpha}\eta^{\mu\beta}h_{\alpha\beta}\,.
\end{align}

The split of the full metric into the background and perturbation is not unique even the background spacetime is Minkowskian due to the gauge choices of $h_{\mu\nu}$. Suppose we make an infinitesimal change of the coordinate system 
\begin{equation}
	x'^{\mu}=x^{\mu}+\xi^{\mu}\,,
\end{equation}
and then in this new coordinate system, the metric becomes
\begin{align}
	g'_{\mu\nu}=\frac{\partial x^{\alpha}}{\partial x'^{\mu}}\frac{\partial x^{\beta}}{\partial x'^{\nu}}\,g_{\alpha\beta}=\eta_{\mu\nu}+h_{\mu\nu}-\xi_{\mu,\,\nu}-\xi_{\nu,\,\mu}\,,
\end{align}
and the metric perturbations will take a new form
\begin{equation}\label{E:tititfkgd}
	h'_{\mu\nu}=h_{\mu\nu}-\xi_{\mu,\,\nu}-\xi_{\nu,\,\mu}\,.
\end{equation}
Apparently the form of metric perturbations depends on the choice of the coordinate system. In other words, the metric perturbations are gauge-dependent. Thus we may choose a suitable gauge to simplify the calculations associated with the metric perturbations.

To the first-order metric perturbation, the Christoffel symbols are given by
\begin{equation}\label{E:rktugusidf}
	\Gamma^{\rho}_{\mu\nu}=\frac{\eta^{\rho\sigma}}{2}\bigl(h_{\mu\sigma,\,\nu}+h_{\nu\sigma,\,\mu}-h_{\mu\nu,\,\sigma}\bigr)\,,
\end{equation}
and then the Riemann tensor $R^{\mu}{}_{\nu\rho\sigma}=\partial_{\rho}\Gamma^{\mu}_{\nu\sigma}-\partial_{\sigma}\Gamma^{\mu}_{\nu\rho}+\Gamma^{\mu}_{\rho\alpha}\Gamma^{\alpha}_{\nu\sigma}-\Gamma^{\mu}_{\sigma\alpha}\Gamma^{\alpha}_{\nu\rho}$ has the form
\begin{align}\label{E:bbcklkfd}
	R_{\mu\nu\rho\sigma}=\frac{1}{2}\bigl(h_{\mu\sigma,\,\nu\rho}-h_{\mu\rho,\,\nu\sigma}+h_{\nu\rho,\,\mu\sigma}-h_{\nu\sigma,\,\mu\rho}\bigr)\,.
\end{align}
Consequently, the Ricci tensor $R_{\mu\nu}$ and the Ricci scalar $R$ become
\begin{align}
	R_{\mu\nu}&=\frac{1}{2}\bigl(h_{\mu}{}^{\alpha}{}_{,\,\nu\alpha}+h_{\nu}{}^{\alpha}{}_{,\,\mu\alpha}-h_{,\,\mu\nu}-\square h_{\mu\nu}\bigr)\,,&R&=h^{\alpha\beta}{}_{,\,\alpha\beta}-\square h\,,
\end{align}
where the trace of $h_{\mu\nu}$ is $h=h^{\alpha}{}_{\alpha}$.

The Einstein equation
\begin{equation}
	G_{\mu\nu}=8\pi T_{\mu\nu}\,,
\end{equation}	
with the Einstein tensor $G_{\mu\nu}=R_{\mu\nu}-\dfrac{1}{2}\,g_{\mu\nu}\,R$, implies that it may be convenient to introduce the trace-reversed form of the metric perturbations, 	
\begin{equation}
	\bar{h}_{\mu\nu}=h_{\mu\nu}-\frac{1}{2}\,\eta_{\mu\nu}\,h\,,
\end{equation}
with $\bar{h}=-h$. By the trace-reversed metric perturbations, the Einstein tensor is greatly simplied into
\begin{equation}
	G_{\mu\nu}=\frac{1}{2}\bigl(\bar{h}_{\mu}{}^{\alpha}{}_{,\,\nu\alpha}+\bar{h}_{\nu}{}^{\alpha}{}_{,\,\mu\alpha}-\square \bar{h}_{\mu\nu}-\eta_{\mu\nu}\,\bar{h}^{\alpha\beta}{}_{,\,\alpha\beta}\bigr)\,.
\end{equation}
Now we see the advantage of using trace-reversed metric perturbations. If we choose the Lorenz gauge $\bar{h}_{\mu\alpha,\,}{}^{\alpha}=0$, then the Einstein equation reduces to
\begin{equation}\label{E:itjireereaa}
	\square \bar{h}_{\mu\nu}=-16\pi T_{\mu\nu}\,.
\end{equation}
It has a form of the wave equation driven by the source on the righthand side. Note that in arriving at this form, we do not need the traceless condition $\bar{h}=0$, as is usually applied to fix the degrees of freedom in the polarization tensor below.

The trace-reversed metric perturbations $\bar{h}_{\mu\nu}$ transform according to
\begin{equation}\label{E:jgueussds}
	\bar{h}'_{\mu\nu}=\bar{h}_{\mu\nu}-\xi_{\mu,\nu}-\xi_{\nu,\mu}+\eta_{\mu\nu}\,\xi^{\alpha}{}_{,\,\alpha}\,.
\end{equation}
Thus under the gauge transformation \eqref{E:tititfkgd}, the Riemann tensor does not change
\begin{equation}
	R'_{\mu\nu\rho\sigma}=R_{\mu\nu\rho\sigma}\,.
\end{equation}
That is, it is gauge-invariant in Minkowski spacetime.

\subsection{polarization tensors and degrees of freedom}
Since the metric perturbation is a symmetric tensor of rank two, it has 10 independent components. The Lorenz gauge gives four constraints, so that the independent degrees of freedom drop to six. However, if the graviton is massless, we expect that only two physical degrees of freedom remain.

A particular convenient choice of gauge in the context of gravitational radiation is to pick $\bar{h}_{\mu\alpha}u^{\alpha}=0$, $\bar{h}_{\alpha}{}^{\alpha}=0$. Together with $\bar{h}_{\mu\alpha,\,}{}^{\alpha}=0$, they form the transverse traceless (TT) gauge, and reduce the degrees of freedom in $\bar{h}_{\mu\nu}$ down to two. If we write
\begin{equation}
    h_{\mu\nu}=\mathsf{e}_{\mu\nu}f(x)+\text{C.C.}\,,
\end{equation}
then in the absence of the source, Eq.~\eqref{E:itjireereaa} implies $\square f(x)=0$, with a plane-wave solution $f(x)=e^{ik\cdot x}$. If we suppose the perturbations propagate along the $z$ direction, then we have two independent polarization tensors $\mathsf{e}_{\mu\nu}$, each of which takes a particular simple form
\begin{align}\label{E:ngkfd}
	\mathsf{e}_{+}&=\begin{pmatrix}0 &\phantom{-}0 &\phantom{-}0 &\phantom{-}0\\0 &\phantom{-}1 &\phantom{-}0 &\phantom{-}0\\0 &\phantom{-}0 &-1 &\phantom{-}0\\0 &\phantom{-}0 &\phantom{-}0 &\phantom{-}0\end{pmatrix}\,,&\mathsf{e}_{\times}&=\begin{pmatrix}0 &0 &0 &0\\0 &0 &1 &0\\0 &1 &0 &0\\0 &0 &0 &0\end{pmatrix}\,.
\end{align}
Thus, the metric perturbations propagate as  transverse waves at the speed of light, with polarizations described by \eqref{E:ngkfd}. It is of interest to note that even though the ``amplitude'' of the wave is small, the gravitational wave can induce quite a substantial curvature for sufficiently short wavelength modes.

\begin{figure}
	\centering
	\includegraphics[width=0.8\textwidth]{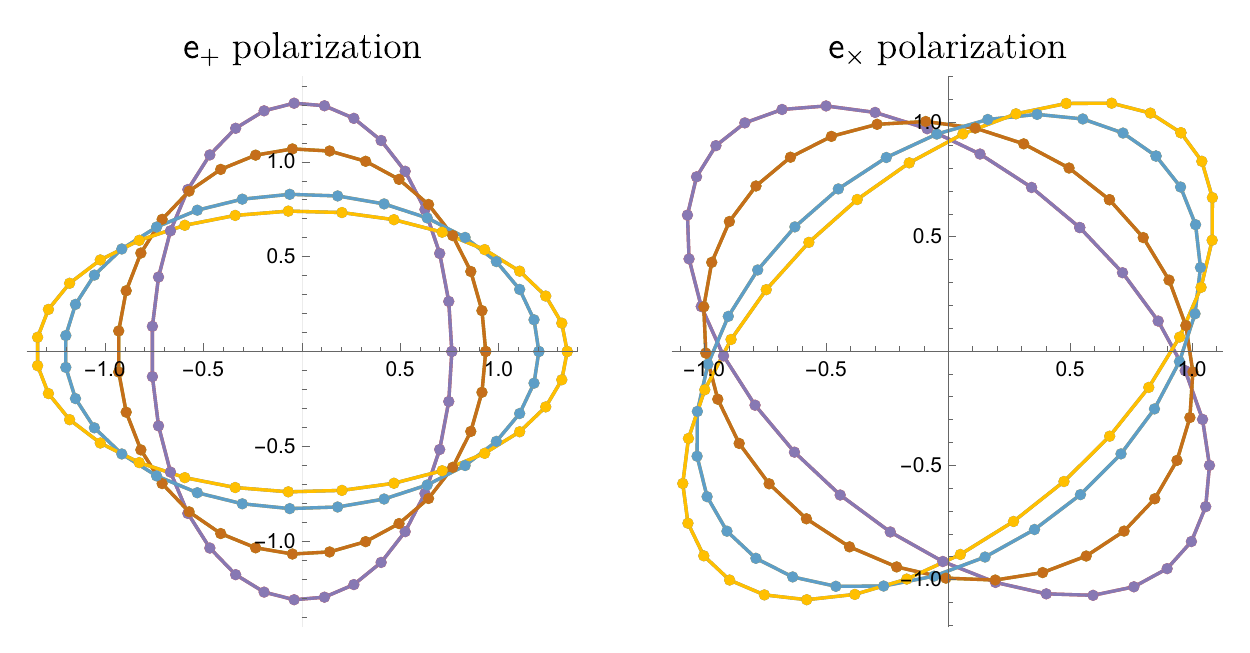}
	\caption{The physical effects of polarizations of the metric perturbations. Different colors correspond to different times.}\label{Fi:polarization2}
\end{figure}

In the transverse traceless gauge there is no difference between $h_{\mu\nu}$ and $\bar{h}_{\mu\nu}$ due to $h=0$, so we will remove the overhead bar hereafter. Clearly the gravitational wave polarizes in a way different from the light waves. Since the metric perturbation changes the physical length scales, it is easier to see its effects via the relative distance. For example, we may arrange a bunch of test particles to form a circle on the $x$--$y$ plane, and examine how their relative positions vary when a plane gravitational wave, traveling along the $z$ direction, passes them. Suppose each particle initially at rest with initial four-velocity $u^{\mu}=(1,0,0,0)$, then the geodesic deviation between the neighboring geodesics of the test particles is described by,
\begin{equation}\label{E:udnfere}
	\frac{d^{2}r^{\mu}}{dt^{2}}=R^{\mu}{}_{\nu\rho\sigma}u^{\nu}u^{\rho}\,r^{\sigma}\,,
\end{equation}
for the Minkowski background, where $r^{\mu}$ gives the relative position orthogonal to the four-velocity $u^\mu$. The neighboring test particles will be displaced from the circle, according to the polarization of the waves, as shown in Fig.~\ref{Fi:polarization2}. For example, if the incoming wave has the $+$ polarization, then the circle of particles will be distorted  with time into ellipses of different eccentricities whose major and minor axes lie along the $x$ and $y$ axes.

\subsection{quantized gravitational wave in Minkowski spacetimes}\label{S:wurgfsb}
Similar to quantizing the electromagnetic wave, we may quantize the gravitational wave and promote $h_{\mu\nu}$ to an operator in a plane-wave expansion,
\begin{equation}\label{E:nnhslrie}
    \hat{h}^{\mu\nu}(x)=\int\!\frac{d^3\bm{k}}{(2\pi)^{\frac{3}{2}}}\frac{1}{\sqrt{2\omega}}\;\sum_{\lambda=1}^2\mathsf{e}_{\lambda\bm{k}}^{\mu\nu}\Bigl[\hat{a}_{\lambda,\,\bm{k}}\,e^{ik\cdot x}+\textsc{H.C.}\Bigr]\,,
\end{equation}
where $\lambda$ labels two independent polarization tensors for each mode $\bm{k}$, $k^{\mu}=(\omega,\,\bm{k})$, and $\hat{a}_{\lambda\bm{k}}$ and its Hermitian conjugate are the standard creation and annihilation operators, satisfying $[\hat{a}_{\lambda,\,\bm{k}}^{\vphantom{\dagger}},\,\hat{a}_{\lambda',\,\bm{k}'}^{\dagger}]=\delta_{\lambda\lambda'}\delta(\bm{k}-\bm{k}')$.

Since here the wave three-vector $\bm{k}$ is not limited to the $z$-direction, it is not straightforward to write down the polarization tensors $\mathsf{e}_{\lambda\bm{k}}^{\mu\nu}$ right off hand. An useful observation is that the polarization tensors in \eqref{E:ngkfd} can be written as a dyadic product of the usual polarization unit vectors $\bm{\mathsf{e}}_1$ and $\bm{\mathsf{e}}_2$ along the $x$ and $y$ axes as
\begin{align}\label{E:ngbeora}
    \mathsf{e}_{+}&=\frac{1}{\sqrt{2}}\bigl(\bm{\mathsf{e}}_1\bm{\mathsf{e}}_1-\bm{\mathsf{e}}_2\bm{\mathsf{e}}_2\bigr)\,,&\mathsf{e}_{\times}&=\frac{1}{\sqrt{2}}\bigl(\bm{\mathsf{e}}_1\bm{\mathsf{e}}_2+\bm{\mathsf{e}}_2\bm{\mathsf{e}}_1\bigr)\,,
\end{align}  
with the triad $\{\bm{\mathsf{e}}_1,\bm{\mathsf{e}}_2,\bm{\kappa}\}$ satisfying $\bm{\kappa}\cdot\bm{\mathsf{e}}_1=0=\bm{\kappa}\cdot\bm{\mathsf{e}}_2$ and $\bm{k}=\omega\,\bm{\kappa}$, such that
\begin{align}\label{E:gbkdeter}
    \mathsf{e}_{+}^{\mu\nu}\mathsf{e}_{+}{}_{\mu\nu}&=1\,,&\mathsf{e}_{\times}^{\mu\nu}\mathsf{e}_{\times}{}_{\mu\nu}&=1\,,&\mathsf{e}_{\times}^{\mu\nu}\mathsf{e}_{+}{}_{\mu\nu}&=0\,.
\end{align}
Thus, for a general $\bm{k}$, if we have two mutually orthogonal unit vectors $\bm{\mathsf{e}}_1$ and $\bm{\mathsf{e}}_2$ and they are also orthogonal to $\bm{k}$, then we can construct two polarization tensors $\mathsf{e}_{\lambda\bm{k}}$ in the same way as shown in \eqref{E:ngbeora}.

There is no unique way to determine $\bm{\mathsf{e}}_1$ and $\bm{\mathsf{e}}_2$ due to the cylindrical symmetry about $\bm{k}$ in their orthogonality conditions with $\bm{k}$. We may start with an arbitrary but fixed unit three-vector $\bm{l}$, not collinear with $\bm{k}=\omega\bm{\kappa}$, to find $\bm{\mathsf{e}}_1$, $\bm{\mathsf{e}}_2$ by 
\begin{align}
    \bm{\mathsf{e}}_1&=\bm{l}\times\bm{\kappa}\,,& \bm{\mathsf{e}}_2&=\bm{\kappa}\times\bm{\mathsf{e}}_1\,.
\end{align}
It is straightforward to check that $\{\bm{\mathsf{e}}_1,\bm{\mathsf{e}}_2,\bm{\kappa}\}$ does form a set orthonormal bases of the spatial section of Minkowski spacetime. The arbitrariness of $\bm{l}$ is the manifest of the aforementioned symmetry.

It will be convenient to put the above triad into a covariant expression
\begin{equation}\label{E:ngbhd}
    \eta_{\mu\nu}e_i{}^{\mu}e_j{}^{\nu}=\delta_{ij}\,,
\end{equation}
with, for example $e_{1}{}^{\mu}=(0,\bm{\mathsf{e}}_{1})$, and $e_{3}{}^{\mu}=(0,\bm{\kappa})$, where the Greek letters run from $0$ to $3$ while the Latin letters range from 1 to 3. Here implicitly we have assumed a preferred frame in which the four velocity $u^{\mu}=(1,\bm{0})$ and $u_{\mu}e_i{}^{\mu}=0$. This implies that together with the triad, we can use this four velocity to form a tetrad $\{e_0{}^{\mu},e_1{}^{\mu},e_2{}^{\mu},e_3{}^{\mu}\}$ with $e_0{}^{\mu}=u^{\mu}$ for Minkowski spacetime with the handedness condition
\begin{equation}
    e_0{}^{\mu}e_1{}^{\nu}e_2{}^{\rho}e_3{}^{\sigma}\varepsilon_{\mu\nu\rho\sigma}=1\,,
\end{equation}
where $\varepsilon_{\mu\nu\rho\sigma}$ is the Levi-Civita tensor in Minkowski spacetime, which gives 1 if $\mu$, $\nu$, $\rho$, $\sigma$ for even permutations of $0$, $1$, $2$, $3$, and $-1$ for odd permutations, and 0 if any two of the tensor indices are the same. Then we can generalize the orthogonality condition of the tetrad from \eqref{E:ngbhd} to
\begin{equation}\label{E:kkefdnd}
    \eta_{\mu\nu}e_{\alpha}{}^{\mu}e_{\beta}{}^{\nu}=\eta_{\alpha\beta}\,,
\end{equation}
which will be valid in any frame. The completeness condition of the tetrad is then formulated as
\begin{equation}
    \sum_{\alpha=0}^4\eta^{\alpha\alpha}e_{\alpha}{}^{\mu}e_{\alpha}{}^{\nu}=\eta^{\mu\nu}\,.
\end{equation}
Recall that the innermost subscript in the notation of the polarization vector $e_{\alpha}{}^{\mu}$ is the label of the polarization vector, not the tensor indices. The polarization tensors $\mathsf{e}_{\lambda\bm{k}}^{\mu\nu}$ then follow accordingly.

It seems that from this point on, we may treat the quantized gravitational wave and its quanta, gravitons, in a way similar to the quantized electromagnetic field and photons, and further discuss its physical effects due to interactions with the other matter fields by the standard quantum field theory techniques. However, it has been argued that if higher-order contributions are systematically included, such a quantized theory is not renormalizable due to the dimensional nature of the coupling strength, the Newton constant. Thus, we will restrict ourselves to the low-energy regime, limited by some energy cutoff scale $\Lambda$ that depends on the configuration at hand, and treat the quantized gravitational wave theory as an effective field theory.

Keeping the above consideration in mind, we will discuss the natures of the quantized gravitation wave in some of the well known Gaussian states, and the corresponding two-point functions of the graviton. Among the latter, the retarded Green's function and the Hadamard function will be the building blocks to discuss the quantum effects of linearized gravity.

\section{Quantum Field Theory: Green's functions and graviton states}

The simplest two-point function of the free quantized gravitational field is the Wightman function, defined by
\begin{equation}
    G_{>,0}^{(h)}{}^{\mu\nu\rho\sigma}(x,x')=i\,\langle\hat{h}^{\mu\nu}(x)\hat{h}^{\rho\sigma}(x')\rangle\,,    
\end{equation}
where $\langle\cdots\rangle$ is the expectation value with respect to a given pure state of the graviton, but when the state is mixed, the notation corresponds to the trace average with respect to that state. In addition, here we have assumed that $\langle\hat{h}^{\mu\nu}(x)\rangle=0$. If not, we simply replace $\hat{h}^{\mu\nu}(x)$ by $\hat{h}^{\mu\nu}(x)-\langle\hat{h}^{\mu\nu}(x)\rangle$ or $\hat{a}_{\lambda\bm{k}}$ by $\hat{a}_{\lambda,\bm{k}}-\langle\hat{a}_{\lambda,\bm{k}}\rangle$ in the definitions of the Green's functions. Hereafter we will use the Heisenberg picture. Apart from the tensor indices, the superscript in a notation like $G_{>,0}^{(h)}{}^{\mu\nu\rho\sigma}(x,x')$ specifies the type of field the Green's function is associated with, and the subscript gives the kind of Green's function. If the subscript contains a zero, it means the associated field is a free field; otherwise it is an interacting field. Occasionally, we will put an additional superscript to indicate the state of the graviton. For example, $\prescript{\beta}{}{G}_{>,0}^{(h)}{}^{\mu\nu\rho\sigma}(x,x')$ tells that this is a Wightman function of the free graviton in its thermal state. This myriad of superscripts and subscripts will be useful when we deal with interacting systems.

The retarded Green's function $G_{R,0}^{(h)}{}^{\mu\nu\rho\sigma}(x,x')$ is given by
\begin{equation}
    G_{R,0}^{(h)}{}^{\mu\nu\rho\sigma}(x,x')=i\,\theta(t-t')\,\langle\bigl[\hat{h}^{\mu\nu}(x),\,\hat{h}^{\rho\sigma}(x')\bigr]\rangle\,,    
\end{equation}
with $x=(t,\bm{x})$ and $\theta(\tau)$ being the Heaviside unit-step function. It will be shown later that this Green's function respects the causal structure of Minkowski spacetime, and relays the causal influence of the source point $x'$ to the field point $x$. It is of interest to note that due to the standard commutation relation, the retarded Green's function of the free quantized gravitational wave is state-independent, so it will have the same form for all quantum states of the graviton. In the context of the in-in or the influence functional formalism outlined in Appendix~\ref{S:eiwt}, this retarded Green's function is often termed the dissipation kernel.

Another Green's function that often accompanies the retarded Green's function is the Hadamard function, also called the noise kernel, which is the expectation value of the anticommutator of the free field operator $\hat{h}^{\mu\nu}(x)$
\begin{equation}
    G_{H,0}^{(h)}{}^{\mu\nu\rho\sigma}(x,x')=\frac{1}{2}\,\langle\bigl\{\hat{h}^{\mu\nu}(x),\,\hat{h}^{\rho\sigma}(x')\bigr\}\rangle\,.
\end{equation}
In contrast to its retarded Green's function counterpart, it is in general state-dependent, and it is nonzero even for the spacelike interval between $x$ and $x'$.

Finally, in the in-out formalism or in the context of graviton scattering, we often come across the Feynman propagator $G_{F,0}^{(h)}{}^{\mu\nu\rho\sigma}(x,x')$, which is given by the expectation value of the time-ordered free field operator $\hat{h}^{\mu\nu}(x)$
\begin{equation}
    G_{F,0}^{(h)}{}^{\mu\nu\rho\sigma}(x,x')=i\,\langle\mathcal{T}\,\hat{h}^{\mu\nu}(x)\hat{h}^{\rho\sigma}(x')\rangle\,,
\end{equation}
where $\mathcal{T}$ denotes time ordering in the sense of $\mathcal{T}\,\hat{A}(t)\hat{B}(t')=\theta(t-t')\,\hat{A}(t)\hat{B}(t')+\theta(t'-t)\hat{B}(t')\hat{A}(t)$ for any pair of bosonic operators $\hat{A}$ and $\hat{B}$. It turns out that the Feynman propagator can be expressed in terms of the previous two Green's functions by
\begin{equation}
    G_{F,0}^{(h)}{}^{\mu\nu\rho\sigma}(x,x')=\frac{1}{2}\Bigl[G_{R,0}^{(h)}{}^{\mu\nu\rho\sigma}(x,x')+G_{A,0}^{(h)}{}^{\mu\nu\rho\sigma}(x,x')\Bigr]+i\,G_{H,0}^{(h)}{}^{\mu\nu\rho\sigma}(x,x')\,,
\end{equation}
where $G_{A,0}^{(h)}{}^{\mu\nu\rho\sigma}(x,x')$ is the advanced Green's function of the free quantized gravitational wave, and can also be expressed as $G_{A,0}^{(h)}{}^{\mu\nu\rho\sigma}(x,x')=G_{R,0}^{(h)}{}^{\mu\nu\rho\sigma}(x',x)$. Thus from now on we will focus on the retarded Green's function and the Hadamard function. By the plane-wave expansion~\eqref{E:nnhslrie}, we can write down these two Green's functions in more explicit forms
\begin{align}
    G_{R,0}^{(h)}{}^{\mu\nu\rho\sigma}(x,x')&=i\,\theta(t-t')\int\!\frac{d^3\bm{k}}{(2\pi)^3}\frac{1}{2\omega}\;V^{\mu\nu\rho\sigma}_{\bm{k}}\,e^{ik\cdot(x-x')}+\text{C.C.}\,,\label{E:ngbgbsw}\\
    G_{H,0}^{(h)}{}^{\mu\nu\rho\sigma}(x,x')&=\frac{1}{2}\int\!\frac{d^3\bm{k}}{(2\pi)^3}\frac{1}{2\omega}\;V^{\mu\nu\rho\sigma}_{\bm{k}}\,\Bigl[\langle\bigl\{\hat{a}_{\lambda,\bm{k}},\,\hat{a}_{\lambda,\bm{k}}\bigr\}\rangle\,e^{ik\cdot(x+x')}+\langle\bigl\{\hat{a}_{\lambda,\bm{k}}^{\vphantom{\dagger}},\,\hat{a}_{\lambda,\bm{k}}^{\dagger}\bigr\}\rangle\,e^{ik\cdot(x-x')}\Bigr]+\text{C.C.}\,,\label{E:ybrtjgf}
\end{align}   
where
\begin{equation}
    V^{\mu\nu\rho\sigma}_{\bm{k}}=\sum_{\lambda=1}^2\mathsf{e}_{\lambda\bm{k}}^{\mu\nu}\mathsf{e}_{\lambda\bm{k}}^{\rho\sigma}\,.
\end{equation}
This in a sense is analogous to the corresponding result in the Coulomb gauge in electromagnetism. It can be shown that for unbounded Minkwoski spacetime (that is, being isotropic and homogeneous), it is convenient to use the Lorenz gauge alone to write down this tensor in a form independent of $\bm{k}$, so that it can be pulled out of the integrals to make the Green's function of the free graviton be written as 
\begin{equation}\label{E:dkfbsdf}
    G_{0}^{(h)}{}^{\mu\nu\rho\sigma}(x,x')=V^{\mu\nu\rho\sigma}\,G_{0}^{(\phi)}(x,x')\,.
\end{equation}
Here we have suppressed the subscript that distinguishes the type of the Green's functions, and dropped the subscript $\bm{k}$ of $V^{\mu\nu\rho\sigma}_{\bm{k}}$. It is computationally simpler to deal with the Green's function of the form \eqref{E:dkfbsdf}, which frees us from complexity of tensor indices. Thus hereafter we will assume the form \eqref{E:dkfbsdf} unless mentioned otherwise.

The two-point function $G_{0}^{(\phi)}(x,x')$ turns out to be the corresponding Green's function of the free massless scalar field $\hat{\phi}(x)$, satisfying $\square\hat{\phi}(x)=0$. In this isotropic case, we often re-write the $\bm{k}$ integrals in \eqref{E:ngbgbsw} into, say,
\begin{equation}
    \int\!\frac{d^3\bm{k}}{(2\pi)^3}\frac{\pi}{\omega^2}\;\varrho(\omega)\,e^{ik\cdot(x-x')}\,.
\end{equation}
The spectral density $\varrho(\omega)$ is equal to $\omega/(2\pi)$ when the Green's function is the vacuum Hadamard function of the free graviton. In other words, the free graviton follows Ohmic dynamics. The introduction of the spectral density is useful when we need to address the non-Ohmic, non-Markovian properties due to the interaction with other systems.

Finally, due to the gauge choice, these Green's functions of $\hat{h}^{\mu\nu}$ are gauge-dependent, so they are typically not physical observables. However, certain combinations of their derivatives can form the Green's functions of the physical observables, say, the Riemann tensor from \eqref{E:bbcklkfd}.

The properties of each Green's function depends on the state of the graviton, and from \eqref{E:ybrtjgf} we see that this information is encoded in the expectation values $\langle\{\hat{a}_{\lambda,\,\bm{k}}^{\vphantom{\dagger}},\,\hat{a}_{\lambda,\bm{k}}^{\vphantom{\dagger}}\}\rangle$ and $\langle\{\hat{a}_{\lambda,\,\bm{k}}^{\vphantom{\dagger}},\,\hat{a}_{\lambda,\bm{k}}^{\dagger}\}\rangle$. Hence we will examine their behaviors for a few well known Gaussian states of the graviton.

\subsection{vacuum state}
In Minkowski spacetime, we have a unique vacuum state $\lvert0\rangle$, annihilated by the operator $\hat{a}_{\lambda,\bm{k}}^{\vphantom{\dagger}}$ associated with the positive-energy solution $e^{ik\cdot x}$ defined by the global Killing vector $\partial_t$
\begin{equation}
    \hat{a}_{\lambda,\bm{k}}^{\vphantom{\dagger}}\,\lvert0\rangle=0\,.
\end{equation}
Following the same line of argument in the quantized electromagnetic field or quantum harmonic oscillator, we can construct a series of Fock states $\lvert n_{\lambda,\bm{k}}\rangle$ for each mode $(\lambda,\bm{k})$ by
\begin{equation}
    \frac{1}{\sqrt{n_{\lambda,\bm{k}}!}}\bigl(\hat{a}_{\lambda,\bm{k}}^{\dagger}\bigr)^{n_{\lambda,\bm{k}}}\,\lvert0\rangle=\lvert n_{\lambda,\bm{k}}\rangle\,,
\end{equation}
and define the number operator $\hat{N}_{\lambda,\bm{k}}$ by
\begin{equation}
    \hat{N}_{\lambda,\bm{k}}=\hat{a}_{\lambda,\bm{k}}^{\dagger}\hat{a}_{\lambda,\bm{k}}^{\vphantom{\dagger}}\,,
\end{equation}
such that the state $\lvert n_{\lambda,\bm{k}}\rangle$ is an eigenstate of the operator $\hat{N}_{\lambda,\bm{k}}$,
\begin{equation}
    \hat{N}_{\lambda,\bm{k}}\,\lvert n_{\lambda,\bm{k}}\rangle=n_{\lambda,\bm{k}}\,\lvert n_{\lambda,\bm{k}}\rangle\,.
\end{equation}
The parameter $n_{\lambda,\bm{k}}$ gives the number of gravitons in the state $\lvert n_{\lambda,\bm{k}}\rangle$ of the mode $(\lambda,\bm{k})$.

The expectation values $\langle\{\hat{a}_{\lambda,\bm{k}}^{\vphantom{\dagger}},\,\hat{a}_{\lambda,\bm{k}}^{\vphantom{\dagger}}\}\rangle$ and $\langle\{\hat{a}_{\lambda,\bm{k}}^{\vphantom{\dagger}},\,\hat{a}_{\lambda,\bm{k}}^{\dagger}\}\rangle$ are then readily obtained
\begin{align}
    \langle\{\hat{a}_{\lambda,\bm{k}}^{\vphantom{\dagger}},\,\hat{a}_{\lambda,\bm{k}}^{\vphantom{\dagger}}\}\rangle&=0\,,&\langle\{\hat{a}_{\lambda,\bm{k}}^{\vphantom{\dagger}},\,\hat{a}_{\lambda,\bm{k}}^{\dagger}\}\rangle&=1\,,
\end{align}
for the vacuum state.

The vacuum state has the characteristic of the minimal uncertainty for each mode. This is most easily seen via the combinations
\begin{align}\label{E:gbheius}
    \hat{\chi}_{\lambda,\bm{k}}&=\frac{1}{\sqrt{2}}\bigl(\hat{a}_{\lambda,\bm{k}}^{\vphantom{\dagger}}+\hat{a}_{\lambda,\bm{k}}^{\dagger}\bigr)\,,&\hat{p}_{\lambda,\bm{k}}&=\frac{1}{i\sqrt{2}}\bigl(\hat{a}_{\lambda,\bm{k}}^{\vphantom{\dagger}}-\hat{a}_{\lambda,\bm{k}}^{\dagger}\bigr)\,.
\end{align}
we immediately find that $[\hat{\chi}_{\lambda,\bm{k}},\,\hat{p}_{\lambda,\bm{k}}]=i$, $\langle\hat{\chi}_{\lambda,\bm{k}}\rangle=0$, $\langle \hat{p}_{\lambda,\bm{k}}\rangle=0$, and 
\begin{align}\label{E:bgbeisasd}
    \langle\Delta\hat{\chi}_{\lambda,\bm{k}}^2\rangle&=\frac{1}{2}\,,&\langle\Delta \hat{p}_{\lambda,\bm{k}}^2\rangle&=\frac{1}{2}\,,&\frac{1}{2}\langle\bigl\{\Delta\hat{\chi}_{\lambda,\bm{k}},\,\Delta\hat{p}_{\lambda,\bm{k}}\bigr\}\rangle&=0\,,
\end{align}
such that the Robertson-Schr\"odinger uncertainty relation is given by
\begin{equation}\label{E:bgsooqs}
    \varsigma_{\lambda,\bm{k}}=\langle\Delta\hat{\chi}_{\lambda,\bm{k}}^2\rangle\langle\Delta \hat{p}_{\lambda,\bm{k}}^2\rangle-\frac{1}{4}\langle\bigl\{\Delta\hat{\chi}_{\lambda,\bm{k}},\,\Delta\hat{p}_{\lambda,\bm{k}}\bigr\}\rangle^2=\frac{1}{4}\,,
\end{equation}
for the vacuum state of mode $(\lambda,\,\bm{k})$, with, for example, $\Delta\hat{\chi}_{\lambda,\bm{k}}=\hat{\chi}_{\lambda,\bm{k}}-\langle\hat{\chi}_{\lambda,\bm{k}}\rangle$. Eq.~\eqref{E:bgsooqs} saturates the lower bound of the uncertainty relation with equal quadratures $\langle\Delta\hat{\chi}_{\lambda,\bm{k}}^2\rangle=\langle\Delta \hat{p}_{\lambda,\bm{k}}^2\rangle=1/2$. Thus in the phase space $\chi_{\lambda,\bm{k}}$--$p_{\lambda,\bm{k}}$, they depict a circular distribution for the Wigner function.

\subsection{coherent state}
The coherent state arises from the inquiry whether the annihilation operator $\hat{a}_{\lambda,\bm{k}}^{\vphantom{\dagger}}$ has an eigenstate,
\begin{equation}
    \hat{a}_{\lambda,\bm{k}}^{\vphantom{\dagger}}\,\lvert\alpha_{\lambda,\bm{k}}^{\vphantom{\dagger}}\rangle=\alpha_{\lambda,\bm{k}}^{\vphantom{\dagger}}\,\lvert\alpha_{\lambda,\bm{k}}^{\vphantom{\dagger}}\rangle\,,
\end{equation}
such that the expectation value of the function $f(\hat{a}_{\lambda,\bm{k}}^{\vphantom{\dagger}})$ in this state will be simply given by substituting the operator $\hat{a}_{\lambda,\bm{k}}^{\vphantom{\dagger}}$ in $f(\hat{a}_{\lambda,\bm{k}}^{\vphantom{\dagger}})$ with the complex c-number $\alpha_{\lambda,\bm{k}}^{\vphantom{\dagger}}$ i.e., $f(\alpha_{\lambda,\bm{k}}^{\vphantom{\dagger}})$. In this sense, $f(\alpha_{\lambda,\bm{k}}^{\vphantom{\dagger}})$, can be identified as the classical counterpart of $f(\hat{a}_{\lambda,\bm{k}}^{\vphantom{\dagger}})$. Thus the expectation value of $\hat{h}^{\mu\nu}(x)$ in the multi-mode coherent state $\lvert\{\alpha_{\lambda,\bm{k}}\}\rangle$ is
\begin{align}
    h^{\mu\nu}(x)&=\int\!\frac{d^3\bm{k}}{(2\pi)^{\frac{3}{2}}}\frac{1}{\sqrt{2\omega}}\;\sum_{\lambda=1}^2\mathsf{e}_{\lambda\bm{k}}^{\mu\nu}\Bigl[\langle\alpha_{\lambda,\bm{k}}^{\vphantom{\dagger}}\vert\hat{a}_{\lambda,\bm{k}}\vert\alpha_{\lambda,\bm{k}}^{\vphantom{\dagger}}\rangle\,e^{ik\cdot x}+\textsc{C.C.}\Bigr]\notag\\
    &=\int\!\frac{d^3\bm{k}}{(2\pi)^{\frac{3}{2}}}\frac{1}{\sqrt{2\omega}}\;\sum_{\lambda=1}^2\mathsf{e}_{\lambda\bm{k}}^{\mu\nu}\Bigl[\alpha_{\lambda,\bm{k}}^{\vphantom{\dagger}}\,e^{ik\cdot x}+\textsc{C.C.}\Bigr]\,,\label{E:ebfjsdsd}
\end{align}
where $\{\alpha_{\lambda,\bm{k}}\}$ is the shorthand notation for a set of $\alpha_{\lambda,\bm{k}}$ for all modes. Eq.~\eqref{E:ebfjsdsd} is exactly the mode sum of the classical gravitational wave introduced in Sec.~\ref{S:bgfkdf} for arbitrary $\lambda$ and $\bm{k}$, and $\alpha_{\lambda,\bm{k}}^{\vphantom{\dagger}}$ is interpreted as the complex amplitude of mode $(\lambda,\bm{k})$. In other words, Eq.~\eqref{E:ebfjsdsd} gives a Fourier representation of an arbitrary gravitational disturbance signal. In addition, the expectation value of the number operator $\hat{N}_{\lambda,\bm{k}}$ in the coherent state is $\langle\hat{N}_{\lambda,\bm{k}}\rangle=\lvert\alpha_{\lambda,\bm{k}}\rvert^2$. Thus the amplitude $\alpha_{\lambda,\bm{k}}$ is related to the average number of gravitons in the coherent state. With increasing $\lvert\alpha_{\lambda,\bm{k}}\rvert$, the discrete nature of the graviton is overshadowed. Eq.~\eqref{E:ebfjsdsd} then looks really like a classical wave, and thus we often identify it as the classical gravitational wave.

However, strictly speaking, the gravitons only behave or look like a classical gravitational wave in the average sense, in particular when $\lvert\alpha_{\lambda,\bm{k}}\rvert$ is large. They are still quantum-mechanical to the core. If we examine the uncertainty relation $\varsigma_{\lambda,\bm{k}}$ for each mode of $\hat{h}^{\mu\nu}(x)$ in the coherent state, we find it also saturates the lower bound $1/4$ as in the vacuum state in \eqref{E:bgsooqs}. That is why sometimes the coherent state is said to be the most classical quantum state. It is still a quantum state because the uncertainty relation is never zero, and in particular does not depend on $\alpha_{\lambda,\bm{k}}$ at all. This is where quantumness lies even though \eqref{E:ebfjsdsd} looks very classical.

It is straightforward to see that the noise kernel associated with $\hat{h}^{\mu\nu}(x)$ in the coherent state has exactly the same form as that in the vacuum state due to the presence of $\hat{h}^{\mu\nu}(x)-\langle\hat{h}^{\mu\nu}(x)\rangle$ in the definition of the noise kernel. All the $\alpha_{\lambda,\bm{k}}$ dependence is canceled out. Thus it is the vacuum fluctuations of the free graviton in the coherent state that will contribute to the quantum fluctuation effects on the model system the graviton is coupled with. This reminiscent quantum nature of the coherent state of the gravitons can still possibly induce tiny but nonzero entanglement between the constituents of the model system, as has been addressed in Brownian motion. Thus, classicality will not emerge unless the decoherence process is at play to wipe off the quantumness.

We can also construct the coherent state $\lvert\alpha_{\lambda,\bm{k}}^{\vphantom{\dagger}}\rangle$ by acting an unitary operator $\hat{D}(\alpha_{\lambda,\bm{k}}^{\vphantom{\dagger}})$ on the vacuum state $\lvert0\rangle$, that is,
\begin{equation}
    \lvert\alpha_{\lambda,\bm{k}}^{\vphantom{\dagger}}\rangle=\hat{D}(\alpha_{\lambda,\bm{k}}^{\vphantom{\dagger}})\,\lvert0\rangle\,.
\end{equation}
The vacuum state can be viewed as a coherent state with $\alpha_{\lambda,\bm{k}}^{\vphantom{\dagger}}=0$. The multi-mode coherent state $\lvert\{\alpha_{\lambda,\bm{k}}\}\rangle$ can thus formally given by
\begin{equation}
    \lvert\{\alpha_{\lambda,\bm{k}}\}\rangle=\prod_{\lambda,\bm{k}}\hat{D}(\alpha_{\lambda,\bm{k}}^{\vphantom{\dagger}})\,\lvert0\rangle\,.
\end{equation}
Here the unitary displacement operator $\hat{D}(\alpha_{\lambda,\bm{k}}^{\vphantom{\dagger}})$ of each mode $(\lambda,\bm{k})$ is defined by
\begin{equation}\label{E:gdjdjjuwer}
    \hat{D}(\alpha_{\lambda,\bm{k}}^{\vphantom{\dagger}})=\exp\Bigl[\alpha_{\lambda,\bm{k}}^{\vphantom{\dagger}}\hat{a}_{\lambda,\bm{k}}^{\dagger}-\alpha_{\lambda,\bm{k}}^{\vphantom{\dagger}*}\hat{a}_{\lambda,\bm{k}}^{\vphantom{\dagger}}\Bigr]\,.
\end{equation}
It is readily seen that $\hat{D}^{\dagger}(\alpha_{\lambda,\bm{k}}^{\vphantom{\dagger}})=\hat{D}(-\alpha_{\lambda,\bm{k}}^{\vphantom{\dagger}})$. With the help of the Baker–Campbell–Hausdorff formula, we can derive an identity of the displacement operator on $\hat{a}_{\lambda,\bm{k}}^{\vphantom{\dagger}}$
\begin{equation}\label{E:gbgsbwo}
    \hat{D}^{\dagger}(\alpha_{\lambda,\bm{k}}^{\vphantom{\dagger}})\hat{a}_{\lambda,\bm{k}}^{\vphantom{\dagger}}\hat{D}(\alpha_{\lambda,\bm{k}}^{\vphantom{\dagger}})=\hat{a}_{\lambda,\bm{k}}^{\vphantom{\dagger}}+\alpha_{\lambda,\bm{k}}^{\vphantom{\dagger}}\,,
\end{equation}
which turns out to be very convenient for calculations involving the coherent state. Eq.~\eqref{E:gbgsbwo} explicitly shows that the similarity transformation of the annihilation operator $\hat{a}_{\lambda,\bm{k}}$ by the displacement operator $\hat{D}(\alpha_{\lambda,\bm{k}}^{\vphantom{\dagger}})$ leads to a displacement by an amount $\alpha_{\lambda,\bm{k}}$.

\subsection{squeezed coherent state}
In the context of cosmological evolution, we are particularly interested in two-mode squeezing where the involved two modes have the opposite momenta due to entangled-particle pair creations. The two-mode squeezed quantum state can be formally obtained by applying the two-mode squeezing operator $\hat{S}_2(\zeta_{\lambda\omega})$
\begin{equation}
	\hat{S}_{2}^{\vphantom{\dagger}}(\zeta_{\lambda\omega}^{\vphantom{\dagger}})=\exp\Bigl[\zeta_{\lambda\omega}^{*\vphantom{\dagger}}\,\hat{a}_{\lambda,\mathbf{k}}^{\vphantom{\dagger}}\hat{a}_{\lambda,\shortminus\mathbf{k}}^{\vphantom{\dagger}}-\zeta_{\lambda\omega}^{\vphantom{\dagger}}\,\hat{a}_{\lambda,\mathbf{k}}^{\dagger}\hat{a}_{\lambda,\shortminus\mathbf{k}}^{\dagger}\Bigr]\,.\label{E:bkeursd}
\end{equation}
to any pure quantum state $\lvert\Psi\rangle$, like the vacuum state or coherent state, by $\hat{S}_{2}^{\vphantom{\dagger}}\,\lvert\Psi\rangle$ or to any mixed state $\hat{\varrho}$ like the thermal state by $\hat{S}_{2}^{\vphantom{\dagger}}\hat{\varrho}\hat{S}_{2}^{\dagger}$. The squeeze parameter $\zeta_{\lambda\omega}\in\mathbb{C}$ is a function of $\omega=\lvert\bm{k}\rvert$ because the squeeze operator \eqref{E:bkeursd} is symmetric with $\pm\bm{k}$.

The physical meaning of quantum squeezing is best seen when we consider a simple system that contains only two modes 1, 2. Let the corresponding annihilation operators be denoted by $\hat{a}_1$ and $\hat{a}_2$. Then in terms of their normal modes $\hat{a}_{\pm}$
\begin{align}
	\hat{a}_{1}&=\frac{\hat{a}_{+}+\hat{a}_{-}}{\sqrt{2}}\,,&\hat{a}_{2}&=\frac{\hat{a}_{+}-\hat{a}_{-}}{\sqrt{2}}\,,
\end{align}
with $[\hat{a}_{+},\hat{a}_{-}]=0$, we can write the two-mode squeeze operator $\hat{S}_{2}$ in this case as
\begin{align}
	\hat{S}_{2}(\zeta)=\exp\Bigl[\zeta^{*}\,\hat{a}_1\hat{a}_2-\zeta\,\hat{a}^{\dagger}_1\hat{a}^{\dagger}_2\Bigr]&=\exp\Bigl[\frac{\zeta^{*}}{2}\,a_{+}^{\vphantom{\dagger}2}-\frac{\zeta}{2}\,a_{+}^{\dagger2}\Bigr]\times\exp\Bigl[-\frac{\zeta^{*}}{2}\,a_{-}^{\vphantom{\dagger}2}+\frac{\zeta}{2}\,a_{-}^{\dagger2}\Bigr]\notag\\
 &=\hat{S}_{a_{+}}(\zeta)\times\hat{S}_{a_{-}}(-\zeta)\,.\label{E:dkgueusd}
\end{align}
Here $\hat{S}_{a_{+}}(\zeta)$ is the single-mode squeeze operator for the normal mode $\hat{a}_+$, and it often appears in the context of quantum optics. That is, in terms of the normal modes, the two-mode squeeze operator can be decomposed into a product of two single-mode squeezed operators of their normal modes.

The action of the single-mode squeeze operator on its the annihilation operator $\hat{a}_+$ is given by
\begin{align}
    \hat{S}_{a_{+}}^{\dagger}(\zeta)\,\hat{a}_{+}\hat{S}_{a_{+}}^{\vphantom{\dagger}}(\zeta)=\cosh\eta\,\hat{a}_{+}^{\vphantom{\dagger}}-e^{i\theta}\sinh\eta\,\hat{a}_{+}^\dagger\,,
\end{align}
if we write the squeeze parameter $\zeta$ in the polar form $\zeta=\eta\,e^{i\theta}$ with $\eta\geq0$ and $0\leq\theta<2\pi$. From this, we then find that 
\begin{align}
	\langle \hat{a}_+\rangle_{\textsc{sv}}&=0=\langle \hat{a}_+^{\dagger}\rangle_{\textsc{sv}}\,,\\
	\langle \hat{a}_+^{2}\rangle_{\textsc{sv}}&=-\frac{e^{+i\theta}}{2}\,\sinh2\eta\,,&\langle\hat{a}_+^{\dagger2}\rangle_{\textsc{sv}}&=-\frac{e^{-i\theta}}{2}\,\sinh2\eta\,,&\langle\hat{a}_+^{\dagger}\hat{a}_+\rangle_{\textsc{sv}}&=\sinh^{2}\eta\,,\label{E:gbkf}
\end{align}
for the single-mode squeezed vacuum state of the normal mode $\hat{a}_+$, such that the quadratures, with $\hat{\chi}_+$, $\hat{p}_+$ defined in a way similar to those in \eqref{E:gbheius}, are given by
\begin{align}
	\langle\hat{\chi}_+^{2}\rangle_{\textsc{sv}}&=\bigl(\cosh2\eta-\cos\theta\,\sinh2\eta\bigr)\langle\hat{\chi}_+^{2}\rangle_{\textsc{v}}\,,&\langle\hat{p}_+^{2}\rangle_{\textsc{sv}}&=\bigl(\cosh2\eta+\cos\theta\,\sinh2\eta\bigr)\langle\hat{p}_+^{2}\rangle_{\textsc{v}}\,,\label{E:tbtbdk}\\
	\frac{1}{2}\langle\bigl\{\hat{\chi}_+,\hat{p}_+\bigr\}\rangle_{\textsc{sv}}&=-\frac{1}{2}\,\sin\theta\,\sinh2\eta\,.\label{E:bgbgbds}
\end{align}
Here the subscript $\textsc{v}$ of the expectation value denotes that it is taken with respect to the vacuum state. Hence the uncertainty principle for the normal mode $\hat{a}_+$ in its single-mode squeezed vacuum state takes the form
\begin{equation}
	\varsigma_+=\langle\hat{\chi}_+^{2}\rangle_{\textsc{sv}}\langle\hat{p}_+^{2}\rangle_{\textsc{sv}}-\frac{1}{4}\langle\bigl\{\hat{\chi}_+,\hat{p}_+\bigr\}\rangle^{2}_{\textsc{sv}}=\langle\hat{\chi}_+^{2}\rangle_{\textsc{v}}\langle\hat{p}_+^{2}\rangle_{\textsc{v}}\,.
\end{equation}
There are two interesting features with this expression: 1) it gives the same value as the corresponding uncertainty principle in the vacuum state; both saturate the lower bound $1/4$. However, 2) the operators $\hat{\chi}_+$ and $\hat{p}_+$ are correlated as seen from \eqref{E:bgbgbds}, in contrast to the vacuum state, and 3) the quadratures $\langle\hat{\chi}_+^{2}\rangle_{\textsc{sv}}$ and $\langle\hat{p}_+^{2}\rangle_{\textsc{sv}}$ are not equal; they can be either far larger or smaller than the respective corresponding vacuum values, depending on the choice of the squeeze parameters $\eta$ and $\theta$. In fact in the phase space $\chi_+$--$p_+$, they map out an oblique elliptic distribution of the Wigner function, depending on $\theta$. We may identify the corresponding major and minor axes of the ellipse via the eigenvalues of the covariance matrix 
\begin{equation}
    \renewcommand\arraystretch{1.8}
    \sigma_+=\begin{pmatrix}\langle\hat{\chi}_+^{2}\rangle_{\textsc{sv}}&\dfrac{1}{2}\langle\bigl\{\hat{\chi}_+,\hat{p}_+\bigr\}\rangle_{\textsc{sv}}\\\dfrac{1}{2}\langle\bigl\{\hat{\chi}_+,\hat{p}_+\bigr\}\rangle_{\textsc{sv}}&\langle\hat{p}_+^{2}\rangle_{\textsc{sv}}\end{pmatrix}\,.
\end{equation}    
They are $e^{\pm2\eta}/2$. Thus, it is clear to see that when co-rotating with the ellipse, we find that squeezing can make the value of one quadrature smaller than that of the vacuum state at the expense of the other quadrature. By tuning $\theta$ we can squeeze any quadrature at our disposal. In other words, we can introduce very large or small fluctuations by the squeezed state to meet our goals.

This observation has been applied to the high-precision experiment in which the quantum noise in the quadrature of interest can be suppressed below the vacuum level.

The two-mode squeezed coherent state of the graviton is particularly interesting in the cosmological setting because the expanding evolution of the universe in terms of the scale factor can induce an extreme amount of squeezing to the gravitational wave, described by the coherent state of the graviton. The two-mode squeezed coherent state $\lvert\zeta_{\lambda\omega},\,\alpha_{\lambda,\shortpm\bm{k}}\rangle$ for the modes $(\lambda,\pm\bm{k})$ is given by
\begin{equation}
    \lvert\zeta_{\lambda\omega},\,\alpha_{\lambda,\shortpm\bm{k}}\rangle=\hat{\Lambda}(\zeta_{\lambda\omega},\alpha_{\lambda,\shortpm\bm{k}})\,\lvert0\rangle\,.
\end{equation}
where $\hat{\Lambda}(\zeta_{\lambda\omega},\alpha_{\lambda,\shortpm\bm{k}})=\hat{S}_2(\zeta_{\lambda\omega})\hat{D}(\alpha_{\lambda,\bm{k}})\hat{D}(\alpha_{\lambda,\shortminus\bm{k}})$. Then the corresponding annihilation operators $\hat{a}_{\lambda,\bm{k}}$ and $\hat{a}_{\lambda,\shortminus\bm{k}}$ transform as
\begin{align}
    \hat{\Lambda}^{\dagger}(\zeta_{\lambda\omega},\alpha_{\lambda,\shortpm\bm{k}})\,\hat{a}_{\lambda,\bm{k}}\,\hat{\Lambda}(\zeta_{\lambda\omega},\alpha_{\lambda,\shortpm\bm{k}})&=\cosh\eta_{\lambda\omega}^{\vphantom{\dagger}}\,\bigl(\hat{a}_{\lambda,\mathbf{k}}^{\vphantom{\dagger}}+\alpha_{\lambda,\mathbf{k}}^{\vphantom{\dagger}}\bigr)-e^{+i\theta_{\lambda\omega}}\sinh\eta_{\lambda\omega}^{\vphantom{\dagger}}\,\bigl(\hat{a}_{\lambda,\shortminus\mathbf{k}}^{\dagger}+\alpha_{\lambda,\shortminus\mathbf{k}}^{*}\bigr)\,,\label{E:bbeirdf}
\end{align}
due to the actions of the two-mode squeeze operator $\hat{S}_2(\zeta_{\lambda\omega})$ and the coherent operators $\hat{D}(\alpha_{\lambda,\shortpm\bm{k}})$, where
\begin{align}
	\hat{S}_{2}^{\dagger}(\zeta_{\lambda\omega}^{\vphantom{\dagger}})\,\hat{a}_{\lambda,\mathbf{k}}^{\vphantom{\dagger}}\,\hat{S}_{2}^{\vphantom{\dagger}}(\zeta_{\lambda\omega}^{\vphantom{\dagger}})&=\cosh\eta_{\lambda\omega}^{\vphantom{\dagger}}\,\hat{a}_{\lambda,\mathbf{k}}^{\vphantom{\dagger}}-e^{+i\theta_{\lambda\omega}}\sinh\eta_{\lambda\omega}^{\vphantom{\dagger}}\,\hat{a}_{\lambda,\shortminus\mathbf{k}}^{\dagger}\,,\label{E:vqwrerw}\\
    \hat{D}^{\dagger}(\alpha_{\lambda,\bm{k}}^{\vphantom{\dagger}})\hat{a}_{\lambda,\bm{k}}^{\vphantom{\dagger}}\hat{D}(\alpha_{\lambda,\bm{k}}^{\vphantom{\dagger}})&=\hat{a}_{\lambda,\bm{k}}^{\vphantom{\dagger}}+\alpha_{\lambda,\bm{k}}^{\vphantom{\dagger}}\,,
\end{align}
by \eqref {E:gbgsbwo}. Eq.~\eqref{E:bbeirdf} implies that
\begin{equation}
    \langle\zeta_{\lambda\omega},\,\alpha_{\lambda,\shortpm\bm{k}}\rvert\,\hat{a}_{\lambda,\mathbf{k}}\,\lvert\zeta_{\lambda\omega},\,\alpha_{\lambda,\shortpm\bm{k}}\rangle=\cosh\eta_{\lambda\omega}^{\vphantom{\dagger}}\,\alpha_{\lambda,\mathbf{k}}^{\vphantom{\dagger}}-e^{+i\theta_{\lambda\omega}}\sinh\eta_{\lambda\omega}^{\vphantom{\dagger}}\,\alpha_{\lambda,\shortminus\mathbf{k}}^{*}\,,
\end{equation}    
and thus apparently we arrive at
\begin{equation}
    \hat{\Lambda}^{\dagger}(\zeta_{\lambda\omega},\alpha_{\lambda,\shortpm\bm{k}})\,\Delta\hat{a}_{\lambda,\bm{k}}\,\hat{\Lambda}(\zeta_{\lambda\omega},\alpha_{\lambda,\shortpm\bm{k}})=\cosh\eta_{\lambda\omega}^{\vphantom{\dagger}}\,\hat{a}_{\lambda,\mathbf{k}}^{\vphantom{\dagger}}-e^{+i\theta_{\lambda\omega}}\sinh\eta_{\lambda\omega}^{\vphantom{\dagger}}\,\hat{a}_{\lambda,\shortminus\mathbf{k}}^{\dagger}\,,\label{E:tbiedfk}
\end{equation}
with $\Delta\hat{a}_{\lambda,\bm{k}}=\hat{a}_{\lambda,\bm{k}}-\langle\hat{a}_{\lambda,\bm{k}}\rangle$ where the expectation value is taken with respect to the two-mode squeezed coherent state $\lvert\zeta_{\lambda\omega},\,\alpha_{\lambda,\shortpm\bm{k}}\rangle$. Eq.~\eqref{E:tbiedfk} thus implies that as far as the fluctuations are concerned, the squeezed coherent state behaves the same as the squeezed vacuum state does. The coherent parameter $\alpha_{\lambda,\bm{k}}$ plays no role. Thus the fluctuations of $\hat{h}_{\mu\nu}(x)$ will retain the quantum nature in squeezing even though its mean value may look quite classical for large coherent parameter.

To be more specific, let us write out the noise kernel of free gravitons in the squeezed coherent state more explicitly
\begin{align}\label{E:btrhter}
    G_{H,0}^{(h)}{}^{\mu\nu\rho\sigma}(x,x')&=\int\!\frac{d^3\bm{k}}{(2\pi)^3}\frac{1}{2\omega}\;e^{i\bm{k}\cdot(\bm{x}-\bm{x}')}\biggl[-\Bigl(\sum_{\lambda=1}^2\mathsf{e}_{\lambda\bm{k}}^{\mu\nu}\mathsf{e}_{\lambda\shortminus\bm{k}}^{\rho\sigma}\Bigr)\,\frac{e^{i\theta_{\lambda\omega}}}{2}\,\sinh2\eta_{\lambda\omega}\,e^{-i\omega(t+t')}+\biggr.\notag\\
    &\qquad\qquad\qquad\qquad\qquad\qquad\qquad+\biggl.\Bigl(\sum_{\lambda=1}^2\mathsf{e}_{\lambda\bm{k}}^{\mu\nu}\mathsf{e}_{\lambda\bm{k}}^{\rho\sigma}\Bigr)\,\frac{1}{2}\,\cosh2\eta_{\lambda\omega}\,e^{-i\omega(t-t')}+\text{C.C.}\biggr]\,,
\end{align}
where
\begin{align*}
    \langle0\vert\hat{S}_{2}^{\dagger}(\zeta_{\lambda\omega}^{\vphantom{\dagger}})\,\hat{a}_{\lambda,\mathbf{k}}^{\vphantom{\dagger}2}\,\hat{S}_{2}^{\vphantom{\dagger}}(\zeta_{\lambda\omega}^{\vphantom{\dagger}})\vert0\rangle&=0\,,&\langle0\vert\hat{S}_{2}^{\dagger}(\zeta_{\lambda\omega}^{\vphantom{\dagger}})\,\hat{a}_{\lambda,\mathbf{k}}^{\vphantom{\dagger}}\hat{a}_{\lambda,\shortminus\mathbf{k}}^{\vphantom{\dagger}}\,\hat{S}_{2}^{\vphantom{\dagger}}(\zeta_{\lambda\omega}^{\vphantom{\dagger}})\vert0\rangle&=-\frac{e^{i\theta_{\lambda\omega}}}{2}\sinh2\eta_{\lambda\omega}\,,\\
    \langle0\vert\hat{S}_{2}^{\dagger}(\zeta_{\lambda\omega}^{\vphantom{\dagger}})\,\hat{a}_{\lambda,\mathbf{k}}^{\vphantom{\dagger}}\hat{a}_{\lambda,\mathbf{k}}^{\dagger}\,\hat{S}_{2}^{\vphantom{\dagger}}(\zeta_{\lambda\omega}^{\vphantom{\dagger}})\vert0\rangle&=\cosh^2\eta_{\lambda\omega}\,,&\langle0\vert\hat{S}_{2}^{\dagger}(\zeta_{\lambda\omega}^{\vphantom{\dagger}})\,\hat{a}_{\lambda,\mathbf{k}}^{\vphantom{\dagger}}\hat{a}_{\lambda,\shortminus\mathbf{k}}^{\dagger}\,\hat{S}_{2}^{\vphantom{\dagger}}(\zeta_{\lambda\omega}^{\vphantom{\dagger}})\vert0\rangle&=0\,.
\end{align*}
Comparing with the vacuum case, we note that the noise kernel of free graviton contains the factors $\sinh\eta_{\lambda\omega}$ or $\cosh\eta_{\lambda\omega}$ everywhere. Thus based on earlier discussions, it may enhance or suppress the noise in the system the graviton interacts with, depending on the squeeze angle $\theta_{\lambda\omega}$.

The squeezed state arises naturally in the evolving linear system if it is initially prepared in any Gaussian state, say, the vacuum state or the thermal state, because the Gaussian nature is preserved during the evolution, and the most general Gaussian state can always be expressed in terms of the actions of the squeezed operator, coherence operator and rotation operator on a thermal-like state. Even though the initial state of a linear system is not Gaussian, squeezing may still emerge because the unitary evolution operator is quadratic in the canonical operators, which are superposition of the creation and annihilation operators. Thus the unitary evolution operator likewise can always written as the product of the aforementioned operators. This characteristics can also be rephrased in terms of the Bogoliubov coefficients in the Heisenberg picture. The evolution of linear quantum system implies that the creation or the annihilation operators at one time can be expressed as a superposition of the counterparts at a different time because they all form complete sets of the description of the system. To be more specific, for example, we have
\begin{equation}\label{E:ghjrbdfdf}
	\hat{b}_{\mathbf{k}}^{\vphantom{\dagger}}=\alpha_{\mathbf{k}}^{\vphantom{\dagger}}\,\hat{a}_{\mathbf{k}}^{\vphantom{\dagger}}+\beta_{\shortminus\mathbf{k}}^{\vphantom{\dagger}*}\,\hat{a}_{\shortminus\mathbf{k}}^{\dagger}\,.
\end{equation}
where $\hat{b}_{\mathbf{k}}$ and $\hat{a}_{\mathbf{k}}$ are annihilation operators of mode $\bm{k}$ at two different times.  The coefficients of superposition $\alpha_{\mathbf{k}}$ and $\beta_{\mathbf{k}}$ are called the Bogoliubov coefficients. The commutation relations of $\hat{b}_{\mathbf{k}}$ and $\hat{a}_{\mathbf{k}}$ then require that the coefficients should satisfy the condition $\lvert\alpha_{\mathbf{k}}\rvert^2-\lvert\beta_{\mathbf{k}}\rvert^2=1$. Note that Eq.~\eqref{E:vqwrerw} gives exactly this.

One interesting application of the squeezed state in this context is that it is generally believed gravitons produced in the very early universe are highly squeezed due to the extreme expansion of the background spacetime. If they can be detected, then the information of squeezing can be extracted from the Hadamard function of the gravitons such as Eq.~\eqref{E:btrhter}. This information  of squeezing can then be used to decipher the quantum gravitational processes in inflationary universe or black-hole dynamics. 

Here is a good place to comment on a few misleading issues on classicality of quantum squeezing. It has been proposed that in the inflationary setting, the quantum cosmological fluctuations may spontaneously decohere to generate the classical cosmological perturbations, which then serve as the seeds to generate the large-scale structure due to gravitational instability. It was argued that in the closed system setting, the rapid expansion of the background spacetime will induce extreme squeezing. The consequence is that if only the leading contributions associated with such exceptionally large squeezing are kept, then the equal-time commutator of the operators of the conjugated pair becomes essentially zero. Thus classicality emerges without resorting to any environmental decoherence mechanism. It happens spontaneously as long as the change of the scale factor is sufficiently large. This conclusion raises decade-long debates. Conceptually, within a closed system where nothing can lose, it is rather unusual that the decoherence, as the consequence of loss of phase information, can in principle happen. In addition, the vanishing equal-time commutator that supports the emergence of classicality can be amended if the subleading contributions are taken into consideration. They are needed to recover what is missing to make the commutator nonzero. More discussions on this matter can be found in the overview~\cite{HHDecCos}.

Nonetheless, the classical wave can still generate the squeezing-like effects on a time-dependent background. This is most clearly seen~\cite{bravo} from the creation of quantum squeezing in the massless quantum field due to parametric particle production over time-dependent background spacetime by the Heisenberg picture. Suppose the initial state is a vacuum state of the field in the static IN regime. After the spacetime undergoes a time-dependent scale changes and rests on another static OUT regime, the final state of the field remains a vacuum state, but due to the evolution of the field operator, quantum squeezing occurs. The amount of squeezing can be solely expressed in terms of two fundamental solutions of the Heisenberg wave equations, e.g. (68)--(70) in~\cite{bravo}. Since the dynamics is linear, the same fundamental solutions satisfy the classical wave equation, and then we can use the same expression for classical squeezing. From this perspective, the squeezing-like behavior can be understood as the "amplitude changes" of the fundamental solution due to evolution of the wave over a time dependent background. However, this does not mean quantum squeezing is classical in the misleading sense that linear quantum dynamics is classical. The state plays a very important role in the complete quantum description of dynamics. Various ubiquitous quantum effects, like quantum interference or entanglement, show up once we sandwich the operator solutions with the states.

\section{Influence of Gravitons: Langevin equation for geodesic separation}

Here we will discuss how the quantum effects of gravitons may enter into the model system the graviton interacts with. One of the most systematic approaches to treat the nonequilibrium interacting systems is to adopt the influence functional formalism, in which the gravitons will be identified as the environment, and their overall effects will be summarized into the influence functional. This entity allows us to either construct the in-in effective action, or to write down the reduced density matrix elements of the system of our interest. Then we can derive the Langevin equation, or master equation to describe the nonequilibrium evolution of the reduced system under the influence of gravitons in a systematic, self-consistent manners.

However, additional complexity arises. Even though perturbative gravity considered here is linear, and the interaction term involves $\hat{h}_{\mu\nu}(x)$ linearly, due to the tensor nature of $\hat{h}_{\mu\nu}(x)$ it will usually couple with the composite operators of the system's canonical variables. This implies that the resulting Langevin equation can become nonlinear and contains   multiplicative graviton noise. Thus it is challenging to seek {long-time} analytic solutions to the equation of motion for the reduced system interacting with a gravitational field environment.


We first look at a model system that has been considered earlier in Sec.~\ref{S:bgfkdf}. We examine the quantum effects of weak gravitational field on the timelike, geodesic congruence in Minkowski spacetime.

The action of a free massive particle in spacetime with a metric tensor $g_{\mu\nu}$ has the form
\begin{equation}\label{E:ghhidfgd}
    S_{\textsc{m}}=-m\int\!\sqrt{-d\tau^2}\;
\end{equation}
where $m$ is the mass and $\tau$ is the proper time. Since we are interested in the deviation between the neighboring geodesics of the particle, it is convenient to use the Fermi normal coordinates $(t,z^i)$ along a given geodesic, by which the metric tensor has the form
\begin{align}\label{E:bvdgsw}
    g_{00}&=-1-R_{k0l0}\,z^kz^l+\cdots\,,&g_{0i}&=-\frac{2}{3}\,R_{0ikl}\,z^kz^l\,,&g_{ij}&=\delta_{ij}-\frac{1}{3}\,R_{ijkl}\,z^kz^l\,,
\end{align}
where the metric perturbation and the Riemann tensor will be evaluated at spatial origin, which is the location of the given geodesic. Clearly such a frame locally look inertial, and since $z^i$ lie on the three-hypersurface orthogonal to the axis of time coordinate $t$, it describes the deviation of the neighboring geodesics. Within the small neighborhood, we may expand the action \eqref{E:ghhidfgd} up to the second order in $z^i$
\begin{equation}
    S_{\textsc{m}}=\int\!dt\,\left[\frac{m}{2}\,\delta_{ij}\dot{z}^i\dot{z}^j+\frac{m}{4}\,\ddot{h}_{ij}z^iz^j\right]\,.
\end{equation}
Here the overhead dot represents the derivative with respect to $t$, and in the transverse traceless gauge the Riemann tensor $R_{i0j0}$ has a simple form $-\dfrac{1}{2}\ddot{h}_{ij}$.

On the gravity side, the action of pure gravity is given by the Hilbert action. If we include contributions of second order in metric perturbations, in the transverse-traceless gauge the action has the form 
\begin{equation}
    S_{\textsc{g}}=\frac{1}{16\pi}\int\!d^4x\sqrt{-g}\;R\simeq-\frac{1}{64\pi}\int\!d^4x\;\bigl(\partial_{\rho}h_{ij}\bigr)\bigl(\partial^{\rho}h^{ij}\bigr)+\cdots\,,
\end{equation}
so that the total action $S_{\textsc{m}}+S_{\textsc{g}}$ becomes
\begin{equation}\label{E:bksjgfdf}
    S_{\textsc{m}}+S_{\textsc{g}}=\int\!dt\;\frac{m}{2}\,\delta_{ij}\dot{z}^i\dot{z}^j+\int\!dt\;\frac{m}{4}\,\ddot{h}_{ij}z^iz^j-\frac{1}{64\pi}\int\!d^4x\;\bigl(\partial_{\rho}h_{ij}\bigr)\bigl(\partial^{\rho}h^{ij}\bigr)\,.
\end{equation}
The second term on the righthand side may be identified as the interaction term between the particle and the metric perturbation.

The action \eqref{E:bksjgfdf} has a form that describes a free particle coupled to a tensor field, and we immediately see a major difference from that in the Brownian particle case, because the interaction term is quadratic in $z^i$ instead of being linear.

If we variate $z^i$, we obtain the geodesic deviation equation \eqref{E:udnfere}
\begin{equation}\label{E:bgswowr}
    \ddot{z}_i(t)=-\frac{1}{2}\,\ddot{h}_{ij}(t,\bm{0})\,z^j(t)\,,
\end{equation}
expressed in the frame \eqref{E:bvdgsw} where $u^{\mu}=(1,0,0,0)$. On the other hand, if we variate $h_{ij}$, we obtain
\begin{equation}\label{E:bgbgbeas}
    \square h^{ij}(t,\bm{z})=-16\pi\,\frac{m}{2}\frac{\partial^2}{\partial t^2}\bigl[z^i(t)z^j(t)\bigr]\,\delta^{(3)}(\bm{z})\,,
\end{equation}
where $\dfrac{m}{2}\dfrac{\partial^2}{\partial t^2}\bigl[z^i(t)z^j(t)\bigr]\,\delta^{(3)}(\bm{z})$ acts like an energy momentum stress tensor evaluated at the given geodesic, $\bm{z}=\bm{0}$. The solution to \eqref{E:bgbgbeas} is formally given by
\begin{equation}\label{E:bgskbgsgf}
    h^{ij}(x)=h_{\mathrm{h}}{}^{ij}(x)-16\pi\int\!d^4x'\;G^{(h)}_{R,0}{}^{ij}{}_{kl}(x,x')\,J^{kl}(x')
\end{equation}
where $x^{\mu}=(t,\bm{z})$, and we have denoted $J^{ij}(x)=\dfrac{m}{2}\dfrac{\partial^2}{\partial t^2}\bigl[z^i(t)z^j(t)\bigr]\,\delta^{(3)}(\bm{z})$. The homogeneous solution $h_{\mathrm{h}}{}^{ij}(x)$ satisfies the free wave equation $\square_x h_{\mathrm{h}}^{ij}(x)=0$, and $G^{(h)}_{R,0}{}^{ijkl}(x,x')$ is the retarded Green's function associated with the free metric perturbation $h_{\mathrm{h}}^{ij}(x)$ in the transverse traceless gauge.

Plugging the formal solution \eqref{E:bgskbgsgf} into the geodesic equation \eqref{E:bgswowr} we arrive at
\begin{equation}\label{E:ewriudsd}
    \ddot{z}^i(t)-4\pi m\,z^j(t)\frac{\partial^2}{\partial t^2}\int^t_{t_a}\!dt'\;G^{(h)}_{R,0}{}^i{}_{jkl}(t-t')\,\frac{\partial^2}{\partial t'^2}\bigl[z^k(t')z^l(t')\bigr]=-\frac{1}{2}\,\frac{\partial^2}{\partial t^2}h_{\mathrm{h}}{}^i{}_j(t)\,z^j(t)\,.
\end{equation}
Here we have suppressed the spatial coordinates because the above equation are evaluated at the spatial origin. Note that we have not manifestly casted this equation of motion in a gauge-invariant form. To make \eqref{E:ewriudsd} gauge invariant, we can rewrite the nonlocal term by integration by parts and with the help of $\ddot{h}_{\mathrm{h}}{}^{ij}(t)=-2R_{\mathrm{h}}{}^{i0j0}(t)$, i.e, via the free Riemann tensor $R_{\mathrm{h}}{}^{i0j0}(t)$ associated with $h_{\mathrm{h}}{}^{ij}$. In so doing, the nonlocal term on the lefthand side of \eqref{E:ewriudsd} can be expressed in terms of the retarded Green function of the free Riemann tensor $R_{\mathrm{h}}{}^{i0j0}(t)$, and the force term on the righthand side by the free Riemann tensor itself. Eq.~\eqref{E:ewriudsd} then has a gauge-invariant form in Minkowski background.

Next, we promote $z^i(t)$ and $h_{\mathrm{h}}{}^{ij}(t,\bm{0})$ to the operators, and write \eqref{E:ewriudsd} into an operator equation as it is. Then we arrive at a quantum Langevin equation of $\hat{z}^i(t)$, describing the quantum evolution of $\hat{z}^i$ under the influence of the gravitons in a self-consistent manners up to the order quadratic in $h_{ij}$. Eq.~\eqref{E:ewriudsd} takes a form slightly different from (35) in~\cite{ChoHu22,ChoHu23} because there the derivatives with respect to $t'$ have been moved to $G^{(h)}_{R,0}{}^i{}_{jkl}(t-t')$.

The expression $h_{\mathrm{h}}^{ij}(t)$ on the righthand side of \eqref{E:ewriudsd} accounts for quantum fluctuations of the free gravitons, and thus we can interpret it as the quantum noise from the graviton environment. However, the strength of the noise also depends on the dynamical state of $z^i(t)$, so we have a multiplicative noise, rather than the additive noise we often have for the Brownian particle. The nonlocal term contains derivatives of the retarded Green's function of free graviton, so it will account for the causal influence of $z^i$ at earlier times in terms of emitted gravitons due to the coupling with $z^i(t)$, in the sense similar to the electromagnetic radiation field due to the moving charge~\cite{HsHu23}. Thus, the (nonlinear) gravitational self-force is also contained in this expression. However, we note since we have discarded the higher-order contributions of the metric perturbation, our configuration implicitly covers only the lower energy regime such that we should have a scale of the energy cutoff in the retarded Green's function. This has two subtle implications~\cite{HAH22,HsHu23}: 1) the dynamics of $z^i$ becomes non-Markovian, and 2) the causal structure is slightly smeared due to the presence of the cutoff scale. In addition, the dynamics described by \eqref{E:ewriudsd} is also highly super-Ohmic due to the presence of the derivatives with respect to $t$ and $t'$ in the nonlocal expression. Finally, from the experience of the Brownian particle, the nonlocal term will give out a cutoff-dependent term that will contribute to the correction or renormalization of the parameters in the equation. But the geodesic deviation, $z^i$, when in the absence of the interaction with the gravitons, behaves like a free particle, so we do not have a suitable parameter to absorb this contribution. Hence the interpretation or the physical effect of this contribution is not yet clear.

The lower limit $t_a$ of the $t'$-integral of the nonlocal term in \eqref{E:ewriudsd} is unspecified yet because in the cosmological setting, we have no information about the initial conditions. They are usually inferred a posteriori from the observation data nowadays. Finally, since \eqref{E:ewriudsd} is nonlinear, it may inherently have an ambiguity in operator ordering. These are the pedagogical issues centered on this Langevin equation, which in summary is a nonlinear integro-differential operator equation with a multiplicative noise. Thus, even though we have cleared up the aforementioned ambiguities, it is next to impossible to analytically solve the dynamics described by \eqref{E:ewriudsd}.

In practice, we usually discard the nonlocal term for the time scale of our interest, because we reasonably assume that its effects are minute due to the small gravitational coupling, weak gravitational field, and slow motion of the reduced system. With this simplification, Eq.~\eqref{E:ewriudsd} reduces to~\cite{DM99, PWZ21b}
\begin{equation}
    \ddot{z}^i(t)+\frac{1}{2}\,h_{\mathrm{h}}{}^i{}_j(t)\,z^j(t)=0\,.
\end{equation} 
This is essentially the equation of motion for a parametric oscillator, except for that $h_{\mathrm{h}}{}^i{}_j(t)$ is not positive-definite and is stochastic, and the dynamics becomes much more manageable.

\section{Graviton Noise: Master equation for gravitational decoherence}

In this section we study a fundamental issue, that of environment-induced decoherence: the transition of a system from quantum to classical due to its interaction with an environment,  and in particular,  gravitational decoherence of a mass system interacting with a gravitational field as its  environment.
Gravitational decoherence  refers to the effects of gravity in actuating the classical appearance of  a quantum system. Because the underlying processes involve issues in general relativity (GR), quantum field theory (QFT) and quantum information (QI),  this topic  has multi-disciplinary theoretical significance. There is a great variety of gravitational decoherence theories, many of them involving physics that diverge from GR and/or QFT. Here we shall stick to these two cornerstones of modern physics. 

Stating our aim forthright,   we want to show that the Hadamard function of the quantum gravitational field introduced earlier which measures the noise of gravitons is instrumental in gravitational decoherence. Technically  the best way to show this is to introduce the influence functional (IF) which captures the influence of the environment on the system in two parts, represented by two nonlocal kernels:  the retarded Green function in the dissipation kernel is responsible for quantum dissipation and the Hadamard function in the noise kernel is responsible for quantum decoherence.  Instead of going through the formal derivation via the IF formalism which we have devoted an appendix for a more systematic presentation, we will put more emphasis on motivating the physics, starting with some familiar simpler cases.  

A proper understanding of dissipation/backreaction and noise/fluctuations requires some basic knowledge of nonequilibrium statistical mechanics, such as the  kinetic equations for quantum transport,  and stochastic processes, such as the master, Langevin or the Fokker-Planck equations.  The quantum versions of these latter topics make up the important emergent field of  open quantum systems \cite{OQS}. These subjects are probably familiar to researchers in quantum optics but perhaps less so for researchers in general relativity and quantum field theory. As this paper is in the nature of a tutorial, we will adopt a somewhat pedagogical rather than formal approach. Also, to avoid excesses, we shall take the path of least action, select out only a handful of illustrative or essential papers to highlight the main ideas and describe the key steps leading to a basic understanding of gravitational decoherence. 

\subsection{Quantum Brownian motion: Decoherence in the configuration space basis }

We begin with a description of quantum decoherence using the generic quantum Brownian motion (QBM) model providing an easier and more familiar pathway towards understanding the effect of quantum noise in the environment on a quantum system. With this as a backdrop, we can then point out the physical differences when we display the master equation for gravitational decoherence. In the spirit of ``taking the path of least action" we suggest three sets of papers for the reader to read: 1) For an easy grasp of the idea behind decoherence we recommend starting with Zurek's 1991 Physics Today essay \cite{ZurPT}. To broaden the perspective, browse over some interesting papers in \cite{Joos}.  To follow the ensuing developments, read the reviews \cite{DecRev}.  The familiar  Caldeira-Leggett (CL) master equation (Markovian)  is invoked to explain the essential physics of quantum dissipation and diffusion, the latter being responsible for quantum decoherence. 2) A more complete nonMarkovian  master equation valid for all bath temperatures and spectral densities is the Hu-Paz-Zhang (HPZ) master equation \cite{HPZ92}. Paz et al \cite{PHZ93} carried out a detailed study of decoherence using the Fokker-Planck version of the HPZ master equation for the Wigner distribution, a derivation of which is shown by Halliwell and Yu \cite{HalYu}. See also the recent paper by Homa et al \cite{Homa23}.  3) A master equation for the study of gravitational decoherence was derived by Anastopoulos and Hu \cite{AHGraDec}, and Blencowe \cite{Blencowe} independently. We shall highlight the key features in them. Learning from these few  papers will help the uninitiated yet strongly motivated reader leapfrog to current frontline research on this subject.  

\subsubsection{Phenomenology: Decoherence of a quantum particle in an Ohmic high temperature bath}

We begin with the von Neumann  equation which should be familiar to the reader from quantum mechanics and  statistical mechanics courses.  It describes the time evolution of the density operator $\hat \rho$ of a closed quantum system, `closed' referring to the fact that the system is isolated from the environment, in the dynamics generated by the Hamiltonian $\hat H$:
\begin{equation}
    \frac{\partial \hat \rho  }{\partial t} = -\frac{i}{\hbar} [ \hat H(t), \hat \rho ] \,.
\end{equation}
For the motion of a free particle of mass $M$  interacting with an (Ohmic) bath of (high) temperature $T$, two additional terms appear in this equation: one for dissipation, another for diffusion.  In a coordinate representation of the density matrix $\rho (x, x', t)$  the master equation for this open quantum system reads:
\begin{eqnarray}
 \frac{\partial \rho (x, x', t) }{\partial t} &=&  -\frac{i}{\hbar}  [ H(t), \rho (x, x, t)] \nonumber\\
    && - \gamma\, (x-x') \Bigl(\frac{\partial }{\partial x}- \frac{\partial }{\partial x'}\Bigr)\rho (x, x', t)     -\frac{2M\gamma k_BT}{\hbar^2}(x-x')^2 \rho (x, x', t) = 0\,.
\end{eqnarray}
where $\gamma$ is the damping constant (note the subtle difference between relaxation and dissipation, see, e.g, Chapter 1 of \cite{CalHu08}), $\tau_{dis} = 1/\gamma$ being the dissipation time scale. In the term following it we see the quantum diffusion constant $D^{CL}_{pp} = \frac{2M\gamma k_BT}{\hbar^2} \equiv \gamma\lambda_{tdB}^{-2}$ where $\lambda_{tdB}$ is the thermal de Broglie wavelength.  We now use this master equation to analyze decoherence effects\footnote{A noteworthy point is, mathematically one can always construct a basis wherein a symmetric matrix is in a diagonal form. The physical reason why the reduced density matrix of a quantum system after interacting with its environment could in time  become diagonal is nontrivial. This is the idea behind Zurek's `pointer-basis', which depends on several factors, most salient  of them all is the form of the interaction Hamiltonian. A good illustration of this point is in the plots of~\cite{PHZ93}.}. After the phenomenological description in \cite{ZurPT} reproduced here, the reader may wish to look up the more quantitative analysis based on numerical calculations in \cite{HPZ92,PHZ93}.

Let $\psi_\pm(x, t)$ be the wave functions of two Gaussian wavepackets  located initially $(t = 0)$ at $x= \pm x_0$ with the same initial spread $\sigma$. Assume that their separation $\Delta x \gg \sigma$.   Now consider a coherent superposition of these two Gaussian wave packets $\psi(x) = \psi_+ (x) + \psi_-(x)$.   The corresponding density matrix $\rho (x,x') = \psi(x) \psi^*(x')$ has four peaks: two on the diagonal  $x=x'$ and two off the diagonal (see the figures in \cite{ZurPT}). The off-diagonal peaks are there because of quantum coherence, they will diminish in time and this signifies that decoherence is in progress. In situations where decoherence is consummated, position would emerge as an approximate preferred basis. Afterwards the classical notion of probability distribution takes hold. It is only when decoherence is complete that one can say that there is an equal probability of finding the particle in either (but not both) of the specific locations where the Gaussian wave packets are centered at.  

Referring to the master equation above,  what is responsible for quantum decoherence is the quantum diffusion term. We will see that all diffusion terms are based on the noise kernel which for a quantum field is rooted in the Hadamard function. Since it is proportional to $(x - x')^2$ its effect on the diagonal peaks $x=x'$ is small but  much stronger on the off-diagonal peaks at $\Delta x^2$ apart.  Quantum coherence  disappears exponentially fast on a decoherence time scale  $\tau_{dec} = \tau_{dis} (\lambda_{tdB}/ \Delta x)^2$.

For macroscopic objects, the decoherence time scale $\tau_{dec}$ is typically orders of magnitude smaller than the dissipation time $\tau_{dis} = \gamma^{-1}$. As an illustration, Zurek estimated that for a system at room temperature $T= 300\,\textrm{K}$ with mass $m = 1\,\textrm{g}$ and separation $\Delta x=1\,\textrm{cm}$, the ratio $\tau_{dec} /\tau_{dis} = 10^{-40}$. To get a much longer decoherence time one needs to work with systems of smaller masses, at the atomic scale, and at very low temperatures.  Quantum coherence being at the heart of quantum science and engineering, and quantum entanglement being the resource for quantum information processing, quantum computing, quantum communications, exploring ways to stretch out the decoherence time of quantum devices becomes a  necessary prerequisite. Largely this is related to finding measures to mitigate the impact of noise in the environment on the quantum system of interest.

\subsubsection{Exact non-Markovian master equation for a quantum system in a general environment}

The master equation used in the above by Zurek to expound these basic physics of open quantum systems is called the Caldeira-Leggett (CL) master equation \cite{CalLeg83}. It is the closest to those in textbooks because it is for quantum systems interacting with a high temperature Ohmic bath, the range where  familiar relations like the Einstein-Kubo formula are valid. Low temperature non-Ohmic baths pose a greater challenge.   Consider as our system (S) a quantum harmonic oscillator with mass $M$ and bare frequency $\Omega$ coupled bilinearly to $n$ oscillators  with mass $m_n$ and frequency $\omega_n$ making up the thermal bath (B).  The total Hamiltonian consists of three parts
\begin{equation}
 \hat{H}=\hat{H}_S+ \hat{H}_B  + \hat{H}_I\,, \nonumber
\end{equation}
with
\begin{align}
 \hat{H}_S&= \frac{\hat{p}^2}{2M} +\frac12 M\Omega^2 \hat{x}^2\,,&\hat{H}_B &= \sum_n \hat{H}_n =\sum_n \frac{\hat{p}_n^2}{2m_n}+\frac12 m_n \omega_n^2 \hat{q}_n^2\,,&\hat{H}_I&= \hat{x}\sum_n C_n \hat{q}_n\,,\label{eq:full_hamiltonian}
\end{align}
where $\hat{x}$, $\hat{p}$ and $\hat{q}_n$, $\hat{p}_n$ are the coordinates and momenta of the system and bath oscillators, respectively. The fact that we are dealing with harmonic  oscillators plus the assumption of a bilinear type of coupling between the system  and the bath oscillators (with strength $C_n$) preserving the Gaussian form makes this problem analytically solvable.  This is indeed the   reason why an exact master equation for the reduced density matrix $\hat{\rho}$ of the open system (obtained by tracing over the environmental degrees of freedom, and where an initially factorizable condition between the system and the bath is assumed) can be found under general conditions, as obtained by Hu, Paz and Zhang \cite{HPZ92}:
	\begin{equation} \label{HPZ}
	 i \hbar \frac{\partial \hat{\rho}}{ \partial t}= [\hat{H}_S,\hat{\rho}]+ \Gamma(t) [\hat{x},\{\hat{p},\hat{\rho}\}] 
        -i D_{pp}(t)[\hat{x},[\hat{x},\hat{\rho}]] 	
	+i (D_{px}(t)+D_{xp}(t))[\hat{x},[\hat{p},\hat{\rho}]]\,,   
	\end{equation}
where $\Gamma(t)$ is the dissipation `constant' and $D_{pp}(t)$ $D_{px}(t)$ are the diffusion `constants'. In truth, they are rather complicated time-dependent functions whose explicit expressions are presented in Eq.~(2.41) of \cite{HPZ92}, or Eq.~(2.46) there in the weak coupling limit. A quick glance at these expressions shows that  $\Gamma$ is related to the dissipation kernel $\mu$ and the diffusion $D$ functions are related to the noise kernel $\nu$.  The corresponding quantities for a quantum field environment are the retarded and Hadamard Green's functions.

The master equation Eq.~\eqref{HPZ} expressed in the position basis representation has the form:
	\begin{eqnarray}
	&&i \hbar \frac{\partial}{\partial t} \rho(x,x',t)=\Big[ - \frac{\hbar^2 }{2M}\left(\frac{\partial^2}{\partial x^2}-
	\frac{\partial^2}{\partial x'^2} \right) + \frac{1}{2} M \Omega_{\rm ren}^2(t)  \left(x^2-x'^2\right) -i \hbar\Gamma(t) (x-x') \left(\frac{\partial}{\partial x}-
	\frac{\partial}{\partial x'} \right) \nonumber \\
		&& \qquad -i D_{pp}(t) (x-x')^2 + \hbar \left(D_{px}(t) + D_{xp}(t)\right) (x-x') \left(\frac{\partial}{\partial x}+	\frac{\partial}{\partial x'} \right) \Big]  \rho(x,x',t). \label{HPZx}
	\end{eqnarray}
where $\Omega_{\rm ren}$  is the renormalized natural frequency of the system oscillator. The functionality of the diffusion functions $D_{pp}$, $D_{px}+ D_{xp}$  responsible for quantum decoherence  can now be seen more clearly:   $D_{pp}$ suppresses spatial coherence from the long-range correlations between spatially separated components of the system wave-function  whereas $D_{px}+ D_{xp}$  suppresses mixed spatial-momentum coherence from the correlations between components in the system wave-function separated both in position and momentum. 

Comparing with the Caldeira-Leggett master equation in form, one notices an extra term $D_{px}+ D_{xp}$ in the HPZ master equation, called anomalous diffusion function. In contrast to the normal diffusion function $D_{pp}$ which has increasing contribution at higher temperatures, the anomalous diffusion contribution increases at lower temperatures (that is why it is absent in the CL equation).  As mentioned above, for the design of quantum devices making use of quantum coherence properties,  the lower the temperature, the better one's chance to see a longer decoherence time (also with supra-Ohmic baths). In that regime the CL equation fails\footnote{See Ref.~\cite{HomaCL} for the deficiencies of the CL equation, e.g., mathematically the density operator being not positive definite. It is worth mentioning an important class  of master equation which is most popular, called the Lindblad-Gorini–Kossakowski–Sudarshan equation \cite{Lindblad,GKS}. The beauty of it is that it is completely positive, but it also has pathologies, e.g., it violates the uncertainty principle at very low temperatures \cite{LinbPatho}.} and  one needs to work with the HPZ master equation \cite{Homa20}.  One can see some examples in \cite{HPZ92} and \cite{PHZ93} of how the decoherence time varies with temperature of the bath, using the Gaussian wave packet example.

\subsubsection{System and bath with time-dependent frequencies, squeezed thermal bath, quantum field}

A third step in the development of an open quantum systems repertoire is to treat a parametric oscillator with time-dependent natural frequency interacting with a bath of oscillators also with time-dependent frequencies in squeezed thermal states. This is useful for quantum optics problems, and,  maybe a bit unexpectedly, also for quantum cosmological problems.  Parametric amplification is a basic mechanism in laser physics, it also is an ubiquitous process in cosmology -- the expansion of the universe squeezes a quantum field,  turning a vacuum or thermal state into a squeezed vacuum or squeezed thermal state. Cosmological particle creation is arguably the most sensational yet consequential vacuum squeezing process \cite{GriSid}. Here we shall only quote the form of the master equation from the results of \cite{HM94}, the third of the QBM trilogy,  after \cite{HPZ92,HPZ93}. 
    
The HPZ master equation for a harmonic oscillator with time-dependent frequency interacting with a bath of harmonic oscillators with time-dependent frequencies was derived in~\cite{HM94}. It reads  
\begin{align}\label{HM}
	i\hbar\,\frac{\partial\hat{\rho}}{\partial t} &= \bigl[\hat{H}_{\text{ren}},\hat{\rho}\bigr] + i\,D_{pp}\bigl[\hat{x},\bigl[\hat{x},\hat{\rho}\bigr]\bigr] + i\,D_{xx}\bigl[\hat{p},\bigl[\hat{p},\hat{\rho}\bigr]\bigr]+	 i\,(D_{px}+D_{xp})\bigl[\hat{x},\bigl[\hat{p},\hat{\rho}\bigr]\bigr] + \Gamma\bigl[\hat{x},\bigl\{\hat{p},\hat{\rho}\bigr\}\bigr] \,.
\end{align}

Finally, instead of a harmonic oscillator bath one can use a scalar field, as was done in \cite{UnrZur}.
Unlike an oscillator bath where one can specify or design a specific spectral density function, for a quantum field the spectral density is fixed, e.g., a scalar field in four-dimensional Minkowski spacetime is Ohmic. The theoretical transition from a parametric oscillator bath to a quantum field is shown in \cite{HM94} where one can find derivations of the Unruh effect for uniformly accelerated oscillators \cite{Unr76}, the Davies-Fulling effect for moving mirrors \cite{DavFul}  as well as  the Hawking effect in black holes \cite{Haw75} and the Gibbons-Hawking effect in de Sitter universe \cite{GibHaw77}. 

\subsection{Gravitational decoherence happens in the energy basis, not in the configuration space basis}

Having expounded the physical essence of decoherence in the generic QBM model with some requisite technicalities, we only need to point out the key differences when a gravitational field acts as the environment. 

\subsubsection{Major differences from Brownian motion with bilinear interactions}

First, our system refers to the masses of particles, mirrors or macroscopic objects. In the harmonic oscillator model we saw earlier, which in quantum optics mimics a harmonic atom, the dynamical variables $x$ or $Q$ are the internal degrees of freedom of the atom, the energy levels of electrons. Here, the quantities of interest are the external or mechanical degrees of freedom -- the movement of particles, the displacement of mirrors, the trajectories of satellites, etc.  For example, if one wants to calculate the effects of a gravitational wave on an atom \cite{BouRot}, one would have to work with the mass quadrupole of the atom, via gravitational interactions, not the electric dipole via electromagnetic (EM) interactions. Naturally the effects of gravitational decoherence are scaled down drastically compared to decoherence via EM interactions.  

Second, the interaction would not be bilinear, but nonlinear,  because the coupling would be between the gravitational radiation (which are the weak perturbations $h_{\mu\nu}$ off of a background spacetime) acting as the environment to the system's stress energy tensor $T_{\mu\nu}$. The system could be as simple as two masses and their geodesic separation would be the system variable. We are interested in whether and how the gravitational field environment decoheres the quantum system. The interaction Hamiltonian has   $h_{\mu\nu}$ multiplied to $T_{\mu\nu}$, which contains a product of the proper velocity $u^\mu u^\nu$ of the mass, thus in gravitational interactions the system variable enters nonlinearly.  

This nonlinearity makes the derivation of the master equation a lot more challenging, prompting the introduction of suitable approximations. Functional perturbative methods have been introduced for nonlinear QBM problems \cite{HPZ93} where the interaction Hamiltonian is of the form $f(x)q^k$,  between any arbitrary function of the system variable $x$ and  a polynomial form of the bath variable $q$.  New ways to fully tackle gravitational interactions have yet to be worked out.  At present, a Markovian master equation has been derived, the so-called ABH master equation of Blencowe \cite{Blencowe}, Anastopoulos and Hu \cite{AHGraDec} mentioned at the beginning, while a nonMarkovian equation is currently being worked out \cite{CHHnM}.  

Finally,  perhaps the most important feature is that gravitational decoherence based on general relativity and quantum field theory occurs in the energy basis, not in the configuration basis, as many alternative quantum or gravitational theories which "uses" gravity to help explain quantum foundation issues such as the `collapse' of the wave function.  We will not dwell on these alternative theories but refer the reader to two articles \cite{HuGraDecSCG,AHVac} where one can find the source references and how they relate to the theories we are expounding here, rooted in GR + QFT. 

\subsubsection{A Markovian master equation for gravitational decoherence}

In the  gravitational decoherence theory of Anastopoulos \& Hu and Blencowe (ABH) the source of decoherence comes as noise  from gravitational waves (classical weak perturbations) or of gravitons (quantized linear perturbations). What counts here is the transverse-traceless components of the metric perturbations which are the dynamical degrees of freedom of a gravitational field,  not the Newtonian force,  which is pure gauge. Marking this difference is essential in the theoretical implications of gravitational entanglement experiments and noteworthy here. We have discussed graviton noise earlier, they are the quantum fluctuations in the gravitational field. The source may be stochastic gravitons of cosmological or astrophysical or structural origins, the last referring to the underlying `textures' of an emergent spacetime (see, e.g., \cite{E/QG} for an explanation). 

A brief description of the methodologies:  Blencowe considered a quantum scalar field interacting with a graviton bath. To make the scalar field look like a mass he conjured a `coherent state ball'. The effects of the environment on the system is captured in the influence action with the dissipation and noise kernels, the latter is responsible for gravitational decoherence. Blencowe's master equation is obtained under the Born and Markov approximations. Anastopoulos and Hu also assumed a quantum scalar field as the system which interacts with the quantized  weak perturbations off the Minkowski spacetime as the gravitational field environment. They  used the canonical quantization method and derived a  master equation  (Markovian, to linear perturbative order in Newton constant $G$) for this system. Projecting it to the single-particle subspace in a Fock space representation, they obtained a master equation for a single particle which simplifies significantly in the non-relativistic regime. Under these approximations the two independently derived  master equations have the same (ABH) form:
\begin{eqnarray}
 \frac{\partial \hat{\rho}}{\partial t} = -\frac{i}{\hbar}  [\hat{H}, \hat{\rho}] - \frac{\tau}{16M^2} (\delta^{ij} \delta^{kl} + \delta^{ik}\delta^{jl}) [\hat{p}_i
\hat{p}_j,[\hat{p}_k\hat{p}_l, \hat{\rho}]] \label{ABHME}
\end{eqnarray}
 where $\tau$ is a constant of dimension time and $\hat{H}=\frac{\hat{p}^2}{2M}$.
The structure is rather transparent:  The $ (\delta^{ij} \delta^{kl} + \delta^{ik}\delta^{jl}) $ is a tensor projection operator, and, comparing with the  exemplary QBM model earlier,  in the interaction Hamiltonian, in place of the system variable $x$ in QBM we have the stress energy tensor of the nonrelativistic particle. This marks one important difference we emphasized before, namely, gravitational decoherence occurs in the energy basis, not in  the  position basis.  Ostensibly gravitational decoherence based on general relativity does not help in explaining the ``collapse of the wavefunction"  in space. 

For motion in one dimension, this ABH master equation simplifies to
 \begin{eqnarray}
 \frac{\partial \hat{\rho}}{\partial t} = -\frac{i}{\hbar}  [\hat{H}, \hat{\rho}] - \frac{\tau}{2} [\hat{H},[\hat{H},\hat{\rho}]], \label{1dim}
 \end{eqnarray}
where $\tau$, the relevant time scale of the process under study, measured in reference to the Planck time $\tau_P = 0.54 \times 10^{-43}\,\textrm{s}$, is given by 
 \begin{eqnarray}
 \tau = \frac{32\pi G \Theta}{9} = \frac{32\pi}{9} \tau_P (\Theta/T_P),
 \end{eqnarray}
where $T_P = 1.4\times 10^{32}\,\textrm{K}$ is the Planck temperature and $\Theta$, a free parameter introduced in \cite{AHGraDec},  is the noise temperature related to the strength of gravitational perturbations in the initial state. $\Theta$ is  a measure of the power $P$ carried by the noise, $P \sim\Theta \Delta \omega$, where $\Delta \omega$ is the band-width of the noise. If we regard the parameter  $\Theta$ as a noise temperature originating from some emergent gravity theory, then it  need not be related to the Planck temperature $T_P$ or any Planck scale, which marks the gravitational field's fundamental quantum structure. Emergent gravity theories permit a broader range of parameters,  enhancing the probability of observing gravitational decoherence effects. 

How do we understand the appearance of a free parameter $\Theta$ in the master equation for gravitational decoherence? Just like in the QBM examples the decoherence rate depends not only on the matter–gravity coupling, the temperature of the bath,  but also on the response properties of the environment. $\Theta$  conveys some coarse-grained (mesoscopic) information reflective of the underlying micro-structures of spacetime, similar to temperature with regard to molecular motion, or the  spectral density function of the environment in Brownian motion.  It is in this sense that  measurements of gravitational decoherence may reveal the underlying ‘textures’ of spacetime, as speculated in~\cite{AHGraDec}.
 

Measurement of the gravitational decoherence rate, if this effect
due to gravity can be cleanly separated from other sources, may thus provide valuable
information about the gravitational ‘noise temperature’ or ‘spectral density’, which Anastopoulos and Hu called  the  ‘textures’ of spacetime, beneath that described by classical general relativity. It is in this  sense that gravitational decoherence may be relevant to quantum gravity, theories about the microscopic structures of spacetime. 

The ABH model can be generalized to photons, where we can obtain a master equation for a general photon state \cite{LagAna}. For a single photon,
\begin{eqnarray}\label{result}
 \frac{\partial \hat{\rho}}{\partial t} &= -\frac{i}{\hbar} \, [ \hat{H } , \hat{\rho}] - \frac{\tau_{ph}}{2}   \left( \delta^{in} \delta^{jm} -\frac{1}{3} \delta^{ij} \delta^{nm} \right) \left[ \frac{\hat{p}_i \, \hat{p}_j}{\hat{p}_0} , \left[ \frac{\hat{p}_n \, \hat{p}_m}{\hat{p}_0}, \hat{\rho}  \right] \right] \, ,
\end{eqnarray}
where $\hat{H} = |\hat{\bf p}|$ and $\tau_{ph} = 4 G \Theta$.
Note also the  work in  Ref. \cite{Oniga}, which  derives a related theory of gravitational decoherence for matter and light.

A brief summary of different types of experiments which can measure gravitational decoherence and a comparison of results obtained in theories based on GR such as the above versus those which are not GR based (such as the Diosi-Penrose model) can be found in \cite{AHVac}, and with more details in \cite{Bassi},  from which the reader can trace the sources of such theories and earlier theoretical developments. 

On the theory side, since the ABH master equation is of the Lindblad form, and as we commented in an earlier footnote, the Lindblad equation has pathologies at very low temperature, there is an obvious need to go beyond, avoid making the Born-Markov approximations,  and obtain a nonMarkovian master equation valid at low temperatures.  This work is in progress \cite{CHHnM}.

For pedagogical reasons we chose to present quantum decoherence and  gravitational decoherence along the environment-induced decoherence \cite{envdec} pathway, using master equations, because from there one can identify the different quantum diffusion terms where environmental noise enters.  There are other pathways to study decoherence phenomena, less explicit in the features of the environments.  This includes the consistent history \cite{conhis} scheme of Griffith and Omnes  and the decoherent history \cite{dechis} scheme of Gell-Mann and Hartle, which advocate that histories are the right way to address foundational issues of quantum mechanics and, in particular, closed systems \cite{Har93} like the universe \cite{DecCos}. Technically, the relation between the decoherent histories formalism and the closed-time-path (CTP)  formalism has been made explicit in the correlation histories formalism of Calzetta and Hu \cite{CalHuCorHis}, and the relation between the CTP or the Schwinger-Keldysh formalism and the Feynman-Vernon influence functional formalism is by now quite well-known. For a brief description of how these different quantum decoherence formalisms are used to address quantum information issues in cosmology, see, e.g., Sec. 2 of \cite{HHDecCos} and \cite{HHEntCos}. The reader can find an abundance of references to earlier works in these papers.   For a description of gravitational decoherence using the decoherent histories/influence functional method, see, e.g., Kanno et al \cite{KaSoTo21,KanGraDec}.

\section{Cosmology: Metric Perturbations in de Sitter Spacetime}

We will follow the same decomposition of the metric tensor, as in Sec.~\ref{S:bgfkdf}, except that now the background metric $\eta_{\mu\nu}$ is not limited to Minkowski spacetime. Our goal in this section is to derive the equation of motion of the metric perturbation in de Sitter background.

Sometimes we use notations $g^{(0)}_{\mu\nu}$ to denote the background metrics and $g^{(1)}_{\mu\nu}$ the corresponding first-order perturbation in metrics, i.e., in this case, $g^{(1)}_{\mu\nu}=h_{\mu\nu}$. Everything discussed in Sec.~\ref{S:bgfkdf} becomes more involved here because we need to be more careful with the order of differentiation. The covariant derivatives in general do not commute.

\subsection{wave equation of the metric perturbation}
The Christoffel symbols due to metric perturbations is given by
\begin{align*}
	 \Gamma^{\rho}_{\mu\nu}=\frac{g^{\rho\sigma}}{2}\Bigl[g_{\mu\sigma,\,\nu}+g_{\nu\sigma,\,\mu}-g_{\mu\nu,\,\sigma}\Bigr]&=\frac{g^{\rho\sigma}}{2}\Bigl\{h_{\mu\sigma;\,\nu}+h_{\nu\sigma;\,\mu}-h_{\mu\nu;\,\sigma}\Bigr\}+\Gamma^{(0)}{}^{\rho}_{\mu\nu}\,,
\end{align*}
where $;$ is the covariant derivative relative to $\eta_{\mu\nu}$, so the correction to the Christoffel symbols is
\begin{equation}\label{E:qddsd}
	 \mathcal{S}^{\rho}_{\mu\nu}=\Gamma^{\rho}_{\mu\nu}-\Gamma^{(0)}{}^{\rho}_{\mu\nu}=\frac{g^{\rho\sigma}}{2}\Bigl\{h_{\mu\sigma;\,\nu}+h_{\nu\sigma;\,\mu}-h_{\mu\nu;\,\sigma}\Bigr\}\,.
\end{equation}
From the definition of the Riemann tensor $R^{\rho}_{\mu\alpha\nu}=\Gamma^{\rho}_{\mu\nu,\,\alpha}-\Gamma^{\rho}_{\mu\alpha,\,\nu}+\Gamma^{\rho}_{\beta\alpha}\Gamma^{\beta}_{\mu\nu}-\Gamma^{\rho}_{\beta\nu}\Gamma^{\beta}_{\mu\alpha}$, we can write the correction to the Riemann tensor in the form
\begin{align}
	 R^{\rho}_{\mu\alpha\nu}-R^{(0)}{}^{\rho}_{\mu\alpha\nu}&=\mathcal{S}^{\rho}_{\mu\nu;\,\alpha}-\mathcal{S}^{\rho}_{\mu\alpha;\,\nu}+\mathcal{S}^{\rho}_{\beta\alpha}\mathcal{S}^{\beta}_{\mu\nu}-\mathcal{S}^{\rho}_{\beta\nu}\mathcal{S}^{\beta}_{\mu\alpha}\,,
\end{align}
and the corresponding correction for the Ricci tensor is obtained by contracting indices $\rho$ and $\alpha$.

Since Eq.~\eqref{E:qddsd} includes all-order corrections, to compute the first-order correction of the Riemann tensor, we only need the first-order contribution from the tensor $\mathcal{S}^{\rho}_{\mu\nu}$,
\begin{equation}
	 \mathcal{S}^{(1)}{}^{\rho}_{\mu\nu}=\frac{\eta^{\rho\sigma}}{2}\Bigl\{h_{\mu\sigma;\,\nu}+h_{\nu\sigma;\,\mu}-h_{\mu\nu;\,\sigma}\Bigr\}\,,
\end{equation}
such that the first-order correction of the Riemann tensor is
\begin{equation}
	 R^{(1)}{}^{\rho}_{\mu\alpha\nu}=\mathcal{S}^{(1)}{}^{\rho}_{\mu\nu;\,\alpha}-\mathcal{S}^{(1)}{}^{\rho}_{\mu\alpha;\,\nu}\,.
\end{equation}
Thus the first-order correction of the Ricci tensor becomes
\begin{align}
	 R^{(1)}_{\mu\nu}&=\mathcal{S}^{(1)}{}^{\alpha}_{\mu\nu;\,\alpha}-\mathcal{S}^{(1)}{}^{\alpha}_{\mu\alpha;\,\nu}=\frac{1}{2}\Bigl[h_{\mu\alpha;\,\nu}{}^{\alpha}+h_{\nu\alpha;\,\mu}{}^{\alpha}-h_{;\,\mu\nu}-\square h_{\mu\nu}\Bigr]\,.\label{E:mdkfd}
\end{align}
Next we consider a little more specific case that the background metric describes spatially flat de Sitter universe, with $\operatorname{diag}\{\eta_{\mu\nu}\}=(-1,a^{2}(t),a^{2}(t),a^{2}(t))$ in the cosmic time $t$ frame for $t\geq0$.

The Einstein equation describing the de Sitter vacuum is given by
\begin{equation}\label{E:dfdaa}
	G_{\mu\nu}+\Lambda g_{\mu\nu}=R_{\mu\nu}-\frac{1}{2}\,Rg_{\mu\nu}+\Lambda g_{\mu\nu}=0\,,
\end{equation}
where $G_{\mu\nu}$ is the Einstein tensor. The positive cosmological constant $\Lambda$ can be related to the Ricci scalar $R$ by contracting \eqref{E:dfdaa}; yielding $R=4\Lambda$. In terms of the Ricci tensor $R_{\mu\nu}$, the Einstein equation takes a more convenient form,
\begin{equation}\label{E:dddaa}
	R_{\mu\nu}-\Lambda g_{\mu\nu}=0\,.
\end{equation}
Now consider the effect of the gravitons on the cosmological constant. Assume small metric perturbation arises around the classical de Sitter metric due to quantum fluctuations of gravitons. Let us choose the transverse traceless gauge, $h_{\mu}{}^{\nu}{}_{;\,\nu}=0$, $h=0$, $h_{0}{}^{\mu}=0$, to fix the degrees of freedom of the perturbed metrics $h_{\mu\nu}$. We systematically expand the Einstein equation \eqref{E:dddaa} with respect to small $h$
\begin{align*}
	 \Bigl[R^{(0)}{}_{\mu\nu}-\Lambda^{(0)}\eta_{\mu\nu}\Bigr]+\Bigl[R^{(1)}{}_{\mu\nu}-\Lambda^{(0)}h_{\mu\nu}\Bigr]+\Bigl[R^{(2)}{}_{\mu\nu}-\Lambda^{(1)}\eta_{\mu\nu}-\Lambda^{(1)}h_{\mu\nu}\Bigr]+\cdots&=0\,,
\end{align*}
where the correction to the cosmological constant starts from the second-order expansions due to the fact that the first-order expansion only describes the gravitational wave perturbation in the background spacetime. Order-by-order comparison shows
\begin{align}
	 R^{(0)}{}_{\mu\nu}-\Lambda^{(0)}\eta_{\mu\nu}&=0\,,\label{E:msdzjj}\\
	 R^{(1)}{}_{\mu\nu}-\Lambda^{(0)}h_{\mu\nu}&=0\,,\label{E:xzdfldfj}\\
	 R^{(2)}{}_{\mu\nu}-\Lambda^{(1)}\eta_{\mu\nu}-\Lambda^{(1)}h_{\mu\nu}&=0\,,
\end{align}
so that the correction to the cosmological constant $\Lambda^{(1)}$ is given by
\begin{equation}\label{E:mxnsu}
	 \Lambda^{(1)}=\frac{1}{4}\,\eta^{\mu\nu}R^{(2)}{}_{\mu\nu}\,,
\end{equation}
where we have used the gauge condition $h=0$. The correction turns out to be of the order $\mathcal{O}(h^{2})$.

To derive the equation of motion for the perturbed metric, we rewrite \eqref{E:mdkfd} by changing the orders of covariant derivatives with the identity
\begin{equation}\label{E:weuwi}
	 Q^{\mu\nu}{}_{;\,\alpha\beta}=Q^{\mu\nu}{}_{;\,\beta\alpha}+R^{\mu}{}_{\sigma\beta\alpha}Q^{\sigma\nu}+R^{\nu}{}_{\sigma\beta\alpha}Q^{\mu\sigma}
\end{equation}
so that we may make use of the gauge conditions $h_{\mu}{}^{\nu}{}_{;\,\nu}=0$ and find
\begin{equation}\label{E:vcn}
	R^{(1)}{}_{\mu\nu}=-\frac{1}{2}\square h_{\mu\nu}+R^{(0)}{}_{\mu\alpha\beta\nu}h^{\alpha\beta}+\frac{1}{2}\,R^{(0)}{}_{\rho\mu}h^{\rho}{}_{\nu}+\frac{1}{2}\,R^{(0)}{}_{\rho\nu}h^{\rho}{}_{\mu}\,.
\end{equation}
Plugging \eqref{E:vcn} into \eqref{E:xzdfldfj} yields
\begin{equation}\label{E:vzzcn}
	\square h_{\mu\nu}-2R^{(0)}{}_{\mu\alpha\beta\nu}h^{\alpha\beta}-R^{(0)}{}_{\rho\mu}h^{\rho}{}_{\nu}-R^{(0)}{}_{\rho\nu}h^{\rho}{}_{\mu}+2\Lambda^{(0)}h_{\mu\nu}=0\,.
\end{equation}
Since the de Sitter spacetime is maximally symmetric, its Riemann tensor takes a rather simple form
\begin{equation}\label{E:mzcdj}
	 R^{(0)}{}_{\mu\rho\alpha\nu}=\mathscr{K}\,\Bigl(\eta_{\mu\alpha}\eta_{\rho\nu}-\eta_{\rho\alpha}\eta_{\mu\nu}\Bigr)\,.
\end{equation}
where $\mathscr{K}$ is determined by contracting \eqref{E:mzcdj} and substituting the resulting Ricci tensor to \eqref{E:msdzjj}, and we find $\mathscr{K}=\Lambda^{(0)}/3$. Thus in the de Sitter space, Eq.~\eqref{E:vcn} is simplified to
\begin{equation}
	R^{(1)}{}_{\mu\nu}=-\frac{1}{2}\square h_{\mu\nu}+\frac{4}{3}\Lambda^{(0)}h_{\mu\nu}\,,
\end{equation}
and the wave equation of the graviton \eqref{E:xzdfldfj} reduces to
\begin{equation}\label{E:afkgjd}
	\square h_{\mu\nu}-\frac{2}{3}\Lambda^{(0)}h_{\mu\nu}=0\,.
\end{equation}
It has an even simpler form if we use the mixed-index metric perturbation tensor in the transverse traceless gauge.

When we apply the tensor d'Alembertian operator $\square$ on $h_{i}{}^{j}$, we obtain
\begin{align*}
	\square h_{i}{}^{j}&=g^{\alpha\beta}\Bigl\{\partial_{\alpha}\left(\nabla_{\beta}h_{i}{}^{j}\right)-\Gamma^{\rho}_{\beta\alpha}\left(\nabla_{\rho}h_{i}{}^{j}\right)-\Gamma^{\rho}_{i\alpha}\left(\nabla_{\beta}h_{\rho}{}^{j}\right)+\Gamma^{j}_{\rho\alpha}\left(\nabla_{\beta}h_{i}{}^{\rho}\right)\Bigr\}\\
    &=g^{\alpha\beta}\Bigl\{\partial_{\alpha}\left(\partial_{\beta}h_{i}{}^{j}\right)-\Gamma^{\rho}_{\beta\alpha}\left(\partial_{\rho}h_{i}{}^{j}\right)-\Gamma^{\rho}_{i\alpha}\left(\partial_{\beta}h_{\rho}{}^{j}\right)+\Gamma^{j}_{\rho\alpha}\left(\partial_{\beta}h_{i}{}^{\rho}\right)\Bigr\}+2\frac{\dot{a}^{2}}{a^{2}}h_{i}{}^{j}\\
	&=\nabla_{\alpha}\left(\partial^{\alpha}h_{i}{}^{j}\right)+2\frac{\dot{a}^{2}}{a^{2}}h_{i}{}^{j}\,,
\end{align*}
where we have used the non-zero Christoffel symbols are
\begin{equation}
	 \Gamma^{(0)}{}^{t}_{ii}=-\frac{1}{2}\,\eta^{tt}\eta_{ii,\,t}=a\dot{a}\,,\qquad\qquad\Gamma^{(0)}{}^{i}_{it}=\frac{1}{2}\,\eta^{ii}\eta_{ii,\,t}=\frac{\dot{a}}{a}\,.
\end{equation}
Since
\begin{align*}
	\nabla_{\alpha}\partial^{\alpha}h_{i}{}^{j}&=\partial_{\alpha}\partial^{\alpha}h_{i}{}^{j}+\Gamma^{\alpha}_{\beta\alpha}\left(\partial^{\beta}h_{i}{}^{j}\right)=\frac{1}{\sqrt{-g}}\,\partial_{\beta}\left(\sqrt{-g}\,g^{\beta\gamma}\,\partial_{\gamma}h_{i}{}^{j}\right)\,,
\end{align*}
and the expression on the righthand side is the scalar d'Alembertian operator $\square_s$ on $h_{i}{}^{j}$, we arrive at
\begin{equation}\label{E:zzsstt}
	\square h_{i}{}^{j}=\square_{s}h_{i}{}^{j}+2\,\frac{\dot{a}^{2}}{a^{2}}h_{i}{}^{j}\,,
\end{equation}
and together with \eqref{E:afkgjd}, we are led to
\begin{equation}\label{E:bgvsures}
	\square_{s}h_{i}{}^{j}=0\,.
\end{equation}
This is the starting point of discussing the quantum effects of gravitons over the cosmological scale.

\subsection{Gravitons in de Sitter spacetime}

In the spatially flat Friedmann-Lema\^itre-Robertson-Walker (FLRW) spacetime in the transverse, traceless gauge, we may expand the general gravitational wave, satisfying \eqref{E:afkgjd} or \eqref{E:bgvsures}, by
\begin{equation}
    h^{\mu\nu}(x)=\sum_{\lambda,\bm{k}}\mathsf{e}_{\lambda\bm{k}}^{\mu\nu}h_{\lambda\omega}(t)\,e^{i\bm{k}\cdot\bm{x}}\,,
\end{equation}
The two independent scalar functions $h_{\lambda\omega}(t)$, with $\lambda=+$ or $\times$ for each $\bm{k}$, then satisfy
\begin{equation}\label{E:nnbbnd}
    \Bigl[\partial_t^2+3H(t)\,\partial_t+\frac{\omega^2}{a^2}\Bigr]h_{\lambda\omega}(t)=0\,,
\end{equation}
from Eq.~\eqref{E:bgvsures}, with the Hubble parameter $H(t)=\dot{a}(t)/a(t)$, in which the overhead dot is the derivative with respect to the cosmic time $t$. However, it turns out much simpler if we switch to the conformal time $\eta$ by $dt=a(\eta)\,d\eta$. The equation of motion \eqref{E:nnbbnd} now has the form
\begin{equation}\label{E:jejswer}
    \Bigl[\partial_{\eta}^2+\frac{2a'(\eta)}{a(\eta)}\,\partial_{\eta}+\omega^2\Bigr]\,h_{\lambda\omega}(\eta)=0\,,
\end{equation}
In particular, in terms of cosmic time $t$, the scale factor $a(t)$ of de Sitter spacetime is given by $a(t)=e^{Ht}$ where $H$ is the Hubble constant, so by the conformal time the scale factor takes an algebraic form $a(\eta)=1/(\shortminus H\eta)$ with $\eta<0$. Here the prime denotes the derivative of the conformal time $\eta$.

Two unnormalized solutions to \eqref{E:jejswer} are
\begin{align}\label{E:doowojj}
	 h_{\lambda\omega}(\eta)&=N\,(\omega\eta)^{\frac{3}{2}}\operatorname{H}_{3/2}^{(1)}(\omega\eta)\,,&&\text{or}&h_{\lambda\omega}(\eta)&=N\,(\omega\eta)^{\frac{3}{2}}\operatorname{H}_{3/2}^{(2)}(\omega\eta)\,,
\end{align}
where $\operatorname{H}_{\nu}^{(1,2)}(x)$ are the Hankel functions. The normalization constant $N$ is determined by the condition $h_{\lambda\omega}^{\vphantom{*}}(\eta)\,\partial_{\eta}h_{\lambda\omega}^*(\eta)-h_{\lambda\omega}^*(\eta)\,\partial_{\eta}h_{\lambda\omega}^{\vphantom{*}}(\eta)=i/a^2(\eta)$, so the normalized mode functions are
\begin{equation}
    h_{\lambda\omega}(\eta)=\Bigl(\frac{\pi}{4H^2}\Bigr)^{\frac{1}{2}}\eta^{\frac{3}{2}}\operatorname{H}_{3/2}^{(1,2)}(\omega\eta)\,,
\end{equation}
and thus the general solution of $h_{\lambda\omega}(\eta)$ is given by
\begin{equation}
    h_{\lambda\omega}(\eta)=c_1\,\Bigl(\frac{\pi}{4H^2}\Bigr)^{\frac{1}{2}}\eta^{\frac{3}{2}}\operatorname{H}_{3/2}^{(1)}(\omega\eta)+c_2\,\Bigl(\frac{\pi}{4H^2}\Bigr)^{\frac{1}{2}}\eta^{\frac{3}{2}}\operatorname{H}_{3/2}^{(2)}(\omega\eta)\,,
\end{equation}
with the coefficients $c_1$, $c_2\in\mathbb{C}$ satisfying $\lvert c_2\rvert^2-\lvert c_1\rvert^2=1$.

The second-quantized metric perturbation becomes an operator of the form
\begin{equation}\label{E:sasguejr}
    \hat{h}^{\mu\nu}(x)=\sum_{\lambda,\bm{k}}\mathsf{e}_{\lambda\bm{k}}^{\mu\nu}\,\hat{a}_{\lambda,\bm{k}}\,h_{\lambda\omega}(\eta)\,e^{i\bm{k}\cdot\bm{x}}+\text{H.C.}\,,
\end{equation}
and the annihilation and the creation operators $\hat{a}_{\lambda,\bm{k}}^{\vphantom{\dagger}}$, $\hat{a}_{\lambda,\bm{k}}^{\dagger}$ obey the standard commutation relations. Here it is interesting to note that these operators and the corresponding vacuum states depend on the choice of $c_1$ and $c_2$, and they are not even equivalent~\cite{Fu73,FP77a,FP77b}. This is one of many ambiguities which arises in quantum field theory in curved space.

Since the asymptotic expansions of the Hankel functions $\operatorname{H}_{\nu}^{(1,2)}(z)$ at large $\lvert z\rvert$ are
\begin{align}
    H_{\nu}^{(1)}(z)&\sim\sqrt{\frac {2}{\pi z}}\,e^{+i(z-\nu \pi/2-\pi/4)}\,,&&\text{for}&-\pi &<\arg z<2\pi\,,\\
    H_{\nu}^{(2)}(z)&\sim\sqrt{\frac {2}{\pi z}}\,e^{-i(z-\nu \pi/2-\pi/4)}\,,&&\text{for}&-2\pi &<\arg z<\pi\,,
\end{align}
a popular choice is $c_1=0$ amd $c_2=1$. Such a solution, denoted by $h_{\lambda\omega}^{(\textsc{bd})}(\eta)$, behaves like a positive-energy wave, $\dfrac{1}{\sqrt{2\omega}}e^{-i\omega\eta}$, at the asymptotic past $\eta\to-\infty$, and the associated vacuum state is called the Bunch-Davies vacuum, $\lvert0\rangle_{\textsc{bd}}$. Thus the operator in \eqref{E:sasguejr} can be equally valid if expanded by
\begin{equation}\label{E:irskf}
    \hat{h}^{\mu\nu}(x)=\sum_{\lambda,\bm{k}}\mathsf{e}_{\lambda\bm{k}}^{\mu\nu}\,\hat{b}_{\lambda,\bm{k}}\,h_{\lambda\omega}^{(\textsc{bd})}(\eta)\,e^{i\bm{k}\cdot\bm{x}}+\text{H.C.}\,,
\end{equation}
with $\hat{b}_{\lambda,\bm{k}}\lvert0\rangle_{\textsc{bd}}=0$.

Both sets of creation and annihilation operators form complete bases, so we can always expand one set by another
\begin{equation}\label{E:gkbhkff}
    \hat{a}_{\lambda,\bm{k}}=\alpha_{\bm{k}}^{\vphantom{*}}\,\hat{b}^{\vphantom{\dagger}}_{\lambda,\bm{k}}+\beta_{\shortminus\bm{k}}^*\hat{b}^{\dagger}_{\lambda,\shortminus\bm{k}}\,.
\end{equation}
Thus, we again get back to the Bogoliubov coefficients discussed in \eqref{E:ghjrbdfdf}, and their linkage to the quantum squeezing. In addition, Eq.~\eqref{E:gkbhkff} implies that the number operator $\hat{N}_{\lambda\bm{k}}^{(a)}=\hat{a}^{\dagger}_{\lambda,\bm{k}}\hat{a}^{\vphantom{\dagger}}_{\lambda,\bm{k}}$ formed by the pair $(\hat{a}^{\vphantom{\dagger}}_{\lambda,\bm{k}},\hat{a}^{\dagger}_{\lambda,\bm{k}})$ has a nonzero expectation value in the Bunch-Davies vacuum
\begin{equation}
    \langle0\vert\hat{N}_{\lambda\bm{k}}^{(a)}\vert0\rangle_{\textsc{bd}}=\lvert\beta_{\shortminus\bm{k}}\rvert^2\,,
\end{equation}
as long as $\beta_{\shortminus\bm{k}}\neq0$. That is, the Bunch-Davies vacuum does not have any quantum created by $\hat{b}^{\dagger}_{\lambda,\bm{k}}$, but it contains opulent quanta with respect to $\hat{a}^{\dagger}_{\lambda,\bm{k}}$. Hence, when $\beta_{\shortminus\bm{k}}\neq0$, two representations of $\hat{h}^{\mu\nu}(x)$ are not equivalent.

\subsection{Infrared Behavior}

The graviton in spatially flat de Sitter space has an outstanding feature that the Green's function associated with the Bunch-Davies vacuum is not well defined; it is infrared divergent. For simplicity let us compute the Wightman function of the graviton in $\lvert0\rangle_{\textsc{bd}}$ associated with one of the polarizations
\begin{equation}
    \hat{h}_{\lambda}(x)=\int\!\frac{d^3\bm{k}}{(2\pi)^\frac{3}{2}}\;\hat{b}_{\bm{k}}\,h_{\lambda\omega}^{(\textsc{bd})}(\eta)\,e^{i\bm{k}\cdot\bm{x}}+\text{H.C.}\,,
\end{equation}
with $\lambda=+$ or $\times$, and then the Wightman function is given by
\begin{align}
    G_>(x,x')=i\,\langle0\vert\hat{h}_{\lambda}(x)\hat{h}_{\lambda}(x')\vert0\rangle_{\textsc{bd}}&=i\,\frac{\pi}{4H^2}\,\bigl(\eta\eta'\bigr)^{\frac{3
}{2}}\int\!\frac{d^3\bm{k}}{(2\pi)^3}\;\operatorname{H}_{3/2}^{(2)}(\omega\eta)\operatorname{H}_{3/2}^{(1)}(\omega\eta')\,e^{i\bm{k}\cdot(\bm{x}-\bm{x}')}\notag\\
&=i\,\frac{\bigl(\eta\eta'\bigr)^{\frac{3
}{2}}}{8\pi H^2R}\int_0^{\infty}\!d\omega\;\omega\,\sin\omega R\,\operatorname{H}_{3/2}^{(2)}(\omega\eta)\operatorname{H}_{3/2}^{(1)}(\omega\eta')\label{E:byiweier}\\
    &=i\,\frac{\left(\frac{1}{4}-\nu^{2}\right)}{16\pi\,H^{-2}}\,\sec\nu\pi\;{}_{2}F_{1}(\frac{3}{2}+\nu,\frac{3}{2}-\nu;2;1-\frac{\sigma^{2}}{4\eta\eta'})\,,\label{E:nksjfal}
\end{align}
where $\sigma^{2}=-(\eta-\eta')^2+\lvert\bm{x}-\bm{x}'\rvert^2$, and we let the order $\nu$ of the Hankel functions be any real number for the moment. The function ${}_{2}F_{1}(a,b;c;z)$ is the hypergeometric function. We immediately note that as $\nu\to3/2$, Eq.~\eqref{E:nksjfal} diverges. Let us explicitly expand \eqref{E:nksjfal} about $\nu=3/2$,
\begin{equation}
	G_>(x,x')=\frac{i}{4\pi^{2}\sigma^{2}}-\frac{i}{8\pi^{2}H^{-2}}\left[\ln\frac{\sigma^{2}}{4H^{-2}}+2\gamma_{\epsilon}-1+\psi(\frac{3}{2}+\nu)+\psi(\frac{3}{2}-\nu)\right]+\bigl(\frac{3}{2}-\nu\bigr)\,f(\sigma^{2})\,\bigg|_{\nu=\frac{3}{2}}\,.
\end{equation}
The function $f(\sigma^2)$ is well behaved when $\nu\to3/2$. In addition, it is a regular function of $\sigma^{2}$ and vanishes in the coincident limit $\sigma\to0$. When $z\to0$, the polygamma function $\psi(z)$ has a pole with residue $-1$, so the infrared divergence is entirely contained in the $\sigma$-independent term proportional to $\psi(3/2-\nu)$.

The source of divergence is most easily seen if we inspect the $z\to0$ limit of the Hankel function $\operatorname{H}_{\nu}^{(2)}(z)$
\begin{align}
    \lim_{z\to0}\operatorname{H}_{\nu}^{(2)}(z)\simeq i\,\frac{2^{\nu}}{\pi}\Gamma(\nu)\,z^{-\nu}+\cdots
\end{align}
for $\nu>0$. Comparing with \eqref{E:byiweier}, we note that the integral over $\omega$ will diverge at the lower limit 0, so this is the infrared divergence of the Bunch-Davies vacuum.

In fact, this is a generic feature of any free, massless, minimally coupled scalar field in spatially flat de Sitter spacetime. There is no de Sitter-invariant vacuum state. The infrared divergence prevents us from writing down a well defined Green's function. However, we can still derive a well behaved Green's function for the quantum state that breaks de Sitter invariance, at the cost that the ultraviolet-renormalized expectation value of $\langle\hat{h}_{\lambda}^2(x)\rangle$ will grow with cosmic time $t$ at late times. Finally we note that this infrared divergence does not forbid us to construct the two-point functions of the quantities which involve the derivatives of $\hat{h}_{\lambda}(x)$ because taking derivatives tends to introduce terms that have higher powers of $\omega$ in the integrand in Eq.~\eqref{E:byiweier}, making the integral well behaved at the infrared end.

For a massless minimally coupled scalar field, the vacuum states which respect or break the de Sitter symmetry were studied in \cite{Allen85,AllFol87}. Unsuccessful attempts had been made to look for physically acceptable \cite{HolWal14} de Sitter-invariant vacuum states which were free of infrared divergences, notably in~\cite{KirGar93}.
Infrared behavior of massless minimally coupled {\it interacting} quantum fields in de Sitter universe has important physical meanings for correlation functions in inflationary cosmology. We mention three important papers using different methods, namely, invoking the zero-mode in Euclidean de Sitter \cite{HuOC86}, using the stochastic approach in real time evolution \cite{StaYok94} and from the perturbative quantum gravity perspective \cite{TsWo94}.  The literature on this topic has mushroomed in the last two decades, some major approaches and important papers are highlighted in this review~\cite{HuIRdS}.

\section{Graviton physics is low-energy, perturbative, not Planck-scale, quantum gravity}
As an epilogue\footnote{The contents here mirror those in Sec.~2 of~\cite{ChoHu22}.},  in view of the increased attention drawn to recent experimental proposals to test the quantum nature of gravitation, it is  perhaps helpful   to add some comments on the differences between {\bf graviton physics} and {\bf quantum gravity} (QG). The former is ultra-low energy QG, directly accessible in experiments at today's energy scale, at least in principle. The latter is QG proper, ultra-high energy physics, operational at the Planck scale: $10^{19}\,\textrm{GeV}$, $10^{-35}\,\textrm{m}$.  Its consequences are unlikely to be detected directly by experiments, but  indirect observations may be possible,  from early universe or black hole quantum processes.  The former is perturbative: gravitons are quantized perturbations off of a background spacectime, at reachable observable scales. The latter is nonperturbative:   QG proper refers to theories  of the fundamental constituents and structures of spacetime at the Planck scale.  In physical terms, QG strives to uncover the `atoms' of spacetime, while quantized linear perturbations off of a background spacetime, the gravitational waves, are like phonons, which are the collective excitations of the elementary structure of spacetime, or spacetime atoms.  Graviton physics and quantum gravity are at two widely separated theoretical levels, structurally and conceptually, physically and philosophically. We shall touch on some philosophical issues at the end.

Laboratory experiments of analogy gravity \cite{AnalogG} have proven very fruitful in the past two decades. With the rise of a new subfield known as gravitational quantum physics \cite{GQP} many proposals of  tabletop experiments for `quantum gravity' have also appeared in recent years \cite{Carney} which attracted immediate attention. While the proposed experiments have their own merits, it is perhaps necessary to  add a quick reminder that in actuality they  \cite{Bose17} are not about quantum gravity proper at the Planck scale, nor about the quantum nature of gravity \cite{Vedral17} -- see the last subsection below.  Now that the term `quantum gravity' is used by a broader range of researchers outside of the  gravitation/cosmology and particles/fields communities, certainly a welcoming development,  it might be helpful to make precise the physical meanings of the key terminology used. For example, even a simple claim like, ``graviton physics is quantum gravity" without qualification can be misleading.  The strict answer would be no, because gravitons are not the fundamental constituents of spacetime, as fundamental strings or loops or simplices or sets or nets are meant to be,  citing a few  examples of many candidate theories of quantum gravity proper.    Gravitons are quantized {\it weak perturbations} of spacetime, which are not the same as the {\it basic constituents} of spacetime, just like phonons are not the same as atoms.  They are categorically different entities -- in fact, as soon as atoms appear phonons cease to exist.   

The terminology `quantum gravity' used often in the particle physics community, e.g., by Feynman in the 50s and Weinberg in the 60s and followers, is actually graviton physics, referring to gravitons as spin-two particles.  There, the background spacetime (e.g., Minkowski, Schwarzschild or Robertson-Walker) dynamics is derived from the Einstein-Hilbert action. The theoretical framework of such activities is quantized {\it perturbative} gravity.  As is well-known,  perturbative quantum gravity is non-renormalizable. That has spurned heroic efforts in the calculation of two and higher loop graviton interactions. On the other hand, dissatisfaction in its nonrenormalizable nature and background spacetime dependence (mostly in the general relativity community) prompted the development of superstring theory, loop quantum gravity, asymptotic safety, causal dynamical triangulation, spin-nets, causal sets, group field theory, in their pursuit of background independent, nonperturbative theories of quantum gravity. 

 The approaches named above are better-known examples of theories of quantum gravity proper,  which target fundamental spacetime structures above the Planck energy $>10^{19}\,\mathrm{GeV}$, less than the Planck length  $<10^{-35}\,\mathrm{m}$.  Theories valid at immediately below the Planck energy include a) dilatonic gravity with the 3-indexed antisymmetric tensor field, which had been proven to be a low energy limit of string theory, and, on the same footing, b) semiclassical gravity, where the Einstein-Hilbert action is augmented by three additional terms ($\Box R$, $R_{\mu\nu}R^{\mu\nu}$, $C_{\mu\nu\lambda\kappa}C^{\mu\nu\lambda\kappa}$, where $R$ is the scalar curvature, $R_{\mu\nu}$ is the Ricci tensor and $C_{\mu\nu\lambda\kappa}$ is the Weyl tensor) resulting from the regularization of the stress energy tensor of quantum matter fields.  One could view all theories derived from quantum gravity proper at energy scales below the Planck  energy as effective field theories (EFT), ranging from semiclassical and stochastic gravity \cite{HuVer20} operative at close to the Planck energy,  to general relativity at today's very low energy \cite{Donoghue,Burgess}. However, EFT is a general, broadband concept,  more specificity in the processes and mechanisms are needed to link up Planck scale quantum gravity with low energy graviton physics. Whether and how a manifold with a metric structure emerges from the interaction of the  basic constituents is a crucial challenge for anyone with a pet quantum gravity theory proposal.  Quantum gravity is nonperturbative and background independent, graviton physics is quantized perturbations off of a given background spacetime. Note that  gravitons can exist at all length scales greater than the Planck length  because once a spacetime manifold exists  one can consider weak perturbations off this classical entity and quantize them.  

The concept of graviton is embedded in the perturbative treatment which requires a background spacetime. In this context the answer to the  question we posed earlier would be yes, if one adds the term `perturbative' to quantum gravity,  because gravitons refer  to the quantized dynamical degrees of freedom of perturbative (linearized) gravity,  which can be seen at today's low energy.  If detected, we can indeed confirm the quantum nature of perturbative gravity,  as a very low energy effective theory of quantum gravity proper,  or as the quantized collective excitation modes of spacetime, similar to phonons (but not atoms).  For this purpose we give some background description below  to  define more clearly the contexts of terms used and their different focuses.  Hopefully this could help mitigate possible confusions from cross-talks over different intended purposes.

\subsection{Classical Gravity: General Relativity. Perturbative Gravity: Gravitational Waves}

Let us agree that when only `gravity' is mentioned it is taken to be classical gravity.  Classical gravity is described with very high accuracy by Einstein's general relativity (GR) theory.  One can distinguish two domains, weak field, such as experiments on earth or in the solar system would fall under, from strong field, such as processes near black holes or neutron stars,  depending on their masses and the proximity of measured events, and in the early universe.

Classical \textit{ `perturbative gravity'}  refers to small perturbations off of a classical gravitational background spacetime.  In the earth's environments, the background spacetime is Minkowski space. While the background spacetime could be strongly curved, perturbative treatments can only consider small amplitude deviations which are weak by proportion. Gravitational waves  are usually treated as perturbations  whose wavelengths can span from the very long of astrophysical or cosmological scales to the very short. Note the crucial difference between \textit{perturbations} and \textit{fluctuations}, the former being a deterministic variable, referring to the small amplitude deviations from the background spacetime, while the latter is a stochastic variable,  referring to the noise. Fluctuations in the gravitons constitute a noise of gravitational origin and of a quantum nature. Such effects at low energies are the target of the present investigation  \cite{PWZ21a,PWZ21b}, not Planck scale physics.  Fluctuations of quantum matter field can also induce metric fluctuations.  The metric fluctuations are governed by the Einstein-Langevin equation of semiclassical stochastic gravity theory \cite{HuVer20}. At the Planck scale they make up the spacetime foams \cite{foam} where topology changes can also enter.  

 \subsection{Quantum Gravity at Planck Energy: Theories for the microscopic constituents of spacetime} 

Quantum gravity (QG) proper refers to theories of the basic constituents of spacetime at above the Planckian energy scale, such as string theory, loop quantum gravity, spin network, causal dynamical triangulation, asymptotic safety, causal sets, group field theory, spacetime foams, etc \cite{Oriti}. Because of the huge  energy scale discrepancy between the Planck scale and the scale of a Earthbound or space laboratory, many such experimental proposals to test Planck scale quantum gravity need to rely on \textit{indirect} implications to high-energy particle phenomenology \cite{QGPhen,Sabine} or, at a lower energy range, analog gravity experiments \cite{AnalogG}. Of the latter kind, many atomic-molecular-fluid, condensed matter-BEC or electro-optical-mechanical experiments can indeed skillfully use analogs to seek indirect implications of quantum gravity. But in terms of \textit{direct} observations, or drawing direct implications, tabletop experiments can only probe weak-field perturbative gravity, nevermind their quantum gravity labels.

 \subsection{Gravitons, even at today's low energy, carry the quantum signature of  perturbative gravity.}

Gravitons are the quantized propagating degrees of freedom of weak perturbations off of  a background metric,   such as the Minkowski spacetime for experiments in the Earth's environment.
Gravitons as~\textit{spin 2 particles} refer  to the high frequency components of weak gravitational perturbations under certain averages (like the Brill-Hartle~\cite{BriHar}-Isaacson~\cite{Isaacson} average), or in the ray representation under the eikonal approximation. As such, beware of the shortcomings when treating gravitons merely as particles in a Fock representation, without taking into account their phase information. This is important when addressing quantum informational issues such as related to quantum coherence and quantum entanglement. Quantum correlations between gravitons are their truly quantum properties.  

 Gravitons can exist at today's very low energy but there is no necessary relation to the basic constituents of spacetime at the Planck energy. One does not need any deeper level theory for their description. Einstein's general relativity theory plus second quantization on weak linear perturbations will do. Note that any linearized degree of freedom in classical systems can be quantized, irrespective of whether it is fundamental or collective. The latter is in abundance in condensed matter physics (e.g., phonons, rotons, plasmons, and many other entities with -on endings).
Seeing the quantum nature of the gravitational field at today's low energy, such as proving the existence of gravitons \cite{Dyson,PWZ21a,PWZ21b},  is certainly of fundamental value, but there is no necessary relation to quantum gravity proper as defined above. Gravitons are the quantized collective excitation modes of spacetime, not the basic building blocks of spacetime. Graviton's existence is predicated on the emergence of spacetime while the building blocks (such as strings, loops, causal sets, spin-nets, etc) are the progenitors of spacetime structure.

These explanations are enough for demarcating graviton physics from quantum gravity.  Since there is increased enthusiasm  in using quantum entanglement of masses to show the quantum nature of gravity we should add one more noteworthy aspect. 

\subsection{Pure gauge says nothing about the quantum nature of gravity}
-- Only the radiative degrees of freedom can testify to the quantum nature of (perturbative) gravity.{}\\

 A more subtle yet serious misconception is attributing quantumness to the pure gauge degrees of freedom (Newton or Coulomb forces) while in truth only the dynamical degrees of freedom (graviton or photons) can act as the signifiers of the quantum nature of the gravitational or electromagnetic  theory.  Experiments measuring the entanglement between two quantum objects \cite{Bose17, Vedral17} through Newtonian gravitational interactions capture only the quantum nature of these objects, not of gravity.  This critique is raised in \cite{AnHucr}.  A more detailed explanation can be found in \cite{2GravCat}. Linearization of the GR field equations around Minkowski spacetime leads to an action similar to the EM one. The true degrees of freedom of the linearized perturbations are their transverse-traceless (TT) components, i.e., gravitational waves. Quantization of the short-wavelength TT perturbations  gives rise to gravitons. Only in the successful detection of gravitons, just like photons in quantum electrodynamics, such as via the Hanbury-Brown Twist experiments, can one make claims pertaining to the quantum nature of gravity, bearing in mind, nevertheless, their fundamental differences from quantum gravity proper as defined above.

\subsection{What does quantization entail in quantum gravity?}

Finally,  we want to end with a conceptual, even philosophical question in our dissection of the relation between graviton physics and quantum gravity. There is a fundamental divide in perspectives,  between viewing gravity as fundamental versus emergent.  And, with it, a deeper question arises, on the functionality of quantization in this context --  Does quantizing a theory effective at low energy necessarily bring forth or reveal the underlying microscopic structures? Not in the case of sound waves versus atoms: Sound waves are  the vibrational modes in the collective motion of atoms. Certainly quantizing sound is not the right way to finding the  atoms.  In addition to the quantum/classical inquiry, there is another, sometimes more relevant,  dimension involved, namely,  macro/micro: given a macroscopic theory such as general relativity how can one decipher the micro structure of spacetime? We will leave the exploration of these deeper issues (see, e.g., \cite{E/QG}) to the probing minds of our readers. \\

\noindent{\bf Acknowledgment}   J.-T. Hsiang is supported by the National Council of  Science and Technology, Taiwan, R.O.C. under Grant No.~NSTC 112-2112-M-011-001-MY3. H.-T. Cho is supported in part by the National Council of Science and Technology, Taiwan, R.O.C. under the Grant NCST 112-2112-M-032-006. B.-L. Hu enjoyed the warm hospitality of Prof.~C.-S. Chu of the National Center for Theoretical Sciences at National Tsing Hua University, and Prof.~K.-W. Ng of the Institute of Physics, Academia Sinica, Taiwan, R.O.C. where a good part of this work was done.

\newpage
\appendix

\section{machinery of the influence functional formalism}\label{S:eiwt}
Consider the alternative version of the action \eqref{E:bksjgfdf}
\begin{align}\label{E:bksjeref}
    S[\bm{z},h]=S_{\textsc{m}}[\bm{z}]+S_{\textsc{g}}[h]&=\int\!dt\;\frac{m}{2}\,\eta_{ij}\dot{z}^i(t)\dot{z}^j(t)+\int\!dt\;\frac{m}{4}\,h_{ij}(t,\bm{0})\,\frac{\partial^{2}}{\partial t^{2}}\Bigl[z^i(t)z^j(t)\Bigr]\notag\\
    &\qquad\qquad\qquad\qquad\qquad\qquad\qquad-\frac{1}{64\pi}\int\!d^4x\;\bigl[\partial_{\rho}h_{ij}(x)\bigr]\bigl[\partial^{\rho}h^{ij}(x)\bigr]\,.
\end{align}
with $x=(t,\bm{z})$, where we have carried out the integration by parts on the interaction term, and ignore any effects of the surface terms for the moment. It is convenient to define the current tensor
\begin{equation}\label{E:rpij}
	J^{ij}(x)=\delta^{(3)}(\bm{z})\,\frac{m}{4}\frac{\partial^{2}}{\partial t^{2}}\Bigl[z^i(t)z^j(t)\Bigr]\,,
\end{equation}
so that the interaction term reduces to
\begin{equation}
	S_{\textsc{int}}[\bm{z},h]=\int\!d^{4}x\;h_{ij}(x)J^{ij}(x)\,.
\end{equation}

In deriving the influence functional, the emergence of the Green's functions endows a clear physical picture of the formalism, but some confusions may arise due to various conventions of the Green's functions used in the literature, so we will leave the explicit notations of the Green's functions to the very last part of the derivations.

We first re-derive the Langevin equation in the same fashion, and this will be handy when compared with the results of the influence functional. 

Varying the classical action \eqref{E:bksjeref} with respect to $z^{i}(t)$ gives
\begin{align}
	\frac{\delta S[\bm{z},h]}{\delta z^{i}(t)}&=0\,,&&\Rightarrow&&-\eta_{ij}\ddot{z}^{j}(t)+\frac{1}{2}\,\ddot{h}_{ij}(t,\bm{0})z^{j}(t)=0\,,
\end{align}
and varying $h_{ij}(x)$, we find
\begin{align}\label{E:ghdbf}
	\frac{\delta S[\bm{z},h]}{\delta h_{ij}(x)}&=0\,,&&\Rightarrow&&\square_{x}h^{ij}(x)=-8\pi m\,\delta^{(3)}(\bm{z})\,\frac{m}{4}\frac{\partial^{2}}{\partial t^{2}}\bigl[z^{i}(t)z^{j}(t)\bigr]\,.
\end{align}
Note that \eqref{E:ghdbf} can be written as $\square_{x}h^{ij}(x)=-32\pi\,J^{ij}(x)$, similar to the Lifshitz equation by a factor of 2.

Thus we arrive a set of equations of motion. The formal solutions to \eqref{E:ghdbf} is
\begin{align}
	h^{ij}(x)&=h_{h}^{ij}(x)-8\pi m\int_{t_{a}}^{t}\!dt'\;\bigl(\square^{-1}_{\textsc{r}})^{ij}{}_{kl}(t,\bm{z};t',\bm{0})\frac{\partial^{2}}{\partial t'^{2}}\bigl[z^{k}(t')z^{l}(t')\bigr]\,,\label{E:kbvkxcd}
\end{align}
where $t_{a}$ is the placeholder for the initial time, $h_{h}^{ij}(x)$ is the homogeneous solution to the wave equation~\eqref{E:ghdbf}, satisfying $\square_{x}h_h^{ij}(x)=0$, so it represents the free metric perturbation. The kernel function $\bigl(\square^{-1}_{\textsc{r}})^{ij}{}_{kl}$ denotes one of the candidates of the inverse to the wave operator $\square_{x}$ such that $\square_{x}\bigl(\square^{-1}_{\textsc{r}})^{ij}{}_{kl}(x,x')=\delta^{ij}{}_{kl}(x,x')$ explicitly. The subscript here highlights the retardation nature of $\bigl(\square^{-1}_{\textsc{r}})^{ij}{}_{kl}(x,x')$ and it will be shown to be related to the retarded Green's function of the free metric perturbation $h_{h}^{ij}(x)$

We substitute \eqref{E:kbvkxcd} to the equation of motion of $z^{i}(t)$
\begin{equation*}
	\ddot{z}^{i}(t)=\frac{1}{2}\,\ddot{h}^{ij}(t,\bm{0})\,z_{j}(t)=\frac{1}{2}\,\ddot{h}_{h}^{ij}(t,\bm{0})\,z_{j}(t)-4\pi m\,z_{j}(t)\,\frac{\partial^{2}}{\partial t^{2}}\int_{t_{a}}^{t}\!dt'\;\bigl(\square^{-1}_{\textsc{r}})^{ij}{}_{kl}(t,\bm{0};t',\bm{0})\frac{\partial^{2}}{\partial t'^{2}}\bigl[z^{k}(t')z^{l}(t')\bigr]\,,
\end{equation*}
or we arrive a generalized Langevin equation
\begin{equation}\label{E:pwoejdnsa}
	\ddot{\hat{z}}^{i}(t)+4\pi m\,\hat{z}_{j}(t)\,\frac{\partial^{2}}{\partial t^{2}}\int_{t_{a}}^{t}\!dt'\;\bigl(\square^{-1}_{\textsc{r}})^{ij}{}_{kl}(t,\bm{0};t',\bm{0})\frac{\partial^{2}}{\partial t'^{2}}\bigl[\hat{z}^{k}(t')\hat{z}^{l}(t')\bigr]=\frac{1}{2}\,\ddot{\hat{h}}_{h}^{ij}(t,\bm{0})\,\hat{z}_{j}(t)\,,
\end{equation}
once we promote $z^{i}(t)$, $h_{ij}(x)$ to the operators $\hat{z}^{i}(t)$, $\hat{h}_{ij}(x)$ respectively. The last step in arriving \eqref{E:pwoejdnsa} is somewhat heuristic in this case because 1) we have treated $z$ in the argument of the delta function as a parameter, rather than an operator, and 2) the solution of $h_{ij}(x)$ in \eqref{E:kbvkxcd} contains the derivatives of $z^{i}(t)$, so when $z^{i}(t)$ is promoted to the operator, there is potential ambiguity in operator ordering in the nonlocal term in \eqref{E:pwoejdnsa}. As commented earlier that this is essentially a nonlinear equation with a multiplicative noise. Similar Langevin equations also appear in the context of the mirror-field interaction or the more sophisticated model of the atom-field interaction.

Next we will derive the influence functional on geodesic deviation $\hat{z}^{i}(t)$ due to gravitons, described by the free component of the metric perturbation $\hat{h}_{ij}(x)$, that is the homogeneous part in \eqref{E:kbvkxcd}. The whole idea of the influence functional is to systematically account for the influence of the environment, gravitons in this case, to the specified order on the evolution of the density matrix operator of the system, say geodesic deviation. Thus, formally, let us denote by $\hat{\rho}$ the density matrix operator of the combined system, and its evolution up to the final time $t_{b}$ of our interest is described by the unitary evolution operator $\hat{U}(t,t_{a})$ of the combined system from the initial time $t_{a}$
\begin{equation}
	\hat{\rho}(t_{b})=\hat{U}(t_{b},t_{a})\,\hat{\rho}(t_{a})\,\hat{U}^{\dagger}(t_{b},t_{a})\,.
\end{equation}
The reduced density operator of the system of our interest, geodesic deviation, can be found if we trace out the degrees of freedom of the environment, i.e., gravitons, 
\begin{equation}\label{E:perijhndf}
	\hat{\varrho}_{\textsc{m}}(t_{b})=\operatorname{Tr}_{\textsc{g}}\bigl\{\hat{U}(t_{b},t_{a})\,\hat{\rho}(t_{a})\,\hat{U}^{\dagger}(t_{b},t_{a})\bigr\}\,.
\end{equation}
This is rather formal, and to make it of better use, we write it in terms of the path integral. In the appropriate $z^{i}$ and $h_{ij}$ basis, the matrix elements of the unitary evolution operator is 
\begin{equation}
	U(z_{b},h_{b},t_{b};z_{a},h_{a},t_{a})=\int_{z_{a}}^{z_{b}}\!\mathcal{D}z\!\int_{h_{a}}^{h_{b}}\!\mathcal{D}h\;\exp\Bigl\{i\,S[\bm{z},h]\Bigr\}\,,
\end{equation}
where the action $S[\bm{z},h]$ of the combined system has been given by \eqref{E:bksjeref}. Hereafter we will suppress the tensor indices to avoid notational cluttering in case there is no confusion with the tensor trace or the vector norm. Thus, the elements of the reduced density matrix $\hat{\varrho}_{\textsc{m}}(t)$ are given by
\begin{align}\label{E:kbvdfsdf}
	\hat{\varrho}_{\textsc{m}}(z_{b},z'_{b};t_{b})&=\int_{-\infty}^{\infty}\!dz_{a}\int_{-\infty}^{\infty}\!dz'_{a}\;\hat{\varrho}_{\textsc{m}}(z_{a},z'_{a};t_{a})\int_{z_{a}}^{z_{b}}\!\mathcal{D}z^{(+)}\!\int_{z'_{a}}^{z'_{b}}\!\mathcal{D}z^{(-)}\;\biggl[\int_{-\infty}^{\infty}\!dh_{a}\int_{-\infty}^{\infty}\!dh'_{a}\;\varrho_{\textsc{g}}(h_{a},h'_{a};t_{a})\biggr.\notag\\
	&\qquad\times\biggl.\int_{-\infty}^{\infty}\!dh_{b}\int_{h_{a}}^{h_{b}}\!\mathcal{D}h^{(+)}\!\int_{h'_{a}}^{h_{b}}\!\mathcal{D}h^{(-)}\;\exp\Bigl\{i\,S[\bm{z}^{(+)},h^{(+)}]-i\,S[\bm{z}^{(-)},h^{(-)}]\Bigr\}\biggr]\,.
\end{align}
The $(\pm)$ superscripts represent the dynamical variables along the forward and the backward time branches associated with $\hat{U}(t_{b},t_{a})$ and $\hat{U}^{\dagger}(t_{b},t_{a})$ respectively. The subscripts $a$, $b$ of a variable indicate the initial or the final value of that variable. For computational simplicity, we assume  the combined system possesses a totally uncorrelated initial system, that is, the initial state is factorizable,
\begin{equation}
	\hat{\rho}(t_{a})=\hat{\varrho}_{\textsc{m}}(t_{a})\otimes\hat{\varrho}_{\textsc{g}}(t_{a})\,.
\end{equation}
Combined with the assumption that the interaction is switched on at the initial time $t_{a}$, this can raise some conceptual ambiguities in the context of gravity, and the more elaborate considerations are beyond the scope of the current notes. On the other hand, it leaves the room for phenomenologically modeling the graviton physics at initial time, e.g., the very early stage of the cosmic evolution.

The trace operation in \eqref{E:perijhndf} is rephrased as the integral expression like
\begin{equation}
	\int_{-\infty}^{\infty}\!dh_{b}\;F(h_{b},h_{b})=\int_{-\infty}^{\infty}\!dh_{b}\int_{-\infty}^{\infty}\!dh'_{b}\;\delta(h_{b}-h'_{b})\,F(h_{b},h'_{b})\,.
\end{equation}
Thus, the influence of the gravitons are completely included in the expressions inside the square brackets of \eqref{E:kbvdfsdf}, and that will constitutes the influence functional of our interest.

Define the so-called in-in effective action
\begin{align}\label{E:gkbfbfd}
	Z[\bm{z}^{(+)},\bm{z}^{(-)}]=\int_{-\infty}^{\infty}\!dh_{a}\int_{-\infty}^{\infty}\!dh'_{a}\;\varrho_{\textsc{g}}(h_{a},h'_{a};t_{a})\oint_{h_{a}}^{h'_{a}}\!\mathcal{D}h\;\exp\biggl\{i\oint_{t_{a}}^{t_{a}}\!d^{4}x\;\mathcal{L}_{\textsc{ig}}[\bm{z},h]\biggr\}\,,
\end{align}
where we have introduced a few shorthand notations
\begin{align}
	\mathcal{L}_{\textsc{ig}}[\bm{z},h]&=h_{ij}(x)J^{ij}(x)-\frac{1}{64\pi}\bigl[\partial_{\rho}h_{ij}(x)\bigr]\bigl[\partial^{\rho}h^{ij}(x)\bigr]\,,\\
	\oint_{h_{a}}^{h'_{a}}\!\mathcal{D}h&\equiv\int_{-\infty}^{\infty}\!dh_{b}\int_{h_{a}}^{h_{b}}\!\mathcal{D}h^{(+)}\!\int_{h'_{a}}^{h_{b}}\!\mathcal{D}h^{(-)}\,,\label{E:peirj}\\
	\oint_{t_{a}}^{t_{a}}\!d^{4}x\;S_{\textsc{ig}}[\bm{z},h]&=\int_{t_{a}}^{t_{b}}\!d^{4}x\;S_{\textsc{ig}}[\bm{z}^{(+)},h^{(+)}]-\int_{t_{a}}^{t_{b}}\!d^{4}x\;S_{\textsc{ig}}[\bm{z}^{(-)},h^{(-)}]\,.\label{E:hgbf}
\end{align}
These notations, in particular, \eqref{E:peirj} and \eqref{E:hgbf}, illustrate why the current formalism is also coined the closed-time-path formalism.

Assuming an initial Gaussian state of gravitons, $\hat{\varrho}_{\textsc{g}}(t_{a})$, we can carry out the path integrals in \eqref{E:gkbfbfd} exactly, and arrive at a compact formal expression
\begin{align}\label{E:gfjgsgg}
	Z[\bm{z}^{(+)},\bm{z}^{(-)}]=\exp\biggl\{-i\,16\pi\oint\!d^{4}x\oint\!d^{4}x'\;J^{ij}(x)\bigl(\square^{-1}\bigr)_{ijkl}(x,x')J^{kl}(x')\biggr\}\,.
\end{align}
In contrast to the conventional in-out effective action, the functional expression of the kernel function $\bigl(\square^{-1}\bigr)_{ijkl}(x,x')$ depends on the initial state $\hat{\varrho}_{\textsc{g}}(t_{a})$, and depends on the time branches in \eqref{E:peirj} or \eqref{E:hgbf}. The exponent is the influence functional $S_{\textsc{if}}[\bm{z}^{(+)},\bm{z}^{(-)}]$
\begin{align}
	S_{\textsc{if}}[\bm{z}^{(+)},\bm{z}^{(-)}]=-16\pi\oint\!d^{4}x\oint\!d^{4}x'\;J^{ij}(x)\bigl(\square^{-1}\bigr)_{ijkl}(x,x')J^{kl}(x')\,.
\end{align}
Clearly all the effects of the gravitons, up to the quadratic order have been included in this functoinal.

With its help, we may write the reduced density matrix $\hat{\varrho}_{\textsc{m}}(z_{b},z'_{b};t_{b})$ in \eqref{E:kbvdfsdf} as
\begin{align}\label{E:gfkdgbf}
	\hat{\varrho}_{\textsc{m}}(z_{b},z'_{b};t_{b})&=\int_{-\infty}^{\infty}\!dz_{a}\int_{-\infty}^{\infty}\!dz'_{a}\;\hat{\varrho}_{\textsc{m}}(z_{a},z'_{a};t_{a})\int_{z_{a}}^{z_{b}}\!\mathcal{D}z^{(+)}\!\int_{z'_{a}}^{z'_{b}}\!\mathcal{D}z^{(-)}\notag\\
	&\qquad\qquad\qquad\qquad\qquad\exp\Bigl\{i\,S_{\textsc{m$_{0}$}}[\bm{z}^{(+)}]-i\,S_{\textsc{m$_{0}$}}[\bm{z}^{(-)}]+i\,S_{\textsc{if}}[\bm{z}^{(+)},\bm{z}^{(-)}]\Bigr\}\,,
\end{align}
with the free action associated with geodesic deviation,
\begin{equation}
	S_{\textsc{m$_{0}$}}[\bm{z}]=\int_{t_{a}}^{t_{b}}\!dt\;\frac{m}{2}\,\eta_{ij}\dot{z}^i(t)\dot{z}^j(t)\,.
\end{equation}
Sometimes we call the coarse grained-effective action
\begin{equation}
	S_{\textsc{cg}}[\bm{z}^{(+)},\bm{z}^{(-)}]=S_{\textsc{m$_{0}$}}[\bm{z}^{(+)}]-S_{\textsc{m$_{0}$}}[\bm{z}^{(-)}]+S_{\textsc{if}}[\bm{z}^{(+)},\bm{z}^{(-)}]\,,
\end{equation}
because this effective action is obtained after we have coarse-grained the environmental degrees of freedom.

To extract the physical meanings of the influence functional, we will use \eqref{E:peirj} and \eqref{E:hgbf} to re-cast \eqref{E:gfjgsgg} in terms of $\bm{z}^{(\pm)}$,
\begin{align}
	Z[\bm{z}^{(+)},\bm{z}^{(-)}]&=\int_{-\infty}^{\infty}\!dh_{a}\int_{-\infty}^{\infty}\!dh'_{a}\;\varrho_{\textsc{g}}(h_{a},h'_{a};t_{a})\oint_{h_{a}}^{h'_{a}}\!\mathcal{D}h\notag\\
	&\qquad\qquad\qquad\qquad\qquad\exp\biggl\{i\oint_{t_{a}}^{t_{a}}\!d^{4}x\;\biggl(h_{ij}(x)J^{ij}(x)-\frac{1}{64\pi}\bigl[\partial_{\rho}h_{ij}(x)\bigr]\bigl[\partial^{\rho}h^{ij}(x)\bigr]\biggr)\biggr\}\label{E:ngfbgr}\\
	&=\exp\biggl\{-i\,16\pi\int\!d^{4}x\int\!d^{4}x'\;\biggl[J_{(+)}^{ij}(x)\bigl(\square^{-1}\bigr)_{ijkl}^{++}(x,x')J_{(+)}^{kl}(x')\biggr.\biggr.\notag\\
	&\qquad\qquad\qquad\qquad\qquad+J_{(+)}^{ij}(x)\bigl(\square^{-1}\bigr)_{ijkl}^{+-}(x,x')J^{(-)}_{kl}(x')+J_{(-)}^{kl}(x)\bigl(\square^{-1}\bigr)_{ijkl}^{-+}(x,x')J_{(+)}^{kl}(x')\notag\\
	&\qquad\qquad\qquad\qquad\qquad\qquad\qquad+\biggl.\biggl.J_{(-)}^{ij}(x)\bigl(\square^{-1}\bigr)_{ijkl}^{--}(x,x')J_{(-)}^{kl}(x')\biggr]\biggr\}\,.\label{E:dbkghd}
\end{align}
This form will be of extreme use to identify the meanings of the kernel function $\bigl(\square^{-1}\bigr)_{ijkl}(x,x')$ via the functional derivatives with respect to the current $J^{ij}(x)$. For this purpose, let us treat $J^{ij}(x)$ as a generic external current source for the moment, rather than the one specified by \eqref{E:rpij}. Formally from \eqref{E:ngfbgr}, in general we have
\begin{align}
	&i^{2}\frac{\delta^{2}Z[\bm{z}^{(+)},\bm{z}^{(-)}]}{\delta J^{ij}(x)\delta J^{kl}(x')}\,\bigg|_{J=0}\notag\\
	&=\int_{-\infty}^{\infty}\!dh_{a}\int_{-\infty}^{\infty}\!dh'_{a}\;\varrho_{\textsc{g}}(h_{a},h'_{a};t_{a})\oint_{h_{a}}^{h'_{a}}\!\mathcal{D}h\;h_{ij}(x)h_{kl}(x')\,\exp\biggl\{-\frac{i}{64\pi}\oint_{t_{a}}^{t_{a}}\!d^{4}x\;\bigl[\partial_{\rho}h_{ij}(x)\bigr]\bigl[\partial^{\rho}h^{ij}(x)\bigr]\biggr\}\notag\\
	&=32\pi\,\operatorname{Tr}_{\textsc{g}}\Bigl\{\hat{\varrho}_{\textsc{g}}(t_{a})\,\mathcal{P}\,\hat{h}_{ij}(x)\hat{h}_{kl}(x')\Bigr\}\,.\label{E:bkdfbd}
\end{align}
This naturally introduces the path-ordered propagator, \eqref{E:bkdfbd}, in the same sense the Feynman propagator is defined by time-ordering. Here the path-ordered propagator has the operators in question arranged by their order $\mathcal{P}$ along the closed time path~\eqref{E:hgbf}. Thus, from \eqref{E:bkdfbd}, the path-ordered propagator is the expectation value $\langle\mathcal{P}\,\hat{h}_{ij}(x)\hat{h}_{kl}(x')\rangle_{\hat{\varrho}_{\textsc{g}}}$ of the path-ordered operators with respect to the initial state $\hat{\varrho}_{\textsc{g}}$. Since the closed time path is composed of the forward $(+)$ and the backward $(-)$ time branches, the expressions of $\langle\mathcal{P}\,\hat{h}_{ij}(x)\hat{h}_{kl}(x')\rangle_{\hat{\varrho}_{\textsc{g}}}$ then depends on the locations of $x$ and $x'$ along these two time branches, $\mathcal{C}^{(+)}$ and $\mathcal{C}^{(-)}$. After some pondering, we realize that
\begin{align}\label{E:rotioer}
	\langle\mathcal{P}\,\hat{h}_{ij}(x)\hat{h}_{kl}(x')\rangle_{\hat{\varrho}_{\textsc{g}}}=\begin{cases}
		+\langle\mathcal{T}\,\hat{h}_{ij}(x)\hat{h}_{kl}(x')\rangle_{\hat{\varrho}_{\textsc{g}}}\,,&x\in\mathcal{C}^{(+)}\,,\;x'\in\mathcal{C}^{(+)}\,,\\
		-\langle\mathcal{T}_{<}\,\hat{h}_{ij}(x)\hat{h}_{kl}(x')\rangle_{\hat{\varrho}_{\textsc{g}}}\,,&x\in\mathcal{C}^{(+)}\,,\;x'\in\mathcal{C}^{(-)}\,,\\
		-\langle\mathcal{T}_{>}\,\hat{h}_{ij}(x)\hat{h}_{kl}(x')\rangle_{\hat{\varrho}_{\textsc{g}}}\,,&x\in\mathcal{C}^{(-)}\,,\;x'\in\mathcal{C}^{(+)}\,,\\
		+\langle\mathcal{T}_{*}\,\hat{h}_{ij}(x)\hat{h}_{kl}(x')\rangle_{\hat{\varrho}_{\textsc{g}}}\,,&x\in\mathcal{C}^{(-)}\,,\;x'\in\mathcal{C}^{(-)}\,.
	\end{cases}
\end{align}
Here $\mathcal{T}$ and $\mathcal{T}_{*}$ represent the time-ordering and anti-time-ordering respectively, so the latter will give the Dyson propagator, meanwhile $\langle\mathcal{T}_{>}\,\hat{h}_{ij}(x)\hat{h}_{kl}(x')\rangle_{\hat{\varrho}_{\textsc{g}}}=\langle\hat{h}_{ij}(x)\hat{h}_{kl}(x')\rangle_{\hat{\varrho}_{\textsc{g}}}$ will give the Wightman function, and $\langle\mathcal{T}_{<}\,\hat{h}_{ij}(x)\hat{h}_{kl}(x')\rangle_{\hat{\varrho}_{\textsc{g}}}=\langle\hat{h}_{kl}(x')\hat{h}_{ij}(x)\rangle_{\hat{\varrho}_{\textsc{g}}}$. The sign difference results from the way $J^{ij}$ appears in the definition of the closed time path~\eqref{E:hgbf}.

On the other hand from Eq.~\eqref{E:dbkghd}, we find
\begin{align}
	&x\in\mathcal{C}^{(+)}\,,\;x'\in\mathcal{C}^{(+)}\,,&\frac{\delta^{2}Z[\bm{z}^{(+)},\bm{z}^{(-)}]}{\delta J_{(+)}^{ij}(x)\delta J_{(+)}^{kl}(x')}\,\bigg|_{J=0}&=-32\pi\, i\,\bigl(\square^{-1}\bigr)_{ijkl}^{++}(x,x')\,,\\
	&x\in\mathcal{C}^{(+)}\,,\;x'\in\mathcal{C}^{(-)}\,,&\frac{\delta^{2}Z[\bm{z}^{(+)},\bm{z}^{(-)}]}{\delta J_{(+)}^{ij}(x)\delta J_{(-)}^{kl}(x')}\,\bigg|_{J=0}&=-32\pi\, i\,\bigl(\square^{-1}\bigr)_{ijkl}^{+-}(x,x')\,,\\
	&x\in\mathcal{C}^{(-)}\,,\;x'\in\mathcal{C}^{(+)}\,,&\frac{\delta^{2}Z[\bm{z}^{(+)},\bm{z}^{(-)}]}{\delta J_{(-)}^{ij}(x)\delta J_{(+)}^{kl}(x')}\,\bigg|_{J=0}&=-32\pi\, i\,\bigl(\square^{-1}\bigr)_{ijkl}^{-+}(x,x')\,,\\
	&x\in\mathcal{C}^{(-)}\,,\;x'\in\mathcal{C}^{(-)}\,,&\frac{\delta^{2}Z[\bm{z}^{(+)},\bm{z}^{(-)}]}{\delta J_{(-)}^{ij}(x)\delta J_{(-)}^{kl}(x')}\,\bigg|_{J=0}&=-32\pi\, i\,\bigl(\square^{-1}\bigr)_{ijkl}^{--}(x,x')\,,
\end{align}
and together with the results in \eqref{E:rotioer} we arrive at
\begin{align}
	\bigl(\square^{-1}\bigr)_{ijkl}^{++}(x,x')&=-i\,\langle\mathcal{T}\,\hat{h}_{ij}(x)\hat{h}_{kl}(x')\rangle_{\hat{\varrho}_{\textsc{g}}}\,,&\bigl(\square^{-1}\bigr)_{ijkl}^{+-}(x,x')&=+i\,\langle\mathcal{T}_{<}\,\hat{h}_{ij}(x)\hat{h}_{kl}(x')\rangle_{\hat{\varrho}_{\textsc{g}}}\,,\\
	\bigl(\square^{-1}\bigr)_{ijkl}^{-+}(x,x')&=+i\,\langle\mathcal{T}_{>}\,\hat{h}_{ij}(x)\hat{h}_{kl}(x')\rangle_{\hat{\varrho}_{\textsc{g}}}\,,&\bigl(\square^{-1}\bigr)_{ijkl}^{--}(x,x')&=-i\,\langle\mathcal{T}_{*}\,\hat{h}_{ij}(x)\hat{h}_{kl}(x')\rangle_{\hat{\varrho}_{\textsc{g}}}\,.
\end{align}
Putting these results back to \eqref{E:dbkghd} and after some algebraic manipulations, we obtain
\begin{align}
	Z[\bm{z}^{(+)},\bm{z}^{(-)}]&=\exp\biggl\{-32\pi\int_{t_{a}}^{t_{b}}\!d^{4}x\int_{t_{a}}^{t_{b}}\!d^{4}x'\;\Delta_{(J)}^{ij}(x)\times\theta(t-t')\langle\bigl[\hat{h}_{ij}(x),\,\hat{h}_{kl}(x')\bigr]\rangle_{\hat{\varrho}_{\textsc{g}}}\times\Sigma_{(J)}^{kl}(x')\biggr.\notag\\
	&\qquad\qquad\quad-\biggl.16\pi\int_{t_{a}}^{t_{b}}\!d^{4}x\int_{t_{a}}^{t_{b}}\!d^{4}x'\;\Delta_{(J)}^{ij}(x)\times\frac{1}{2}\langle\bigl\{\hat{h}_{ij}(x),\,\hat{h}_{kl}(x')\bigr\}\rangle_{\hat{\varrho}_{\textsc{g}}}\times\Delta_{(J)}^{kl}(x')\biggr\}\,,
\end{align}
where we have introduced
\begin{align}\label{E:bvthrt}
	\Sigma_{(J)}^{ij}(x)&=\frac{J^{ij}_{(+)}(x)+J^{ij}_{(-)}(x)}{2}\,,&\Delta_{(J)}^{ij}(x)&=J^{ij}_{(+)}(x)-J^{ij}_{(-)}(x)\,,
\end{align}
and used the identities among the various propagators and commutators of the generic quantum field $\hat{\phi}$ 
\begin{align}
	\langle\mathcal{T}\,\hat{\phi}(x)\hat{\phi}(x')\rangle-\langle\mathcal{T}_{<}\,\hat{\phi}(x)\hat{\phi}(x')\rangle-\langle\mathcal{T}_{>}\,\hat{\phi}(x)\hat{\phi}(x')\rangle+\langle\mathcal{T}_{*}\,\hat{\phi}(x)\hat{\phi}(x')\rangle&=0\,,\\
	\langle\mathcal{T}\,\hat{\phi}(x)\hat{\phi}(x')\rangle+\langle\mathcal{T}_{<}\,\hat{\phi}(x)\hat{\phi}(x')\rangle+\langle\mathcal{T}_{>}\,\hat{\phi}(x)\hat{\phi}(x')\rangle+\langle\mathcal{T}_{*}\,\hat{\phi}(x)\hat{\phi}(x')\rangle&=2\langle\bigl\{\hat{\phi}(x),\,\hat{\phi}(x')\bigr\}\rangle\,,\\
	\langle\mathcal{T}\,\hat{\phi}(x)\hat{\phi}(x')\rangle-\langle\mathcal{T}_{<}\,\hat{\phi}(x)\hat{\phi}(x')\rangle+\langle\mathcal{T}_{>}\,\hat{\phi}(x)\hat{\phi}(x')\rangle-\langle\mathcal{T}_{*}\,\hat{\phi}(x)\hat{\phi}(x')\rangle&=2\theta(t-t')\langle\bigl[\hat{\phi}(x),\,\hat{\phi}(x')\bigr]\rangle\,,\\
	\langle\mathcal{T}\,\hat{\phi}(x)\hat{\phi}(x')\rangle+\langle\mathcal{T}_{<}\,\hat{\phi}(x)\hat{\phi}(x')\rangle-\langle\mathcal{T}_{>}\,\hat{\phi}(x)\hat{\phi}(x')\rangle-\langle\mathcal{T}_{*}\,\hat{\phi}(x)\hat{\phi}(x')\rangle&=2\theta(t'-t)\langle\bigl[\hat{\phi}(x'),\,\hat{\phi}(x)\bigr]\rangle\,.
\end{align}
Here $\theta(t-t')$ is the Heaviside unit-step function, and $[\dots]$, $\{\dots\}$ represent the commutator and the anti-commutator respectively.

Once we restore the definition of the current source by \eqref{E:rpij}, the influence functional $S_{\textsc{if}}[\bm{z}^{(+)},\bm{z}^{(-)}]$ becomes 
\begin{align}
	S_{\textsc{if}}[\bm{z}^{(+)},\bm{z}^{(-)}]&=i\,32\pi\int_{t_{a}}^{t_{b}}\!d^{4}x\int_{t_{a}}^{t_{b}}\!d^{4}x'\;\Delta_{(J)}^{ij}(x)\times\theta(t-t')\langle\bigl[\hat{h}_{ij}(x),\,\hat{h}_{kl}(x')\bigr]\rangle_{\hat{\varrho}_{\textsc{g}}}\times\Sigma_{(J)}^{kl}(x')\notag\\
	&\qquad\qquad\quad+i\,16\pi\int_{t_{a}}^{t_{b}}\!d^{4}x\int_{t_{a}}^{t_{b}}\!d^{4}x'\;\Delta_{(J)}^{ij}(x)\times\frac{1}{2}\langle\bigl\{\hat{h}_{ij}(x),\,\hat{h}_{kl}(x')\bigr\}\rangle_{\hat{\varrho}_{\textsc{g}}}\times\Delta_{(J)}^{kl}(x')\,.
\end{align}
The first kernel functions in this expression is of the retarded nature due to the presence of the unit-step function, and is understood as the dissipation kernel. The second kernel function is called the noise kernel because by construction it is related to the correlation function of the noise force on the righthand side of the Langevin equation~\eqref{E:pwoejdnsa}.

Now we would like to relate these two kernel functions with the suitable Green's functions. We first observe that for a generic linear quantum field $\hat{\phi}(\bm{x},t)$, whose equal-time commutation relation is given by $[\hat{\phi}(t,\bm{x}),\,\hat{\phi}(t,\bm{x})]=i\,\delta^{(3)}(\bm{x}-\bm{x}')$, we have
\begin{equation}
	\square\,\langle\mathcal{T}\,\hat{\phi}(x)\hat{\phi}(x')\rangle=i\,\delta^{(4)}(x,x')\,,
\end{equation}
with $x=(t,\bm{x})$ here. If we would like to define the retarded Green's function $G_{R,0^{(\phi)}}(x,x')$ and the Feynman propagator $G_{F,0^{(\phi)}}(x,x')$ of this free field in such a way that $\square_{x}G_{\textsc{r},0}^{(\phi)}(x,x')=\delta^{(4)}(x,x')$, then it implies we may define the corresponding retarded Green's functions of the free gravitons by
\begin{equation}
	G_{R,0}^{(h)}{}_{\,ijkl}(x,x')=-i\,\theta(t-t')\langle\bigl[\hat{h}_{ij}(x),\,\hat{h}_{kl}(x')\bigr]\rangle_{\hat{\varrho}_{\textsc{g}}}\,,
\end{equation}
and the Hadamard function of the free gravitons by
\begin{equation}
	G_{H,0}^{(h)}{}_{\,ijkl}(x,x')=\frac{1}{2}\langle\bigl\{\hat{h}_{ij}(x),\,\hat{h}_{kl}(x')\bigr\}\rangle_{\hat{\varrho}_{\textsc{g}}}\,.
\end{equation}
In terms of these two Green's function, the influence functional takes the form
\begin{align}\label{E:gbksdt}
	S_{\textsc{if}}[\bm{z}^{(+)},\bm{z}^{(-)}]&=-32\pi\int_{t_{a}}^{t_{b}}\!d^{4}x\int_{t_{a}}^{t_{b}}\!d^{4}x'\;\Delta_{(J)}^{ij}(x)\,G_{R,0}^{(h)}{}_{\,ijkl}(x,x')\,\Sigma_{(J)}^{kl}(x')\notag\\
	&\qquad\qquad\quad+i\,16\pi\int_{t_{a}}^{t_{b}}\!d^{4}x\int_{t_{a}}^{t_{b}}\!d^{4}x'\;\Delta_{(J)}^{ij}(x)\,G_{H,0}^{(h)}{}_{\,ijkl}(x,x')\,\Delta_{(J)}^{kl}(x')\,, 
\end{align}
and thus the coarse-grained effective action $S_{\textsc{cg}}[\bm{z}^{(+)},\bm{z}^{(-)}]$ takes the form
\begin{align}
	S_{\textsc{cg}}[\Delta_{(z)},\Sigma_{(z)}]&=\int_{t_{a}}^{t_{b}}\!dt\;m\,\eta_{ij}\,\dot{\Delta}^{i}_{(z)}(t_{b})\dot{\Sigma}^{j}_{(z)}(t_{b})-32\pi\int_{t_{a}}^{t_{b}}\!d^{4}x\int_{t_{a}}^{t_{b}}\!d^{4}x'\;\Delta_{(J)}^{ij}(x)\,G_{R,0}^{(h)}{}_{\,ijkl}(x,x')\,\Sigma_{(J)}^{kl}(x')\notag\\
	&\qquad\qquad\quad+i\,16\pi\int_{t_{a}}^{t_{b}}\!d^{4}x\int_{t_{a}}^{t_{b}}\!d^{4}x'\;\Delta_{(J)}^{ij}(x)\,G_{H,0}^{(h)}{}_{\,ijkl}(x,x')\,\Delta_{(J)}^{kl}(x')\,,
\end{align}
similar to \eqref{E:bvthrt}, we define
\begin{align}
	\Sigma_{(z)}^{i}(t)&=\frac{z^{i}_{(+)}(t)+z^{i}_{(-)}(t)}{2}\,,&\Delta_{(z)}^{i}(t)&=z^{i}_{(+)}(t)-z^{i}_{(-)}(t)\,.
\end{align}
Then the reduced density matrix elements of geodesic deviation \eqref{E:gfkdgbf} is reduced to
\begin{equation}\label{E:bvtrerhrt}
	\hat{\varrho}_{\textsc{m}}(z_{b},z'_{b};t_{b})=\int_{-\infty}^{\infty}\!dz_{a}\int_{-\infty}^{\infty}\!dz'_{a}\;\hat{\varrho}_{\textsc{m}}(z_{a},z'_{a};t_{a})\int_{z_{a}}^{z_{b}}\!\mathcal{D}z^{(+)}\!\int_{z'_{a}}^{z'_{b}}\!\mathcal{D}z^{(-)}\;\exp\Bigl\{i\,S_{\textsc{cg}}[\bm{z}^{(+)},\bm{z}^{(-)}]\Bigr\}\,.
\end{equation}
Since the coarse-grained effective action in the exponential is cubic in $z^{i}_{(\pm)}$, the path integrals in \eqref{E:bvtrerhrt} can only evaluated perturbatively. This will be the starting expression to study the influence of gravitons on geodesic deviation, or to derive the master equation of geodesic deviation.

In practice, when the metric perturbation can be identified with two copies of minimally coupled scalar fields, we may write the influence functional \eqref{E:gbksdt} by purely scalar quantities without any tensorial contractions involved. To do so, we need the results about the polarization decomposition in Sec.~\ref{S:wurgfsb}. Suppose generically the quantized gravitational wave takes the form
\begin{equation}\label{E:eppwe}
	h_{\mu\nu}(t,\bm{x})=\frac{1}{V}\sum_{\bm{k},\lambda}\mathsf{e}_{\mu\nu}^{(s)}(\bm{k})\,h_{s}(\bm{k},x^{\mu})\,,
\end{equation}
in the Minkowski background, where $\lambda$ labels the polarizations and assumes the values $+$, $\times$. In addition, we suppose that $h_{s}(\bm{k},x^{\mu})$ has the form
\begin{equation}\label{E:jdgjsd}
	h_{\lambda}(\bm{k},x^{\mu})=e^{i\bm{k}\cdot\bm{z}}h_{\lambda}(\bm{k},t)\,.
\end{equation}
Then we can write the action
\begin{equation}
	S=-\frac{1}{64\pi}\int\!d^{4}x\;\partial_{\alpha}h_{\mu\nu}(x)\partial^{\alpha}h^{\mu\nu}(x)
\end{equation}
into
\begin{align}\label{E:etihe}
	S\&=-\frac{1}{64\pi V^{2}}\int\!dt\;\sum_{\bm{k},\lambda}\sum_{\bm{k}',\lambda'}\mathsf{e}_{\mu\nu}^{(\lambda)}(\bm{k})\,\mathsf{e}^{(\lambda')\,\mu\nu}(\bm{k}')\int\!d^{3}z\;\partial_{\alpha}h_{\lambda}(\bm{k},x^{\mu})\,\partial^{\alpha}h_{\lambda'}(\bm{k}',x^{\mu})\,.
\end{align}
The spatial integral in \eqref{E:etihe} has the generic form
\begin{align}
	\int\!d^{3}z\;A_{\lambda}(\bm{k},x^{\mu})B_{\lambda'}(\bm{k}',x^{\mu})&=(2\pi)^{3}\,\delta(\bm{k}+\bm{k}')\,A_{\lambda}(\bm{k},t)\,B_{\lambda'}(\bm{k}',t)\,,
\end{align}
if we assume that each factor can be decomposed into 
\begin{align}\label{E:ebgires}
	A_{\lambda}(\bm{k},x^{\mu})&=e^{i\bm{k}\cdot\bm{z}}\,A_{\lambda}(\bm{k},t)\,, &&\text{and} &B_{\lambda'}(\bm{k}',x^{\mu})&=e^{i\bm{k}'\cdot\bm{z}}\,B_{\lambda'}(\bm{k}',t)\,.
\end{align}
In addition, from \eqref{E:gbkdeter}, we have
\begin{equation}
	\mathsf{e}_{\mu\nu}^{(\lambda)}(\bm{k})\,\mathsf{e}^{(\lambda')\,\mu\nu}(\bm{k})=\delta_{\lambda\lambda'}\,.
\end{equation}
Then Eq.~\eqref{E:etihe} becomes
\begin{align}
	S&=-\frac{1}{64\pi}\int\!dt\;\sum_{\lambda,\lambda'}\int\!\frac{d^{3}\bm{k}}{(2\pi)^{3}}\;\mathsf{e}_{\mu\nu}^{(\lambda)}(\bm{k})\,\mathsf{e}^{(\lambda')\,\mu\nu}(-\bm{k})\,h_{\lambda,\alpha}^{\vphantom{*}}(\bm{k},t)\,h_{\lambda',}{}^{\alpha}(-\bm{k},t)\,,
\end{align}
where $h_{\lambda,\alpha}(\bm{k},t)$ is the formal notation for the expression we obtained after we have computed $\partial_{\alpha}h_{\lambda}(\bm{k},x^{\mu})$ and made the decomposition like \eqref{E:ebgires}. If we suppose $h_{\mu\nu}(x^{\mu})$ is real, and the polarizations are  linear, i.e., $+$ and $\times$, then we have $\mathsf{e}_{\mu\nu}^{(\lambda)}(\bm{k})=\mathsf{e}_{\mu\nu}^{(\lambda)*}(\bm{k})$, and from \eqref{E:eppwe} and \eqref{E:jdgjsd} we find
\begin{align}
	h^{*}_{\mu\nu}(x^{\mu})=\sum_{\bm{k},\lambda}\mathsf{e}_{\mu\nu}^{(\lambda)*}(\bm{k})\,h_{\lambda}^{*}(\bm{k},x^{\mu})=\sum_{\bm{k},\lambda}\mathsf{e}_{\mu\nu}^{(\lambda)}(\bm{k})\,h_{\lambda}(-\bm{k},x^{\mu})=\sum_{\bm{k},\lambda}\mathsf{e}_{\mu\nu}^{(\lambda)}(-\bm{k})\,h_{\lambda}(\bm{k},x^{\mu})=h_{\mu\nu}(x^{\mu})\,,
\end{align}
so that $\mathsf{e}_{\mu\nu}^{(\lambda)}(-\bm{k})=\mathsf{e}_{\mu\nu}^{(\lambda)}(\bm{k})$. It implies that
\begin{align}\label{E:rbkgkrt}
	S&=-\frac{1}{64\pi}\int\!d^{4}x\;\frac{1}{V}\sum_{\bm{k},\lambda}h_{\lambda,\alpha}^{\vphantom{*}}(\bm{k},t)\,h_{\lambda,}^{*}{}^{\alpha}(\bm{k},t)\,,
\end{align}
where $V$ is integration volume and can be identified with $V=\displaystyle\int\!d^{3}z$, and
\begin{equation}
	\sum_{\bm{k}}=V\int\!\frac{d^{3}\bm{k}}{(2\pi)^{3}}\,.
\end{equation}
If we define a scalar field
\begin{equation}\label{E:brvdfs}
	h_{\lambda}(x^{\mu})=\frac{1}{V}\sum_{\bm{k}}h_{\lambda}(\bm{k},x^{\mu})\,,
\end{equation}
and also assume that $h_{\lambda}(\bm{k},x^{\mu})=e^{i\bm{k}\cdot\bm{z}}\,h_{\lambda}(\bm{k},t)$, then we can show that the action $S'$
\begin{equation}
	S'=\frac{1}{4}\int\!d^{4}x\;\sum_{\lambda}\partial_{\alpha}h_{\lambda}(x^{\mu})\,\partial^{\alpha}h_{\lambda}(x^{\mu})\,,
\end{equation}
can be cast into
\begin{align}
	S'&=-\frac{1}{64\pi}\int\!d^{4}x\;\frac{1}{V}\sum_{\bm{k},\lambda}h_{\lambda,\alpha}(\bm{k},t)\,h_{\lambda,}^{*}{}^{\alpha}(\bm{k},t)\,.
\end{align}
It is identical to \eqref{E:rbkgkrt}, so both actions $S$ and $S'$ are equivalent and indeed $h_{\mu\nu}(x^{\mu})$ can be identified with two scalar fields $h_{\lambda}(x^{\mu})$ with $\lambda=+$ and $\times$.

The interaction term generically has the form
\begin{equation}
	S_{\textsc{int}}=\int\!d^{4}x\;J^{\mu\nu}(x^{\mu})\,h_{\mu\nu}(x^{\mu})\,.
\end{equation}
We take the plane-wave expansion
\begin{align}
	J^{\mu\nu}(x^{\mu})=\frac{1}{V}\sum_{\bm{k}}J^{\mu\nu}(\bm{k},x^{\mu})=\frac{1}{V}\sum_{\bm{k}}e^{i\bm{k}\cdot\bm{z}}J^{\mu\nu}(\bm{k},t)\,,
\end{align}
and arrive at
\begin{align}\label{E:peorhds}
	S_{\textsc{int}}&=\int\!d^{4}x\;\frac{1}{V}\sum_{\bm{k},\lambda}h_{\lambda}(\bm{k},t)\,J_{\lambda}^{*}(\bm{k},t)\,,
\end{align}
where we assume the current $J_{\mu\nu}(x^{\mu})$ is real and let
\begin{equation}
	J^{*}_{\lambda}(\bm{k},t)=\mathsf{e}_{\mu\nu}^{(\lambda)}(\bm{k})\,J^{\mu\nu}(-\bm{k},t)\,.
\end{equation}
Eq.~\eqref{E:peorhds} can be likewise obtained from the interaction
\begin{equation}\label{E:bskbdf}
	S'_{\textsc{int}}=\int\!d^{4}x\;\sum_{\lambda}J_{\lambda}(x^{\mu})\,h_{\lambda}(x^{\mu})
\end{equation}
with
\begin{equation}
	J_{\lambda}(x^{\mu})=\frac{1}{V}\sum_{\bm{k}}e^{i\bm{k}\cdot\bm{z}}J_{\lambda}(\bm{k},t)=\int\!\frac{d^{3}\bm{k}}{(2\pi)^{3}}\;\mathsf{e}_{\mu\nu}^{(\lambda)}(\bm{k})\,J^{\mu\nu}(\bm{k},t)\,e^{i\bm{k}\cdot\bm{z}}\,,
\end{equation}
and $h_{\lambda}(x^{\mu})$ given by \eqref{E:brvdfs}. This can be seen as follows
\begin{align}
	S'_{\textsc{int}}=\frac{1}{V^{2}}\int\!dt\sum_{\lambda}\sum_{\bm{k},\bm{k}'}\int\!d^{3}x\;e^{i(\bm{k}+\bm{k}')\cdot\bm{z}}h_{\lambda}(\bm{k},t)\,J_{\lambda}(\bm{k}',t)&=\int\!d^{4}x\;\frac{1}{V}\sum_{\bm{k},\lambda}h_{\lambda}(\bm{k},t)\,J_{\lambda}^{*}(\bm{k},t)\,.
\end{align}
This is exactly Eq.~\eqref{E:peorhds}.

In the gauge we choose, the interaction action \eqref{E:bskbdf} can be rewritten as
\begin{align}\label{E:eiugvsda}
	S'_{\textsc{int}}&=\int\!d^{4}x\;\sum_{\lambda}\int\!\frac{d^{3}\bm{k}}{(2\pi)^{3}}\;\mathsf{e}_{ij}^{(\lambda)}(\bm{k})\,J^{ij}(\bm{k},t)\,e^{i\bm{k}\cdot\bm{z}}\,h_{\lambda}(x^{\mu})\notag\\
	&=\frac{m}{4}\int\!d^{4}x\;\sum_{\lambda}h_{\lambda}(x^{\mu})\int\!\frac{d^{3}\bm{k}}{(2\pi)^{3}}\;\frac{\partial^{2}}{\partial t^{2}}\Bigl[\mathsf{e}_{ij}^{(\lambda)}(\bm{k})\,z^i(t)z^j(t)\Bigr]e^{i\bm{k}\cdot\bm{z}}\,.
\end{align}
Since according to \eqref{E:rpij}, we can write $J^{ij}(x^{\mu})$ as
\begin{align}
	J^{ij}(x^{\mu})=\int\!\frac{d^{3}\bm{k}}{(2\pi)^{3}}\;e^{i\bm{k}\cdot\bm{z}}\,\frac{m}{4}\frac{\partial^{2}}{\partial t^{2}}\Bigl[z^i(t)z^j(t)\Bigr]\,,
\end{align}
we find
\begin{align}
	J^{ij}(\bm{k},t)=\frac{m}{4}\frac{\partial^{2}}{\partial t^{2}}\Bigl[z^i(t)z^j(t)\Bigr]=\mathcal{J}^{ij}(t)\,,
\end{align}
independent of $\bm{k}$. Then Eq.~\eqref{E:eiugvsda} can be expressed as
\begin{equation}
	S'_{\textsc{int}}=\frac{m}{4}\int\!d^{4}x\;\mathcal{J}^{ij}(t)\sum_{\lambda}h_{\lambda}(x^{\mu})\int\!\frac{d^{3}\bm{k}}{(2\pi)^{3}}\;\mathsf{e}_{ij}^{(\lambda)}(\bm{k})\,e^{i\bm{k}\cdot\bm{z}}\,,
\end{equation}
and this serves as the starting point for an alternative expression of the influence functional (9) in~\cite{{ChoHu23}}. For example, the noise kernel $N_{ijkl}(t,t')$ is given by
\begin{align}
	N_{ijkl}(t,t')=\biggl(\frac{m}{4}\biggr)^{2}\!\int\!d^{3}z\!\int\!d^{3}z'\!\int\!\frac{d^{3}\bm{k}}{(2\pi)^{3}}\!\int\!\frac{d^{3}\bm{k}'}{(2\pi)^{3}}\;\,e^{i\bm{k}\cdot\bm{z}+i\bm{k}'\cdot\bm{z}'}\sum_{\lambda}\mathsf{e}_{ij}^{(\lambda)}(\bm{k})\,\mathsf{e}_{kl}^{(\lambda)}(\bm{k}')\,G_{H,0}^{(\lambda)}(x^{\mu},x'^{\mu})\,,
\end{align}
where 
\begin{equation}
	G_{H,0}^{(\lambda)}(x^{\mu},x'^{\mu})=\frac{1}{2}\operatorname{Tr}\Bigl(\varrho^{(\lambda)}\bigl\{h_{\lambda}(x^{\mu}),\,h_{\lambda}(x'^{\mu})\bigr\}\Bigr)	
\end{equation}
is the Hadamard function of the free scalar field $h_{\lambda}(x^{\mu})$. Here $\varrho^{(\lambda)}$ is the density operator of the initial state of the scalar field $h_{\lambda}(x^{\mu})$. The advantage of this expression is that we can express the dissipation and the noise kernels of gravitons respectively in terms of the dissipation and the noise kernels of the scalar fields.

 \end{document}